\documentclass[english,rmp,aps,superscriptaddress,showkeys,showpacs,prepri,nofootinbib]{revtex4-1}

\usepackage{booktabs}
\usepackage{siunitx}

\pdfoutput=1
\usepackage[T1]{fontenc}
\usepackage[letterpaper]{geometry}
\usepackage{tikz}
\usepackage{tikz-layers}
\usetikzlibrary{calc,trees,positioning,arrows,chains,shapes.geometric,math,%
decorations.pathreplacing,decorations.pathmorphing,shapes,%
matrix,shapes.symbols,plotmarks,decorations.markings,shadows}
\usepackage{units}
\usepackage{bbding}
\usepackage{amsmath}
\usepackage{amssymb}
\usepackage{fixmath}
\usepackage{graphicx}
\usepackage{esint}
\usepackage{tensor}
\usepackage[colorinlistoftodos,prependcaption,textsize=tiny]{todonotes}

\usepackage[utf8]{inputenc}
\usepackage{lmodern} 

\makeatletter

\AtBeginDocument{
\heavyrulewidth=.08em
\lightrulewidth=.05em
\cmidrulewidth=.03em
\belowrulesep=.65ex
\belowbottomsep=0pt
\aboverulesep=.4ex
\abovetopsep=0pt
\cmidrulesep=\doublerulesep
\cmidrulekern=.5em
\defaultaddspace=.5em
}

\@ifundefined{textcolor}{}
{%
 \definecolor{BLACK}{gray}{0}
 \definecolor{WHITE}{gray}{1}
 \definecolor{RED}{rgb}{1,0,0}
 \definecolor{GREEN}{rgb}{0,1,0}
 \definecolor{BLUE}{rgb}{0,0,1}
 \definecolor{CYAN}{cmyk}{1,0,0,0}
 \definecolor{MAGENTA}{cmyk}{0,1,0,0}
 \definecolor{YELLOW}{cmyk}{0,0,1,0}
}
\usepackage{hyperref}

\makeatother

\usepackage{babel}

  \def\refeq#1{Eq.~(\ref{#1})}
  \def\refeqs#1{Eqs.~(\ref{#1})}
  \def\reffig#1{Figure~\ref{#1}}
  \def\reftab#1{Table~\ref{#1}}
  \def\refsec#1{Section~\ref{#1}}

  \newcolumntype{C}{>{$}c<{$}}
  \newcolumntype{L}{>{$}l<{$}}
  \newcolumntype{R}{>{$}r<{$}}

  \newenvironment{eqaligned}
    {\begin{equation}
    \begin{aligned}
    }
    { 
    \end{aligned}
    \end{equation}\ignorespacesafterend
    }

  \newcommand{\vect}[1]{ \mathbold{#1}}  

  \newcommand{\lp}{\left(}
  \newcommand{\rp}{\right)}
  \newcommand{\lb}{\left[}
  \newcommand{\rb}{\right]}

  \newcommand{\vx}{\vect{x}}
  
  \newcommand{\vB}{\vect{B}}
  \newcommand{\vBp}{\vect{B}^\prime}
  \newcommand{\vBhtp}{\hat{\vect{B}}^\prime}
  \newcommand{\vJ}{\vect{J}}
  \newcommand{\vJp}{\vect{J}^\prime}
  \newcommand{\vJhtp}{\hat{\vect{J}}^\prime}
  \newcommand{\rhop}{\rho^\prime}
  \newcommand{\rhohtp}{\hat{\rho}^\prime}
  \newcommand{\vA}{\vect{A}}
  \newcommand{\vE}{\vect{E}}
  
  \newcommand{\vEL}{\vect{E_L}}
  \newcommand{\vEp}{\vect{E}^\prime}
  \newcommand{\vEhtp}{\hat{\vect{E}}^\prime}
  \newcommand{\vD}{\vect{D}}
  \newcommand{\vDp}{\vect{D}^\prime}
  \newcommand{\vp}{\vect{p}}
  \newcommand{\vP}{\vect{P}}
  \newcommand{\vH}{\vect{H}}
  \newcommand{\vHp}{\vect{H}^\prime}
  \newcommand{\vM}{\vect{M}}
  \newcommand{\vV}{\vect{V}}

  \newcommand{\vv}{\vect{v}}
  \newcommand{\vF}{ \vect{F} }

  \newcommand{\eps}{\epsilon}
  
  \newcommand{\epsbr}{\bar{\epsilon}}
  \newcommand{\epso}{\epsilon_0}
  \newcommand{\muo}{\mu_0}
  
  \newcommand{\mubr}{\bar{\mu}}

  \newcommand{\rbr}{\bar{\rho}}
  \newcommand{\rht}{\hat{\rho}}
  \newcommand{\rdt}{\dot{\rho}}
  
  \newcommand{\vBbr}{\bar{B}}
  \newcommand{\vJbr}{\bar{J}}
  \newcommand{\vAbr}{\bar{A}}
  \newcommand{\vEbr}{\bar{E}}
  \newcommand{\vETbr}{\bar{E}_T}
  \newcommand{\vELbr}{\bar{E}_L}

  \newcommand{\phibr}{\bar{\phi}}
  
  \newcommand{\tht}{\hat{t}}

  \newcommand{\rhoht}{\hat{\rho}}
  \newcommand{\phiht}{\hat{\phi}}
  \newcommand{\vxht}{\hat{\vect{x}}}
  \newcommand{\xph}{\vxht^\prime}				
  
  \newcommand{\vBht}{\hat{\vect{B}}}
  \newcommand{\vHht}{\hat{\vect{H}}}
  \newcommand{\vAht}{\hat{\vect{A}}}
  \newcommand{\vJht}{\hat{\vect{J}}}
  \newcommand{\vJdt}{\dot{\vect{\vJht}}}
  \newcommand{\vEht}{\hat{\vect{E}}}

  \newcommand{\vDht}{\hat{\vect{D}}}
  \newcommand{\vVht}{\hat{\vect{V}}}
  
  \newcommand{\lunit}{L}
  \newcommand{\xunit}{\lunit}
  \newcommand{\tunit}{T}
  \newcommand{\vxbr}{\bar{L}}
  \newcommand{\xbr}{\vxbr}
  \newcommand{\lbr}{\bar{L}}
  \newcommand{\tbr}{\bar{T}}
  \newcommand{\qunit}{Q}
  \newcommand{\munit}{M}
  \newcommand{\Eunit}{\lb{E}\rb}
  \newcommand{\Bunit}{\lb{B}\rb}
  \newcommand{\Junit}{\lb{J}\rb}
  \newcommand{\runit}{\lb{\rho}\rb}
  \newcommand{\kounit}{\lb{k_1}\rb}
  
  \newcommand{\khunit}{\lb{k_3}\rb}
  
  \newcommand{\aht}{\hat{\alpha}}
  \newcommand{\abr}{\bar{\alpha}}

  \newcommand{\bigO}[1] {\mathcal{O}\lp #1 \rp}
  \newcommand{\EoB}{ \frac{\vEbr}   {\vBbr}}
  \newcommand{\BoE}{ \frac{\vBbr} {\vEbr}}
  \newcommand{\EToB}{ \frac{\vETbr}   {\vBbr}}
  \newcommand{\ELoB}{ \frac{\vELbr}   {\vBbr}}

  \newcommand{\BoET}{ \frac{\vBbr} {\vETbr}}
  
  \newcommand{\roJ}{ \frac{c \rbr}  {\vJbr}}
  \newcommand{\Jor}{ \frac{\vJbr}   {c \rbr}}
  \newcommand{\poA}{ \frac{\phibr}  {\vAbr}}
  \newcommand{\Aop}{ \frac{\vAbr} {\phibr}}
  
  \newcommand{\cocm}{ \frac{c^2}{c_m^2}}
  \newcommand{\tauem}{ \tau_{EM}}
  \newcommand{\taue}{ \tau_{E}}
  \newcommand{\taum}{ \tau_{M}}

  \newcommand{\betaosc}{\beta_{\text{osc}}}
  \newcommand{\betainrt}{\beta_{\text{inertial}}}
  \newcommand{\Nosc}{N_{\text{osc}}}

  \newcommand{\xp}{\vx^\prime}				
  \newcommand{\tp}{t^\prime}				
  \newcommand{\thtp}{\hat{t}^\prime}				

  \newcommand{\rbret}{\rb_{\text{ret}}}
  \newcommand{\dSp}{d\vect{S}^\prime}				
  \newcommand{\dVp}{d^3x^\prime}				
  				
  \newcommand{\vRht}{\hat{\vect{R}}}
  
  \newcommand{\dthp}[1]{\frac{\partial #1}{\partial \thtp}}
  \newcommand{\vbeta}{\vect{\beta}}
  \newcommand{\vbetb}{\frac{\vbeta}{\beta}}
  \newcommand{\vl}{ \vect{l} }
  \newcommand{\vS}{ \vect{S} }

  \newcommand{\rhopdep}{\rho\lp\xp,t\rp}
  \newcommand{\vJpdep}{\vJ\lp\xp,t\rp}
  \newcommand{\vJdpdep}{\dot{\vJ}\lp\xp,t\rp}
  \newcommand{\vJTpdep}{\vJ_T\lp\xp,t\rp}

  \newcommand{\grad}{\vect{\nabla}}
  \newcommand{\gradh}{\hat{\vect{\nabla}}}
  \newcommand{\gradp}{\grad^\prime}				
  \newcommand{\gradhtp}{\gradh^\prime}				
  \newcommand{\curl}[1]{\grad \times #1 }
  \newcommand{\curlh}[1]{\hat{\grad} \times #1 }
  \newcommand{\curlp}[1]{\gradp \times #1 }
  
  \newcommand{\dive}[1]{\grad \cdot #1 }
  \newcommand{\divep}[1]{\gradp \cdot #1 }
  \newcommand{\diveh}[1]{\hat{\grad} \cdot #1 }
  
  \newcommand{\gradsq}{\nabla^2}
  \newcommand{\gradsqh}{\hat{\nabla}^2}

  \newcommand{\dt}[1]{\frac{\partial #1}{\partial t}}
  \newcommand{\dth}[1]{\frac{\partial #1}{\partial \hat{t}}}
  \newcommand{\Dt}[1]{\frac{d #1}{dt}}

  \newcommand{\dtsq}[1]{\frac{\partial^2 #1}{\partial t^2}}
  \newcommand{\dtsqh}[1]{\frac{\partial^2 #1}{\partial \hat{t}^2}}

  \newcommand{\vR}{\vect{R}}

\begin{document}

\title{The Three Quasi-Static Limits of the Maxwell Equations}

\author{S. E. Kruger}
\affiliation{Tech-X Corporation, 5621 Arapahoe Ave. Boulder, CO 80303, USA}

\date{\today}

\begin{abstract}

It is shown that the Galilean limit ($V \ll c$, or $\lunit/\tunit \ll c$)) of the
Maxwell equations admits three different limits: the magneto-quasi-static,
electro-quasi-static, and electromagnetic-quasi-static limits, in addition to
the two obvious static limits.  The first two quasi-static limits have been
previously identified as Galilean Electromagnetics, while the latter is also
known as the Darwin approximation.  Using a perturbation expansion, a
generalization of Rappetti and Rousseaux [Applied Numerical Mathematics, {\bf
79}, 92] orders the vacuum Maxwell equations and obtains all three limits. To
order the equations, the dimensionless version of the Maxwell equations are
derived using a modification of Jackson's review of EM unit systems [Jackson,
{\em Classical Electrodynamics}, Wiley, 1999, 3rd ed.] The perturbation
expansion is repeated for the potential form of the Maxwell equations to
emphasize the importance of gauge conditions.  The integral solutions of the
potentials are derived for the three limits, and the generalized Coulomb and
Biot-Savart equations are derived from these solutions.  It is shown that
although the forms are the same as the static equations, the quasi-static forms
of the Maxwell equations are recovered.  The induction term is recovered when
the time derivative of the vector potential is kept.  The displacement current
is recovered when the Lorenz gauge is used.  The equivalence of this approach
and Jackson's derivation [Amer.\ J.\ of Phys., {\bf 70}, 917 (2002)] of the
Darwin approximation is shown.  The regions of applicability of the quasi-static
forms of the Maxwell equations are discussed in terms of macroscopic media.  

\end{abstract}



\maketitle

\tableofcontents

\section{Introduction}
\label{sec:introduction}

The most common interaction people have with electromagnetic fields is turning
on a light switch.  When the switch is closed, a current at \SI{60}{Hz} (in the
U.S.) flows with an EM wavelength \SI{5E6}{m}, or more than $3/4$ of an Earth
radius.  The light produced by the light bulb has multiple frequencies around
\SI{600}{THz}, or a wavelength of \SI{5E-7}{m}.  The Maxwell equations describe
the physics of both the circuit and the light propagation, but in practice we
use approximations for each extreme of this 13 orders of magnitude range.  For
the high-frequency regime, an eikonal approximation yields the ray-tracing
equations.  These equations are so well-understood that they serve as the basis
for the multi-billion dollar video game and computer animation industries.  At
the other extreme is the quasi-static regime.  Although the traditional realm of
electrical engineering, this regime is less well understood, as evidenced by the
confusion around the different quasi-static limits.

One of the first derivations of quasi-static limits was published by Einstein
and Laub in a series of
papers~\cite{EinsteinLaub1908a,EinsteinLaub1908b,EinsteinLaub1908c,EinsteinLaub1908d}
motivated by several experiments of rotating media.  Some of the more prominent
examples of this are those showing the  Roentgen-Eichenwald effect
\cite{Roentgen1888,Roentgen1890,Eichenwald1903,Eichenwald1904}, the Wilson-Wilson
effect~\cite{wilson1913}, and Barnett effect~\cite{barnett1915}.  The
latter reference is a thorough review of all the experiments related to the
Barnett effect.  Rousseaux~\cite{Rousseaux:2013ft} provides a review that gives
a modern review of these experiments in the context of quasi-static
electromagnetics.  Because these experiments used rotating media in a
laboratory, the rotation speeds were much less than the speed of light.  The
Einstein-Laub papers published in 1908 served as a primary reference for ``low
velocity limits'' for physicists.  In 1920, Darwin~\cite{darwin} derived another
quasi-static limit by truncating the potentials of the far-field in studying
particle motion using Lagrangian mechanics.  In 1927, the relationship between
the Maxwell equations and circuit equations was elucidated by
Carson~\cite{carson1927} using the ratio of system size to wavelength as a small
parameter which, as will be shown, is closely related to quasi-static limits.

The first discussion of two distinct quasi-static limits, the
electro-quasi-static (EQS) and magneto-quasi-static (MQS) regimes, was published
by engineers in 1968~\cite{woodson-melcher1968}.  The first physics discussion
of these two regimes by physicists was published by Le Bellac and L\'evy-Leblond
(LBLL)~\cite{LBLL} in 1973.  They were motivated by a simple question:  If
mechanics has a limit where the equations become Galilean covariant, do the
Lorentz-covariant Maxwell equations also have a limit where they are Galilean
covariant?  Their ``Galilean Electromagnetics'' identified two incompatible
limits: the time-like limit with $E \gg c B$ (EQS), and the space-like limit
with $E \ll c B$ (MQS).  In more recent years, there has been a resurgence of
interest in Galilean electromagnetics.  A derivation using group theory was made
by de Montigny~\cite{demontigny2003}.  A derivation using an order of magnitude
analysis was given by Rousseax~\cite{rousseaux2003,rousseaux2004}.  More formal
derivations using a perturbation analysis with macroscopic media was made by
Manfredi~\cite{manfredi2013} and Rapetti and
Rousseaux~\cite{Rapetti:2011uq,Rapetti:2014ha}.  A 2013 review paper by
Rousseaux~\cite{Rousseaux:2013ft} provides an excellent overview of the
literature to that date, including both the equations and the applications to
experiments of rotating media.

These formal Galilean limits do not appear in the most popular physics textbooks.  In
Griffith's undergraduate textbook~\cite{Griffiths}, only the MQS limit is
given, and it is discussed as {\em the} quasi-static limit.  This is consistent
with the Einstein-Laub papers discussing only the MQS limit.
Jackson~\cite{JacksonSecond,JacksonThird} discusses the MQS limit, although
referred to as the quasi-static approximation, as well as the Darwin
approximation.  Rapetti and Rousseaux~\cite{Rapetti:2014ha} speculate that the
Darwin approximation is an electromagnetic quasi-static (EMQS) limit, but do
not include a formal derivation.   Thus, there is no comprehensive review of all
three quasi-static limits.

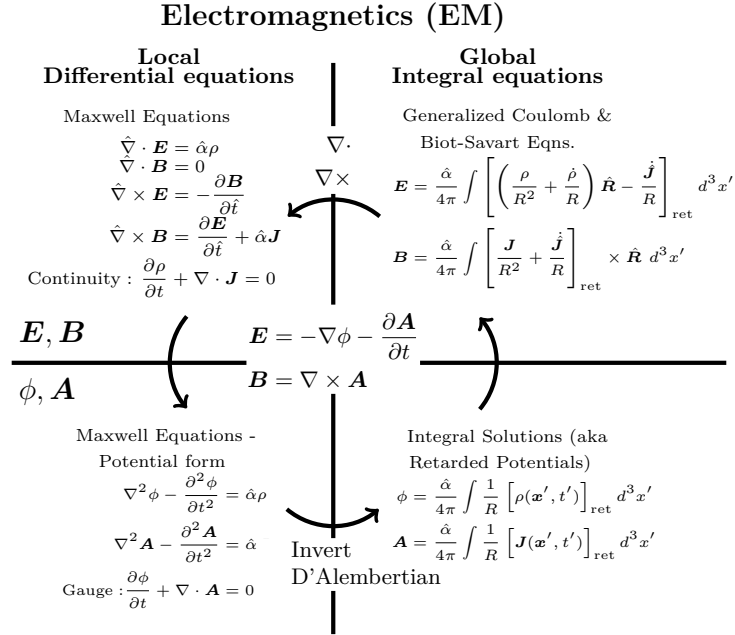
\begin{figure}
\begin{center}
\begin{tikzpicture}[thick,scale=0.95, every node/.style={scale=0.95}]

  \node (n99) [above] at (0, 4.500)%
               { \large \bf
                   Electromagnetics (EM)
               };
  \draw[ultra thick] (-4.5,0) -- (5.5,0);
  \draw[ultra thick] (0,-3.8) -- (0,4.2);

  \node (n11) [anchor=south east] at (-0.6, 0.8)%
               {
                 \begin{minipage}{98pt}
                       {\scriptsize Maxwell Equations}
                       \vspace{-0.09in}
                        \begingroup\makeatletter\def\f@size{7}\check@mathfonts
                        \def\maketag@@@#1{\hbox{\m@th\large\normalfont#1}}%
                     \begin{flalign}
                       \diveh{\vE} &= \aht \rho
                                                                  \nonumber \\[-2\jot]  
                       \diveh{\vB} &=0
                                                                  \nonumber \\[-2\jot]
                       \curlh{\vE} &= - \dth{\vB}
                                                                  \nonumber \\[-1\jot]
                       \curlh{\vB} &=   \dth{\vE} + \aht \vJ
                                                                  \nonumber \\[-1\jot]
                       {\rm Continuity:\ } \dt{\rho}&+ \dive{\vJ} = 0
                                                                      \nonumber 
                   \end{flalign}\endgroup
                 \end{minipage}
               };
  \node (n12) [anchor=north east] at (-0.6, -0.8)%
               {
                 \begin{minipage}{92pt}
                       {\scriptsize Maxwell Equations - Potential form}
                       \vspace{-0.1in}
                        \begingroup\makeatletter\def\f@size{6}\check@mathfonts
                        \def\maketag@@@#1{\hbox{\m@th\large\normalfont#1}}%
                     \begin{flalign}
                       \gradsq\phi - \dtsq{\phi} &= \aht \rho
                                                                  \nonumber \\
                       \gradsq\vA - \dtsq{\vA} &= \aht \vJ
                                                                  \nonumber \\
                       {\rm Gauge:} \dt{\phi}+ \dive{\vA} &= 0
                                                                  \nonumber
                   \end{flalign}\endgroup
                 \end{minipage}
               };
  \node (n13) [anchor=north west] at (0.7, -0.8)%
               {
                 \begin{minipage}{92pt}
                       {\scriptsize  Integral Solutions (aka Retarded Potentials) }
                       \vspace{-0.1in}
                        \begingroup\makeatletter\def\f@size{6}\check@mathfonts
                        \def\maketag@@@#1{\hbox{\m@th\large\normalfont#1}}%
                     \begin{flalign}
   \phi&= \frac{\aht}{4\pi} \int \frac{1}{R}\lb \rho(\xp,\tp) \rbret \dVp
                                                                  \nonumber \\
   \vA &= \frac{\aht}{4\pi} \int \frac{1}{R}\lb \vJ(\xp,\tp) \rbret \dVp 
                                                                  \nonumber
                   \end{flalign}\endgroup
                 \end{minipage}
               };
  \node (n14) [anchor=south west] at (0.7, 0.9)%
               {
                 \begin{minipage}{92pt}
                       {\scriptsize  Generalized Coulomb \& Biot-Savart Eqns. 
                        }
                       \vspace{-0.1in}
                        \begingroup\makeatletter\def\f@size{6}\check@mathfonts
                        \def\maketag@@@#1{\hbox{\m@th\large\normalfont#1}}%
                     \begin{flalign}
                       \vE & = \frac{\aht}{4\pi} \int \lb 
                                \lp \frac{\rho}{R^2}
                                  + \frac{\rdt}{R} \rp  \vRht
                                   - \frac{\vJdt}{R} 
                                   \rbret \dVp
                                   \nonumber \\
                       \vB & = \frac{\aht}{4\pi} \int
                                 \lb \frac{\vJ}{R^2} 
                                 + \frac{\vJdt}{R} \rbret 
                                  \times \vRht\  \dVp
                                                                  \nonumber
                   \end{flalign}\endgroup
                 \end{minipage}
               };

  \draw [->,ultra thick] (-2.05, 0.65) arc [radius=1.0, start angle=140, end angle= 220];
  \draw [->,ultra thick] ( 2.05,-0.65) arc [radius=1.0, start angle=-40, end angle=  40];
  \draw [->,ultra thick] ( 0.65, 2.05) arc [radius=1.0, start angle=50,  end angle= 130];
  \draw [->,ultra thick] (-0.65,-2.05) arc [radius=1.0, start angle=230, end angle= 310];

  \draw [fill=white,white] (-1.20,-0.42) rectangle (1.2,0.8);
  \node (n00) [] at (0.0, 0.29)%
               {
                 \begin{minipage}{0pt}
                   \begin{align}
                       \vE & = -\grad \phi - \dt{\vA}
                                                                  \nonumber \\
                       \vB & = \curl{\vA}
                                                                  \nonumber 
                   \end{align}
                 \end{minipage}
               };
  \draw [fill=white,white] (-0.28, 3.30) rectangle (0.28,2.38);
  \node (n00) [] at (0.0, 3.00)%
               {
                 \begin{minipage}{0pt}
                   \begin{align}
                     \dive{} 
                                                                  \nonumber \\
                     \curl{}
                                                                  \nonumber 
                   \end{align}
                 \end{minipage}
               };
  \draw [fill=white,white] (-1.06, -2.47) rectangle (0.50,-3.18);
  \node (n00) [] at (-0.59, -2.80)%
               {
                 \begin{minipage}{0pt}
                   Invert \\
                   D'Alembertian
                 \end{minipage}
               };

  \node (n00) [] at (-4.4, 0.40)%
               {
                 \begin{minipage}{0pt}
                   \large
                     $\vE, \vB$
                 \end{minipage}
               };
  \node (n00) [] at (-4.4, -0.40)%
               {
                 \begin{minipage}{0pt}
                   \large
                     $\phi, \vA$
                 \end{minipage}
               };
  \node (n980) [] at (-2.3, 4.30)%
               {
                   \bf Local
               };
  \node (n980) [] at (2.3, 4.30)%
               {
                   \bf Global
               };
  \node (n98) [] at (-2.3, 4.00)%
               {
                   \bf
                   Differential equations
               };
  \node (n99) [] at ( 2.3, 4.00)%
               {
                   \bf
                   Integral equations
               };

\end{tikzpicture}

\end{center}
\caption{ 
  The study of electomagnetics requires understanding the different, yet
  mathematically equivalent, forms of Maxwell equations.  The best form
  depends on the application of the study, and this applies to the quasistatic
  regime as well.  Table~\ref{table:summary} summarizes these different forms
  for the three Galilean limits of Maxwell equations.
}
\label{fig:EMstudy}
\end{figure}

The reviews have also been limited in scope.  The study of electrodynamics
involves studying multiple mathematical forms of the Maxwell equations, as
illustrated in \reffig{fig:EMstudy}.  The best form to use for the
electrodynamic equations depends on the application at hand, and the solution
technique employed, either analytic or numerical.  For example, charged particle
motion in an electrodynamic field may best be studied using the potential
formulation if using Lagrangian mechanics, the generalized Coulomb and
Biot-Savart methods if using Newtonian mechanics, the field-form of the Maxwell
equations if using the FDTD numerical method~\cite{yee-fdtd}, or the retarded
potential solutions if using the radiation far field.  
In all of the previous derivations of the quasi-static limits, the complete set
of the forms of the Maxwell equations are not given for each quasi-static limit.
Here, a unified approach to the derivation to the EQS, MQS, and EMQS equations
is presented using a quasi-static perturbation expansion for the vacuum Maxwell
equations.  Ignoring the effects of macroscopic media provides a simpler method
and recovers the elegance found in the original LBLL paper.  

Arguably, the most common encounter of the quasi-static limits in modern physics
is in the area of plasma physics.  Because the electrons are free to move and
shield the slower moving ions, a process known as Debye shielding, the
quasi-neutral approximation is used.  This approximation is equivalent to using
the MQS limit for the bulk of the plasma.  Near the edge of laboratory plasmas,
a thin boundary layer exists, known has the plasma sheath, where the EQS limit
applies.   Because of this transition, and yet still quasi-static nature of the
laboratory experiments, the Darwin approximation is also used in plasma physics.
Despite the widespread use, the relationship of plasma equations to the
quasi-static equations are often not clear, and errors such as using MQS
equations with gauges other the Coulomb are frequently made.  Focusing on the
vacuum equations provides a useful framework for understanding the
approximations before dealing the complications of plasma theory.

The framework for this review will be a perturbation analysis of the Maxwell
equations.  A perturbation analysis uses the expansion of the variables in a
small parameter to study the properties of the equations to a given order -- in
this case, only the first-order equations are studied.  In order to do this, we
must first cast the Maxwell equations in a dimensionless form so that the
relative sizes of terms can be determined appropriately.  While this process is
easy to understand if using a specific unit system such as MKS units, it is
helpful to discuss it for arbitrary unit systems, and we start with a full
discussion of the different unit systems for the Maxwell equations.  After this
background, we order the equations, take the appropriate limits, and show that
only three first-order, quasi-static limits are possible.  We then revisit the
derivation in terms of potentials, and relate our derivation of the Darwin
approximation to the more typical approximation given by Jackson in both his
book and more recent paper~\cite{JacksonSecond,JacksonThird,Jackson2002}.  From
the potential form of the Maxwell equations, integral solutions and generalized
Coulomb and Biot-Savart equations are easily derived.  Finally, we relate this
this derivation to the results of Rapetti and Rousseaux for macroscopic media is
shown.  Frequent references to the textbooks of Griffiths~\cite{Griffiths}
and Jackson~\cite{JacksonSecond,JacksonThird} are given to make this review
easier to follow.
\section{Units and dimensional analysis of the Maxwell equations}
\label{sec:dimensions}

The Maxwell equations have engendered considerable confusion when it comes to
units.  While the trend has been towards using $MKSA$ units, the scientific
literature uses many different systems.  Our technique for understanding the
various
quasi-static limits is to perform a perturbation analysis that analyzes the
terms that become small as the ordering parameter becomes small.  The relative
``smallness'' of a term must be identified independent of units so that units do not
distort which terms appear small.  For the Maxwell equations, this is especially
important because $c$, a key factor, is large, and its location changes
depending on the system of units.

The key issue in dimensional analysis is the choice of fundamental, or base,
units.  As review, for all coherent unit systems, there are fundamental, or
base, units, and the units of every other quantity are then expressed in terms
of these units.  There is freedom in choosing the base units.  For the Maxwell
equations, the natural base units are Length-Time-Charge, while most analyses
are in terms of the traditional Length-Mass-Time units from mechanics (e.g., cgs
and MKS).  In other words, the Maxwell equations themselves do not have mass as
a quantity, so including it confuses the analysis.  The most systematic analysis
of the EM unit system in a textbook is the Appendix of Jackson.  In this paper,
the analysis of Jackson is repeated using a more traditional Rayleigh-Buckingham
dimensional analysis approach with Length-Time-Charge as the base units. The
relationship to the mechanical base units is discussed after this analysis.

The fundamental units of common EM unit systems are given in
\reftab{table:funUnits}.  The Maxwell equations are equations for 4 variables
($\vE, \vB, \rho, \vJ$) with 3 fundamental units ($\xunit, \tunit, Q$) where $Q$
is the unit of charge.  According to the Buckingham $\Pi$ theorem, there will be
(4 variables)$-$(3 units) = (1 dimensionless constant) to describe the system.
Our goal is therefore to derive a dimensionless version of the Maxwell equations
with this dimensionless number, and the expression for it for each system of
units under consideration.  Here, an algebraic approach is used as opposed to
the (admittedly more standard) matrix-power approach of
Heras~\cite{Heras:2008dj} and Rapetti and Rousseaux~\cite{Rapetti:2014ha}.  The
methods are equivalent, but the algebraic approach is useful in understanding
the perturbation analysis in the next section.

\begin{table}
\extrarowheight=1pt
\begin{tabular}{lcccl}
\toprule
\textbf{Base Units}    &\ Length ($\lunit$)\ &\ Time ($\tunit$)\ &\ Charge ($\qunit$)\ 
                       &\ Relationship to standard base units
                                                                  \\ \midrule
Electrostatic (esu)    & centimeter    & second    & statcoulomb &  
gram (\munit):\ \ $\qunit = \lb \munit \lunit^3 \tunit^{-2} \rb^{1/2}$
                                                                  \\
Electromagnetic (emu)  &               &           &     &      
gram (\munit):\ \ $\qunit = \lb \munit \lunit \tunit \rb^{1/2}$
                                                                  \\
Gaussian               &               &           &     &       
gram (\munit):\ \ $\qunit = \lb \munit \lunit^3 \tunit^{-2} \rb^{1/2}$
                                                                  \\
Heaviside-Lorentz      &               &           &     &       
gram (\munit):\ \ $\qunit = \lb \munit \lunit^3 \tunit^{-2} \rb^{1/2}$
                                                                  \\ \midrule
Rationalized MKSA      & meter         & second    & coulomb      & ampere
(I): $\qunit=[I\tunit]$
                                                      \\ \bottomrule
\end{tabular}
\caption{Summary of the base units for each of the unit systems.
      The analysis presented is in terms of the Maxwell base units
      ($\lunit\tunit\qunit$).  The last column shows the fundamental base unit for each 
      system and its relationship to the charge unit.
     }
 \label{table:funUnits}
\end{table}

We start by writing the Maxwell equations with each term on the right-hand side
prefixed by a (rationalized) constant:
\begin{subequations}
   \label{MaxwellEquations-Unit}
   \begin{align}
      \text{Gauss's Law} \ \ \ \  \ &
      \dive{\vE}= 4 \pi k_1 \rho                 \label{GaussLawUnit}
												\\
      \text{No magnetic monopoles} \ \ \ \  \ &
      \dive{\vB}= 0                              \label{DivBUnit}
												\\
      \text{Faradays's Law} \ \ \ \  \ &
      \curl{\vE}= -k_3 \dt{\vB}                  \label{FaradaysLawUnit}
												\\
      \text{Maxwell-Amp\`ere Law} \ \ \ \  \ &
      \curl{\vB}=  K_4 \dt{\vE} + 4 \pi K_2 \vJ. \label{AmperesLawUnit}
    \end{align}
\end{subequations}
It is obvious that any unit system can be described in terms of the four
constants, $k_1, K_2, k_3,$ and $K_4$.  The notation follows that of the
appendix of Jackson, with the capitalized constants being the factors that
differ from Jackson to simplify the new approach.  Our goal is to determine a
minimal set of constants to describe the Maxwell equations. The easiest method
is to determine the capitalized constants in the Maxwell-Amp\`ere law.

Taking the divergence of the Maxwell-Amp\`ere law,
\refeq{AmperesLawUnit}, and inserting Gauss's law, \refeq{GaussLawUnit}, gives
the continuity equation
\begin{equation}
        \frac{\partial \rho}{\partial t} + \frac{K_2}{K_4 k_1} \dive{J}=0.
  \label{eqn:ContinuityEqUnit}
\end{equation}
The factor in front of the divergence term must be unity to match the
conventional relationship between the units of $\rho$ and $\vJ$
that all unit systems obey.  This yields the elimination equation
\begin{equation}
         K_2 = K_4 k_1,
  \label{eqn:constraint1}
\end{equation}
which relates the constant in the Maxwell-Amp\`ere law to the constant $k_1$
in Gauss's Law.  Taking the curl of the source-free form of the
Maxwell-Amp\`ere law, and inserting Faraday's Law, gives the wave
equation:
\begin{equation}
       \frac{\partial^2 \vB}{\partial t^2} - \frac{1}{K_4 k_3} \nabla^2 \vB   = 0.
   \label{eqn:waveEqnUnit}
\end{equation}
This equation describes the propagation of light waves, and thus the factor in front
of the Laplacian must be equal to the inverse square of the speed of light.  This yields 
the second elimination equation:
\begin{equation}
      K_4 = \frac{1}{k_3 c^2},
  \label{eqn:constraint2}
\end{equation}
which relates a scale factor in the Maxwell-Amp\`ere law to the constant $k_3$
in Faraday's Law.

Using these relations, we rewrite the Maxwell-Amp\`ere law as
\begin{equation}
      \curl{\vB} = \frac{1}{c^2 k_3} \lb \dt{\vE} + 4 \pi k_1 \vJ \rb.
  \label{AmperesLawUnit2}
\end{equation}
The constant $4 \pi k_1$, which also prefixes the source term in Gauss's law
(\refeq{GaussLawUnit}), is related to the single dimensionless constant
($\aht$), predicted by the Buckingham $\Pi$ theorem as shown below.

The constant $k_3$ is related to unit relationships.  To see this, the units of
${\vE, \vB, \rho, \vJ}$ are denoted as $\Eunit, \Bunit, \runit, \Junit$.  The
units of space and time are denoted $\lunit$ and $\tunit$ respectively.
Examination of the units of Faraday's law,
\refeq{FaradaysLawUnit}, and the source-free version of the Maxwell-Amp\`ere
law \refeq{AmperesLawUnit2}, gives the same result:
\begin{equation}
        \frac{\Eunit}{\Bunit} = \khunit \frac{\xunit}{\tunit}
   \label{eqn:ebunit2}
\end{equation}
From this we see that $k_3$ serves to adjust the units of $\vB$ relative to
$\vE$, with two reasonable choices being that either $k_3$ is unitless and the
ratio is given by a velocity unit, or that $k_3$ is $c^{-1}$ and the units of $\vE$ and
$\vB$ are the same (see \reftab{table:eqsystem}).

The derived units of the Maxwell equations follow from the above relations.
Starting with the units for $\rho$ that are the same in all unit systems
(equivalently, the units of $\vJ$ could be used because the continuity
equation is the same in all unit systems), the rest of the units are derived
straightforwardly from the above equations:
\begin{eqaligned}
  \runit  &= \qunit \xunit^{-3}        &  \Junit &= \qunit \xunit^{-2} \tunit^{-1}
                                                     \\
  \Eunit &= \kounit \qunit \xunit^{-2} &  \Bunit &=\frac{\kounit}{\khunit}
                                                      \qunit \xunit^{-3} \tunit
   \label{eqn:maxwellUnits}
\end{eqaligned}
The units for $\vE$ and $\vB$ are determined once the units for $k_1$ and $k_3$
are specified for a given unit system.  Similar to Jackson, a table summarizing
the different equation systems is shown in \reftab{table:eqsystem}.  If $k_1$
contains a factor of inverse $4 \pi$, then the unit system is termed a
\emph{rationalized unit system}.  The Heaviside-Lorentz unit system is the
rationalized Gaussian unit system.

Another recent dimensional analysis of the Maxwell equations is the
``$\alpha\beta\gamma$-units'' equations of Heras and
B{\'a}ez~\cite{Heras:2008dj}.  The relationship between our system and that
description is $\alpha = 4 \pi k_1$, $\beta = 4 \pi k_1/(k_3 c^2)$, and $\gamma=
k_3$.  The form of the dimensional Maxwell equations given by
\refeqs{MaxwellEquations-Unit} with either $k_3=1$ or $k_3=1/c$, uses a single
dimensional constant ($k_1$) that by the Buckingham $\Pi$ theorem is the form
for the Maxwell equations with the minimal number of constants.  It is important
to distinguish what is necessary ($k_1$) and what is useful for scaling units
($k_3$)~\footnote{One could imagine yet another constant to enable a difference
      in units between $\rho$ and $\vJ$, which would allow the charge continuity
      equation to be written differently. Fortunately, no unit system has yet
been so created; i.e., all unit systems developed have the same form of the
continuity equation.}.

We are now ready to de-dimensionalize the equations.  To do so, we write
 \begin{eqaligned}
     \vx &= \xbr \vxht    &     t &= \tbr \tht 
                                                     \\
     \vE(\vx,t)  &= \vEbr \vEht(\vx,t)  &    \vB(\vx,t) &= \vBbr \vBht(\vx,t)
                                                     \\
     \rho(\vx,t) &= \rbr \rht(\vx,t)    &    \vJ(\vx,t) &= \vJbr \vJht(\vx,t)
 \end{eqaligned}
where the bar denotes a constant with units, and hat denotes a unitless, but
spatially and temporally varying, quantity.  We use $\xbr$ as our
characteristic length scale such that the units are $[\vx]=[\xbr]=\xunit$.  The
dimensionless spatial vector $\vxht$ should not be confused with a unit vector.
Substituting these into the equations above yields:
\begin{eqaligned}
   \label{MaxwellsEquations-dedim1}
     \diveh{\vEht} &=\lp 4 \pi k_1 \frac{\rbr \xbr}{\vEbr} \rp \rht 
												\\
     \diveh{\vBht} &=0
												\\
     \curlh{\vEht} &= -\frac{\xbr}{c \tbr} \frac{k_3 c \vBbr}{\vEbr} \dth{\vBht}
												\\
     \curlh{\vBht} &=  \frac{\xbr}{c \tbr} \frac{\vEbr}{c k_3 \vBbr}
                       \lb \dth{\vEht} 
                          +4 \pi k_1 \frac{\rbr \xbr}{\vEbr} 
                            \frac{\vJbr}{c \rbr} \frac{c \tbr}{\xbr} \vJht
                       \rb
\end{eqaligned}
Equations~\ref{eqn:ebunit2} and \ref{eqn:maxwellUnits} can be used to
demonstrate that these equations are dimensionless.  

The scale length ratios in these equations provide both freedom and constraints.
We use the freedom to the relationships to set to unity the ratios involving
pure field terms or pure source terms.  This is the process of
de-dimensionalization.  The constraints force the ratios of field magnitudes to
source magnitudes, in other words, the constraints force the factors involving
$k_1$ to have a specified value determined by $\aht$.  In the
relativistic limit, setting these factors to unity is clear:  there is no
separation between electric and magnetic fields, between charge density and
current density, and between space and time.  The quasi-static
(non-relativistic) limit allows these quantities to be ordered separately as
discussed in the next section.

The dimensionless forms of the equations can then be simply written as:
\begin{subequations}
   \label{MaxwellsEquations-dedimensional}
   \begin{align}
     \diveh{\vEht} &= \aht \rht
                                          \label{GaussLaw} \\
     \diveh{\vBht} &=0
                                          \label{DivBzero} \\
     \curlh{\vEht} &= - \dth{\vBht}
                                          \label{FaradaysLaw} \\
     \curlh{\vBht} &=   \dth{\vEht} + \aht \vJht
                                          \label{AmperesLaw}
   \end{align}
\end{subequations}
with the single dimensionless constant given by
\begin{equation}
   \aht =4 \pi k_1 \frac{\rbr \xbr}{\vEbr}.
   \label{eqn:alpha}
\end{equation}
The hat serves to distinguish this alpha from Jackson's and indicates that it
is dimensionless.  Manfredi (Ref.~\cite{manfredi2013}) denotes this factor as
$1/\alpha$.
This parameter is to the Maxwell equations as the Reynolds number is to the
Navier-Stokes equations.  

This form of the Maxwell equations is useful for numerical work because the
variables will be of order unity, which is useful to prevent numerical round-off
issues.  To solve, one would choose useful charge and current density characteristic
scales ($\rbr$ and $\vJbr$) to de-dimensionalize the equations, and also set the
dimensionless parameter $\aht$.  One can then numerically solve the equations.
The conversions to dimensional units for $\vE$ and $\vB$ are given by $\vEbr = 4
\pi k_1 \rbr \xbr/\aht$ and $\vBbr = (4 \pi k_1/c^2/k_3) (\xbr \vJbr)/\aht$.


To summarize the results so far, the arbitrary unit version of the Maxwell
equations are given by \refeqs{MaxwellEquations-Unit}.  Specifying a
unit system requires choosing the constants $k_1$ and $k_3$, with $k_3$ being chosen
as a convenient normalization between $\vE$ and $\vB$.  The dimensionless
form of the Maxwell equations, \refeqs{MaxwellsEquations-dedimensional}, is
obviously the same for any unit system, and the dimensionless number is
specified by \refeq{eqn:alpha}.  A table summarizing the
different equation systems is shown in \reftab{table:eqsystem}.

\begin{table}
\begin{center}

\extrarowheight=1pt
\begin{tabular}{LL}
\toprule
\textbf{Dimensional}  & \textbf{Dimensionless}
                                                                  \\ \midrule
\dive{\vE}=4 \pi k_1 \rho    & 
\dive{\vE} = \aht \rho
												\\
\dive{\vB}=0                 &
\dive{\vB}=0
												\\
\curl{\vE}=-k_3 \dt{\vB}     &
\curl{\vE} = - \dt{\vB}
												\\
\curl{\vB} = \frac{1}{c^2 k_3} \lb \dt{\vE} + 4 \pi k_1 \vJ \rb\ \ \ \ &
\curl{\vB} =   \dt{\vE} + \aht \vJ\ \ \ \ 
                                                      \\ \bottomrule
\end{tabular}
\end{center}
\begin{center}
\extrarowheight=2pt
\begin{tabular}{lCCCCCC}
\toprule
\textbf{Unit System}  &\ \ \mathbf{k_1} \ \ 
                      &\ \ \mathbf{[k_1]} \ \ 
                      &\ \ \mathbf{k_3} \ \ 
                      &\ \ \mathbf{[k_3]} \ \ 
                      &\ \ \mathbf{k_1 c^{-2} k_3^{-1}} \ \ 
                      &\ \ \aht \ \ 
                                                                  \\ \midrule
Electrostatic (esu)   & 1            
                      & --
                      & 1 
                      & --
                      & c^{-2}                 
                      & 4 \pi \frac{\rbr \xbr}{\vEbr}
                                                                  \\
Electromagnetic (emu) & c^2 
                      & \frac{\lbr^2}{\tbr^2}
                      & 1 
                      & --
                      & 1 
                      & 4 \pi c^2 \frac{\rbr \xbr}{\vEbr}
                                                                  \\
Gaussian              & 1  
                      & --
                      &c^{-1}   
                      & \frac{\tbr}{\lbr}
                      & 1  
                      & 4 \pi \frac{\rbr \xbr}{\vEbr}
                                                                  \\
Heaviside-Lorentz     & \frac{1}{4 \pi}  
                      & --
                      &c^{-1}  
                      & \frac{\tbr}{\lbr}
                      & \frac{1}{4 \pi c}  
                      & \frac{\rbr \xbr}{\vEbr}
                                                                  \\
Rationalized MKSA     & \frac{1}{4 \pi \epso}
                      & \frac{\munit\lbr^3}{I^2\tbr^4}
                      & 1  
                      & --
                      & \frac{\muo}{4 \pi} \left[ \frac{\munit\lbr}{I^2\tbr^2}
                                           \right]
                      & \frac{\rbr \xbr}{\epso \vEbr}
                                                      \\ \bottomrule
\end{tabular}
\end{center}
\caption{Summary of the different unit systems used for Maxwell equations. 
         Placing $k_1$ and $k_3$ into the equations give the specific forms.
         The units of $k_1$ and $k_3$ can be combined with 
         \refeq{eqn:maxwellUnits} to determine the units of $\vE$ and $\vB$.
         The dimensionless Maxwell equations number for the different unit
         systems is unit system specific.
   }
\label{table:eqsystem}
\end{table}

As discussed earlier, unit systems such as $cgs$ or $MKS$ refer to the
mechanical base units (Length-Mass-Time).  To relate the EM base units to the
mechanical base units, the Lorentz force law in our dimensional units is
written as
\begin{equation}
         \vF = q \lp \vE + k_3 \vv \times \vB \rp.
      \label{LorentzForceLaw-dimensional}
\end{equation}
Using $\vect{F}=m \vect{a}$ (or the relativistic version thereof), considering
only the electric field, and using \refeqs{eqn:maxwellUnits}, one can show that
\begin{equation}
      \label{qunit}
      \qunit = \sqrt{\frac{\munit \lunit^{3}}{\tunit^{2} \kounit} }.
\end{equation}
Substituting in the values of $\kounit$ for the different unit systems gives the
final column of \reftab{table:funUnits}.

The MKSA system is unique in that the constants used in the Maxwell equations,
$\eps_0$ and $\mu_0$, have units containing mass as seen in
\reftab{table:eqsystem}.  As discussed in the Appendix of Jackson, this is
because the determination of the constant involves Newton's law.  This is an
unfortunate property of the MKSA unit system; a true Rayleigh-Buckingham
approach to dimensional analysis necessitates the inclusion of an additional
equation.  Some authors (e.g., ~\cite{Heras:2010hi,Rousseaux:2013ft}) have made
note of the difference between a $c$ used for units ($c_u$) and the fundamental
constant denoting the speed of light ($c$).  These different $c$'s are a
historical artifact of how the different unit systems developed, and define
their fundamental constants.  As seen in the minimal form of the dimensional
equations above, $c$ is fundamental to the Maxwell equations, appearing in the
Maxwell-Amp\`ere law regardless of the value of $k_1$ or $k_3$.

The use of $k_3$ is helpful in understanding what makes a good engineering set
of units.  In the third edition of Jackson~\cite{JacksonThird}, the first 10
chapters use MKSA units, while the remaining chapters covering relativistic
mechanics use cgs units.  What makes cgs a good base unit system for studying
relativity is that $\vB$ has the same units as $\vE$ in the same way that $x_0 =
c t$, the zeroth component of the space-time 4-vector, has the same units as the
remaining components of the 4-vector.  Just as Newtonian mechanics uses $t$ and
not $x_0$, everyday experiences with electromagnetism work better with $k_3=1$:
$c \vB$ the same units as $\vE$.  MKSA being a good set of ``engineering units''
has less to do with the use of mass in defining its units (which causes
confusion), and more to do with convenience in the Galilean limit.
\section{Low velocity limit of the Maxwell Equations}
\label{sec:theory}

\subsection{Quasi-static limits and the charge continuity equation}
\label{sec:galilean} 

This paper focuses on the ``quasi-static limits'' of
the Maxwell equations where we are interested in the forms of the equations when
the following parameter is small:
\begin{equation} 
      \beta = \frac{\lbr}{\tbr c}.
\label{order} 
\end{equation} 
The particular interpretation of this parameter is problem specific.  For
example, the Galilean limit of the Lorentz transformation is taken by assuming
that the speed of the inertial frame is small: $V=\xbr/\tbr \ll c$, where $c$ is
the speed of light.  The small parameter in this instance will be 
\begin{equation} 
      \betainrt = \frac{V}{c} = \frac{\lbr}{\tbr c}
      \text{  and  }
      \vbeta_\text{inertial} = \frac{\vV}{c},
\label{order-inertial} 
\end{equation} 
where the subscript inertial implies that this is the small parameter for
inertial transformations, and the vector is used to denote the direction of the
inertial velocity.

For the common occurrence of a source oscillating at a
single frequency, $f$, this small parameter can be written as 
\begin{equation} \betaosc =
      \frac{\lbr}{\tbr f \lambda} = \frac{1}{\Nosc} \lbr/\lambda \ \ ,
   \label{order-osc} 
\end{equation} 
where $\Nosc$ is the number of oscillations of interest and $\lbr/\lambda$ is
the {\em electrical length}~\footnote{The electrical
length is electromagnetic and not electrical, and is not a length.  We use the
traditional term regardless.}
in a vacuum.  For example, consider a parallel plate
capacitor that is 1 mm wide as part of an AC circuit oscillating at \SI{60}{Hz}.
At \SI{60}{Hz}, the wavelength is \SI{5E6}{m} gives an electrical length of
\num{2E-10}.  If this capacitor is part of an $RC$ circuit with a charging time
of $1/6$ sec, then one charging time corresponds to 10 oscillations, and hence a
$\betaosc$ = \num{2E-11}.  An expansion in this case is so good that one could
reasonably believe that a static approximation would be satisfactory.  This is
discussed in \refsec{sec:bound}.  The number of periods of interest, $\Nosc$,
depends on details of the specific problem, and this will be discussed
throughout the paper.  The electrical length being much less than one serves as
a conservative estimate for whether we are in the quasi-static regime;
$\Nosc=1$ is the minimum number that one would want to study when oscillating
sources are present.

Other important examples of $\beta$ as a small parameter are the laboratory
electrodynamic experiments of rotating
media~\cite{wilson1905,barnett1915,wilson1913,weber1997,deMontigny:2007kg,Rousseaux:2013ft}.
In this case, the rotation speed is much slower than the speed of light, but the
transformation is non-inertial.  Finally, quasi-static limits can be useful in
studying particle motion.  Darwin~\cite{darwin} developed a quasi-static form by
using an ordering of the ratio of the particle velocity to the speed of light.

To understand how we use this small parameter to derive the quasi-static, or
Galilean limits, we first review the kinematic and dynamic limits from
relativistic mechanics.  We then consider the charge continuity equation to show
how electrodynamics gives different behavior than classical mechanics.

\begin{figure}
\begin{center}
\begin{tikzpicture}[ ]

  \draw[<->,ultra thick] (-4.5,0) -- (4.5,0);
  \draw[<->,ultra thick] (0,-4.5) -- (0,4.5);
  \draw[ultra thick] (0,0) -- (4.5, 4.5);
  \draw[ultra thick] (0,0) -- (4.5,-4.5);
  \draw[ultra thick] (0,0) -- (-4.5, 4.5);
  \draw[ultra thick] (0,0) -- (-4.5,-4.5);
  \begin{scope}[on background layer]
     \fill[fill=gray!30!white] (0,0)--(-4.3, 4.3)--( 4.3, 4.3)--(0,0);
  \end{scope}
  \begin{scope}[on background layer]
     \fill[fill=gray!30!white] (0,0)--(-4.3,-4.3)--( 4.3,-4.3)--(0,0);
  \end{scope}

  \node (n00) [anchor=south west] at (4.2, 0.15)%
               {
                     $\mathbf{x}$
               };
  \node (n01) [anchor=south west] at (0.2, 4.20)%
               {
                     $\mathbf{ct}$
               };

  \node (n11) [anchor=south west] at (-3.0, 3.1)%
               {
                  \Large Time-like
               };
  \node (n12) [anchor=south west] at (-4.1, 1.0)%
               {
                  \Large Space-like
               };
  \node (n13) [anchor=north east, rotate=45] at (3.2,3.2)%
               {
                  \Large Light Cone
               };
  \node (n14) [anchor=north west, rotate=45] at (-4.2,-3.4)%
               {
                  \Large Light Cone
               };
\end{tikzpicture}
\end{center}
\caption{ 
      Space-time is separated into regions called space-like, where causality does not apply,
      and time-like, where causility applies. The boundary is called the light
      cone.
}
\label{fig:spacetime}
\end{figure}
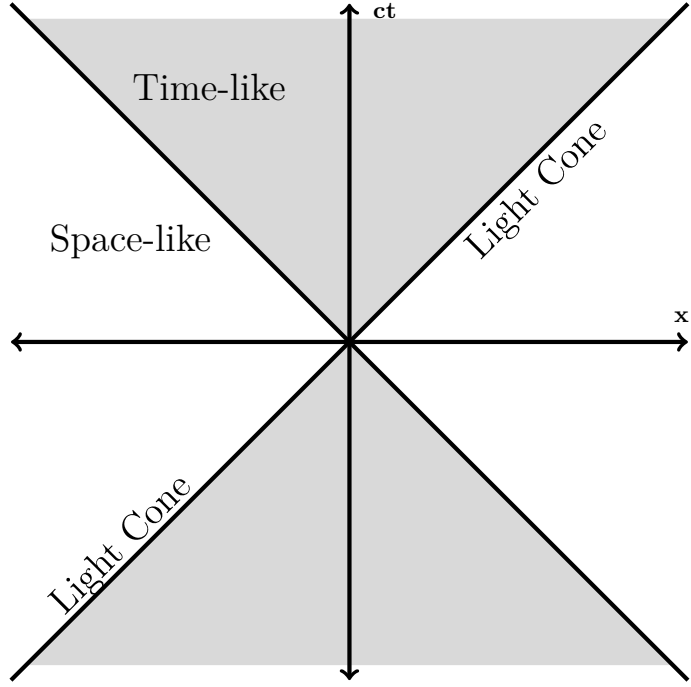

A powerful concept in special relativity is the invariance of 4-vector
products under Lorentz transformations.  For example, the ``length'' of the
spacetime vector
\begin{equation}
    x^\mu x_\mu = c^2 t^2 - \vx \cdot \vx
    \label{spacetimeInvariant}
\end{equation}
is invariant under any inertial transformation.  If there are two events, and we
place one event at the origin of the spacetime grid, then the separation between
the two events corresponds to an invariant quantity.  When this
invariant is positive, the separation is ``time-like''; when the
invariant is negative, the separation is ``space-like''; and when the invariant
is zero, the separation is light-like, as shown in Figure~\ref{fig:spacetime}.

De-dimensionalizing the above spacetime invariant gives
\begin{equation}
      \tht^2 - \frac{\lbr^2}{\tbr^2 c^2} \vxht \cdot \vxht = 
      \tht^2 - \beta^2 \vxht \cdot \vxht = \text{invariant}.
    \label{spacetimeInvariantDeDim}
\end{equation}
Thus between any two inertial reference frames, this invariant is the same:
\begin{equation}
      \tht^{\prime 2} - \beta^2 \vxht^\prime \cdot \vxht^\prime = 
      \tht^2 - \beta^2 \vxht \cdot \vxht,
    \label{spacetimeInvariantDeDimFrames}
\end{equation}
where the prime denotes a different inertial reference frame, and the origin of
both frames is the same to simplify the algebra.  In the limit $\beta
\rightarrow 0$, $\tht^\prime = \tht = $ constant.  Graphically speaking, the
light cone collapses to the $x$ axis, and the hyperbolic lines defining
simultaneity become horizontal lines.  This is the Galilean limit.

This can also be derived from the Lorentz transformation directly. The Lorentz
transformation of the spacetime 4-vector is~%
\footnote{See, for example: Jackson (Ref.~\cite{JacksonSecond,JacksonThird}),
Sec.~11.3.}
 \begin{eqaligned}
       \tht^\prime   = & \gamma (\tht - \beta \vbeta \cdot \vxht)
                                                                \\
       \vxht^\prime   = & \vxht - \gamma \frac{\vbeta}{\beta} \tht
        + (\gamma - 1) \frac{1}{\beta^2} \vbeta (\vbeta \cdot \vxht)
                                                                \label{Lorentz}
 \end{eqaligned}
where $\gamma = (1-\beta^2)^{-1/2}$.  The $\beta$ and $\vbeta$ here are the
inertial versions presented in \refeq{order-inertial}, but the
subscripts ``inertial'' are dropped for convenience.
The Galilean limits are given by the transformation equations (written here with
dimensional quantities):
 \begin{eqaligned}
       \tp   = & t
                                                                \\
       \xp   = & \vx - \vV t.
                                                                \label{Galilean}
 \end{eqaligned}
Thus, the Galilean limit has only the causal time-like behavior with an
event being everywhere simultaneous.
In this dimensionless form, the inertial frame velocity is 
unity.~\footnote{The so-called ``Carroll limit''~\cite{levy1965} is not included
here because it is second-order in the expansion parameter.  In contrast to
Heras~\cite{Heras:2010jw}, our definition of quasi-static is that it be only
first order in $\beta$.}

Conservation of momentum-energy arises because of the invariance of the
4-vector
\begin{equation}
      p^\mu p_\mu = \frac{E^2}{c^2} - \vp \cdot \vp = \text{invariant}.
    \label{momenergyInvariant}
\end{equation}
In the $\beta \rightarrow 0$ limit, the time-like and space-like
solutions decouple, and we have energy and momentum conservation simultaneously.
That is, Newtonian mechanics is the quasi-static limit of Special Relativity, with
identical time-like and space-like limits.

Now consider the current 4-vector.  It has the same invariance properties as
above:
\begin{equation}
      J^\mu J_\mu = c^2 \rho^2 - \vJ \cdot \vJ = \text{invariant}.
\end{equation}
For the current 4-vector, the key physics equation is not from the invariance,
but rather charge continuity:
\begin{equation}
      \partial^\mu J_\mu = \dt{\rho} + \dive{\vJ}=0.
\end{equation}
De-dimensionalizing this equation (while keeping the scale lengths) gives
 \begin{equation}
  \label{ContinuityEquation-ordered}
     \lp \frac{\lbr}{\tbr c}\roJ \rp \dth{\rhoht} + \diveh{\vJht}  
   = \lp \beta \roJ  \rp \dth{\rhoht} + \diveh{\vJht}  = 0.
 \end{equation}
As  $\beta$ becomes small, the ratio of $c \rbr/\vJbr$ can have different
relative orderings, and those different orderings have different physical
consequences.
Two possible regimes are of interest as $\beta \rightarrow 0$:
\begin{align}
      \label{rhoJlimits}
      \text{Time-like: } \ \ & \roJ \sim \frac{1}{\beta} & \dth{\rhoht} + \diveh{\vJht}  = 0,
												\nonumber  \\
      \text{Space-like: } \ \ &  \roJ \sim \beta           & \diveh{\vJht}  = 0.
												\nonumber 
\end{align}
That is, in the quasi-static limit, the charge density and current density
decouple in the same way that space and time decouple.  These two limits are
equivalent to the absolute value of the invariant tensor product being large;
i.e., $\mid J^\mu J_\mu \mid \gg 0$.  The middle ordering of the scale length
ratio $c \rbr/\vJbr \sim\bigO{1}$ might appear to be a ``light-like'' ordering;
however, ``light-like'' corresponds to the fully relativistic case where there
is no separation between $\rho$ and $\vJ$ (which is why these ratios were set to
unity to get the dimensionless equations as discussed after
\refeq{MaxwellsEquations-dedim1}).  In the low-velocity limit, $c \rbr/\vJbr$
being of order unity is related to the static limits as discussed in
\refsec{sec:bound}.  One could expect that the ordering of the sources will lead
to a large value, either positive (time-like) or negative (space-like), in the
Lorentz invariant of the electromagnetic field-strength tensor (with $k_3 =
1/c$):
\begin{equation}
      F^{\mu\nu} F_{\nu\mu} = E^2 - B^2.
\end{equation}
This will be shown in the next section.
%

In summary, our small parameter is the ratio of the length to time scale that is
small relative to the speed of light.  For the space-time vector $x_\mu$, only
one limit exists, and this is the Galilean transformation limit.  It is
time-like, $c^2 t^2 \gg x^2$, and is equivalent to an event being simultaneous
across all reference frames.  For the momentum 4-vector, there is no difference
between the time-like and space-like limits: there is only one Newtonian limit
where energy and momentum conservation are decoupled.  However, the charge
continuity equation admits two limits.  In the time-like limit, $c \rbr \gg
\vJbr$ and full charge continuity exists.  The current density is causal:  a
changing charge density implies a change in current density.   In the space-like
limit, $c \rbr \ll \vJbr$, the current density is only a (possibly
time-dependent) divergence-free field.

\subsection{Perturbation expansion of the Maxwell equations}
\label{sec:MEorder}

As discussed in the previous section, the sources and fields can be expected to
separate into either ``space-like'' or ``time-like'' components in the Galilean
limits; thus the relative magnitudes of $\vE$ to $\vB$, and $\rho$ to $\vJ$ will
be included in the ordering of the Maxwell equations.  Starting with
\refeqs{MaxwellsEquations-dedim1} and specifying $k_3=1/c$, to make
the ratio of electric and magnetic field dimensionless, yields
 \begin{eqaligned}
   \label{MaxwellsEquations-ordered}
     \diveh{\vEht}& = \aht \rht
												\\
     \diveh{\vBht}& = 0
												\\
     \curlh{\vEht}& =-\lp \beta \BoE \rp \dth{\vBht}
												\\
     \curlh{\vBht}& = \lp \beta \EoB \rp \dth{\vEht} +
                       \lp \EoB \Jor \rp     \aht \vJht.
 \end{eqaligned}
\noindent
In the relativistic, Lorentz covariant limit, $\beta$ is not a small parameter
and it can be set to unity as well as the ratios $\vEbr/\vBbr$ and 
$c \rbr/\vJbr$ to give the purely dimensionless form of the Maxwell
equations.


We are interested in the limit of $\aht \sim \bigO{1}$ and $\beta \ll 1$.  
Examination of the last two ordered equations shows an apparent contradiction: the time
derivative terms cannot both be of order unity simultaneously.  When one term is
of order unity, the other term will be of order $\beta^2$.  Similarly, keeping
the current term in the last equation implies an inverse relationship between
$\vEbr/\vBbr$ and  $\vJbr/(c \rbr)$.  Intuitively, when the EM tensor
is (time,space)-like ($\vEbr \gg \vBbr$, $\vEbr \ll \vBbr$ or equivalently
$F^{\mu\nu} F_{\nu\mu} \gg 0$, $F^{\mu\nu} F_{\nu\mu} \ll 0$), the current 
4-vector must also be (time,space)-like, ($c \rbr \gg \vJbr$, $c \rbr \ll \vJbr$).  For
the case of $\vEbr \sim \vBbr$, we recover the trivial cases of the
electrostatic and magnetostatic limits.  While these are obviously Galilean
limits, they are not of interest here although we will revisit these limits in
\refsec{sec:bound}.
 
More formally, the quasi-static limits of the Maxwell equations can be derived
by expanding the fields in terms of $\beta$ as the small parameter: 
\begin{eqaligned}
   \label{variableOrdering}
     \vE &= \vE_0 + \beta \vE_1 + \dotsb
												\\
     \vB &= \vB_0 + \beta \vB_1 + \dotsb
												\\
     \rho &= \rho_0 + \beta \rho_1 + \dotsb
												\\
     \vJ &= \vJ_0 + \beta \vJ_1 + \dotsb   .
\end{eqaligned}
\noindent
These are placed into the form of Maxwell's equations with the fields as
dimensional, the operators dimensionless, and the $\beta$ parameter kept for the
ordering:
\begin{eqaligned}
   \label{MaxwellsEquations-ordered2}
     \diveh{\vE}& = \abr \rho
												\\
     \diveh{\vB}& = 0
												\\
     \curlh{\vE}& =-\beta \dth{\vB}
												\\
     \curlh{\vB}& = \beta \dth{\vE} + \frac{\abr}{c} \vJ,
\end{eqaligned}
where $\abr = \aht \vEbr/\rbr$ has been introduced for convenience.
Following the previous discussion, there are two limits:  the time-like limit
when $c \rbr \gg \vJbr$, and the space-like limit when $c \rbr \ll \vJbr$.
The first ordering is equivalent to $\rho_0 \neq 0$ and $\vJ_0 = 0$ such that 
$c \rbr/\vJbr \sim \bigO{1/\beta}$.  Similarly, the second limit is given by
$\rho_0 = 0$ and $\vJ_0 \neq 0$ such that $c \rbr/\vJbr \sim \bigO{\beta}$.

For the time-like limit,  $\vJ_0 = 0$ and Maxwell-Amp\`ere law implies that the
$\curlh{\vB_0} = 0$.  Combined with $\diveh{\vB_0}=0$ implies that $\vB_0=0$,
assuming appropriate boundary conditions.  Faraday's law gives $\curlh{\vE_0} =
0$, but Maxwell-Amp\`ere law keeps all terms that are of order $\beta$:
\begin{equation}
      \beta \left[\curlh{\vB_1} = \dth{\vE_0} + \aht \vJ_1 \right]
\end{equation}
Likewise the full Gauss's law is kept and it is of order unity.  
Thus, we have that $c \rbr/\vJbr \sim \bigO{1/\beta}$ implies that 
$\vEbr/\vBbr \sim \bigO{1/\beta}$.  Because the electric field is much larger
than the magnetic field, this is called the electro-quasi-static (EQS) limit.

For the space-like limit, $\rho_0 = 0$ and Gauss's law implies that
$\dive{\vE_0}=0$.  Faraday's law implies that $\curlh{\vE_0}=0$.  Having both
the divergence and curl of $\vE_0$ be zero implies that $\vE_0 = 0$ (assuming
appropriate boundary conditions.  This will be discussed further
later).  Maxwell-Amp\`ere law gives  $\curlh{\vB_0} = \aht/c \vJ_0$. Faraday's law keeps
all terms that are of order $\beta$:
\begin{equation}
      \beta \left[ \curlh{\vE_1} =- \dth{\vB_0} \right].
\end{equation}
Likewise the full Gauss's law is kept and it is of order beta.
Thus, we have that $c \rbr/\vJbr \sim \bigO{\beta}$ implies that 
$\vEbr/\vBbr \sim \bigO{\beta}$.  Because the magnetic field is much larger
than the electric field, this is called the magneto-quasi-static (MQS) limit.

In summary, the resultant equations for each regime are
\begin{center}
\begin{tabular}{LL}
 \text{Time-like}  & \text{Space-like}
                                                                  \\ 
 \text{Electro-Quasi-static}  & \text{Magneto-Quasi-static}
                                                                  \\ 
  \vEbr \gg \vBbr, c \rbr \gg \vJbr & \vEbr \ll \vBbr, c \rbr \ll \vJbr 
                                                                  \\ 
  \vJ_0=0                      & \rho_0 = 0
                                                                  \\ 
  \diveh{\vB_0}=0;\ \curlh{\vB_0}=0 \ \ \ \ \ \ \ \ \ \ \ \ 
                               & \diveh{\vE_0}=0;\ \curlh{\vE_0}=0
                                                                  \\ 
  \diveh{\vE_0}=\aht \rho_0 & \diveh{\vE_1}=\abr \rho_1 
                                                                  \\ 
  \diveh{\vB_1}=0 & \diveh{\vB_0}=0 
                                                                  \\ 
  \curlh{\vE_0}=0 & \curlh{\vE_1}= - \dth{\vB_0} 
                                                                  \\ 
  \curlh{\vB_1}= \dth{\vE_0} + \frac{\abr}{c} \vJ_1 \ \ \ \ & 
  \curlh{\vB_0}= \frac{\abr}{c} \vJ_0
                                                      \\ 
  \dth{\rho_0} + \diveh{\vJ_1}=0 & \diveh{\vJ_0} = 0
                                                      \\ 
\end{tabular}
\end{center}
\noindent
If the boundary conditions are such that there are no strong, static external
fields, then we can ignore $(\vB_0,\vE_0)$ in the (EQS,MQS) limit.  The use of
subscripts naturally shows the LBLL regimes of validity of $\vEbr \gg \vBbr$ for
the EQS regimes and $\vEbr \ll \vBbr$ for the MQS regime.  The two regimes are
merely the dropping of relevant time derivative terms.  It will be shown in the
next section that these equations are Galilean covariant.

Choosing $k_3 = 1/c$, the Lorentz force law,
\refeq{LorentzForceLaw-dimensional}, can be written as
\begin{equation}
         \vF = q \lp \vE + \vbeta \times \vB \rp.
      \label{LorentzForceLaw-beta}
\end{equation}
Because $\beta$ is the ordering parameter, this shows that for the EQS regime,
only the electric field contributes to the Lorentz force, and the force is zeroth
order.  For the MQS regime, the force is first order, and the zeroth-order
magnetic field and the first-order electric field contribute.  If the boundary conditions are such
that there {\em are} strong, static external fields, then these fields can be
added as a linear superposition of the fields.  The implications of how the
fields act on particles is discussed by LBLL~\cite{LBLL}, who proposed solving
the equations in both regimes and then combining.  


As can be seen, the key feature of these limits is the dropping of one of the
appropriate time-derivative terms. This changes the Maxwell equations from
second-order in time to first-order.  As an intuitive consideration of this
behavior, consider driven waveguide modes where the electrical length in
\refeq{order-osc} is less than unity.  In this case, the modes decay axially
rather than propagate: a change from a wave solution of a second-order-in-time
equation to an exponentially-decaying solution of a first-order-in-time
equation.  This confined-system case provides a useful method for studying the
validity of the quasi-static equations~\cite{melcher-haus1989}.

These two regimes of ``Galilean Electrodynamics'' do not include the Darwin
approximation that is discussed in textbooks and widely used.  This third
approximation will make use of the 
Helmholtz decomposition, discussed in Appendix B. It decomposes the electric
field into its longitudinal (irrotational) and transverse (solenoidal) parts:
\begin{subequations}
   \label{LongitudinalTransverse}
\begin{equation}
   \vE =\vE_L+\vE_T, \text{ where the terms are defined by}
\end{equation}
\begin{align}
     \dive{\vE_T}&=0,
												\\
     \curl{\vE_L}&=0.
\end{align}
\end{subequations}
To motivate why the EQS and MQS regimes might not be a complete ordering, 
we rewrite the Maxwell equations using this decomposition:
\begin{eqaligned}
   \label{MaxwellsEquations-LT}
     \diveh{\vEht_L}&=\aht \rht
												\\
     \diveh{\vBht}&=0
												\\
     \curlh{\vEht_T}&=-\dth{\vB}
												\\
     \curlh{\vBht}&=\dth{\vEht_L} + \dt{\vEht_T} +\aht \vJht
\end{eqaligned}
The electromagnetic wave equation is formed by combining Faraday's law and the
Maxwell-Amp\`ere law.  A quasi-static approximation eliminates light waves, so it
is reasonable to ask if there is a method of eliminating only the transverse
component in the Maxwell-Amp\`re law.  To investigate this
question, the scale lengths in our ordering procedure are generalized to
\begin{eqaligned}
   \label{MaxwellsEquations-orderedLT}
     \diveh{\vEht}& = \aht \rht
												\\
     \diveh{\vBht}& = 0
												\\
     \curlh{\vEht}& =-\lp \beta \BoET \rp \dth{\vBht}
												\\
     \curlh{\vBht}& = \beta \ELoB \dth{\vEht_L} + \beta \EToB \dth{\vEht_T} +
                       \lp \EoB \Jor \rp     \aht \vJht
 \end{eqaligned}
 The constraint for the magneto-quasi-static limit that the induction term be of
 order unity forces only the transverse electric field to be small, and does
 not constrain the longitudinal component.  There exists another
 ordering of $\vETbr/\vBbr \sim \bigO{\beta}$ and  $\vELbr/\vBbr \sim
 \bigO{1/\beta}$.  This mixed ordering eliminates light-waves, to give a
 quasi-static ordering, and gives
 \begin{subequations}
   \label{MaxwellsEquations-darwin}
   \begin{align}
     \diveh{\vEht}& = \aht \rht
                                            \label{GaussLaw-darwin}			\\
     \diveh{\vBht}& = 0
                                            \label{DivBzero-darwin}			\\
     \curlh{\vEht}& =-\dth{\vBht}
                                            \label{FaradaysLaw-darwin}			\\
     \curlh{\vBht}& = \dth{\vEht_L} + \aht \vJht.
                                            \label{AmperesLaw-darwin}
   \end{align}
\end{subequations}
This is known as the Darwin approximation, and because it includes both the
induction term and part of the displacement current, it will be referred to as
the electromagnetic quasi-static (EMQS) regime.  The relationship of this
form of the Darwin approximation and to that given in Jackson is presented in
\refsec{sec:intsol} when the retarded potentials are discussed.

The EMQS approximation, like the EQS and MQS equations, does not support wave
propagation, as was our goal.   The EMQS equations also yields the full
continuity equation, like the EQS limit.  The MQS limit gives a continuity
equation of $\dive{\vJ}=0$.  A summary of these results are included at the end
of this article in \reftab{table:summary}.

\subsection{Galilean Covariance}
\label{sec:galileanCovariance}

The original LBLL paper derived the EQS and MQS equations by first deriving the
Galilean transformation of fields, and then asking what equations were needed to
satisfy those fields.  Here the inverse procedure is taken, whereby we
demonstrate that the EQS and MQS equations derived previously are Galilean
covariant.  This allows us then to discuss the transformation properties of the
EMQS approximation.

The Lorentz transformation of the fields can be written as (in dimensional
form with $k_3=1/c$):
\footnote{See, for example: Ref.~\cite{Griffiths}, Sec.~12.3.2; 
Ref.~\cite{JacksonSecond,JacksonThird}, Sec.~11.10.}
\begin{eqaligned}
       \vEp   = & \gamma \lp \vE + \vbeta \times \vB  \rp 
                - \frac{\gamma^2}{\gamma + 1} \vbeta (\vbeta \cdot \vE)
                                                                \\
       \vBp   = & \gamma \lp \vB - \vbeta \times \vE  \rp 
                - \frac{\gamma^2}{\gamma + 1} \vbeta (\vbeta \cdot \vB)
                                                                \\
       \vJp   = & \vJ - \gamma c \rho \vbeta
                - \frac{\gamma^2}{\gamma + 1} \vbeta (\vbeta \cdot \vJ)
                                                                \\
       \rhop   = & \gamma \lp \rho - \vbeta \cdot \vJ/c \rp
  \label{LorentzTransformFields}
\end{eqaligned}
These relations are for inertial transformations and thus the $\vbeta$ in these
relationships are the same as that as defined in \refeq{order-inertial} but we
will not use the {\em inertial} subscript in this section.  The Galilean
transformations will have two different limits because of the different limits
for the magnitudes as $\beta$ goes to zero.  After de-dimensionalizing these
equations and taking the limits similar to the previous discussions, the
Galilean transformations are
\begin{center}
\begin{tabular}{LL}
  \text{Electro-Quasi-static}\ \ \ \   & \text{Magneto-Quasi-static}
                                                                        \\ 
       \vEhtp   =  \vEht & \vEhtp   = \vEht + \vVht \times \vBht
                                                                        \\ 
       \vBhtp   = \vBht - \vVht \times \vEht & \vBhtp   = \vBht
                                                                        \\ 
       \rhohtp   = \rht & \rhohtp   = \rhoht - \vVht \cdot \vJht
                                                                        \\ 
       \vJhtp = \vJht - \vVht \rht &  \vJhtp = \vJht.
                                                                        \\ 
\end{tabular}
\end{center}
The $\vVht$ in these relationships is equivalent to $\vbeta/\beta$.  In
``engineering dimensional units''  with $k_3=1$, no $c$ will appear in the
equations as expected for the Galilean limit.  For ``relativistic dimensional
units'' with $k_3=1/c$, the $c$ that appears is strictly to get the units
correct; that is, $\vV$ is de-dimensionalized by $c$ for relativistic units.  As
pointed out by LBLL~\cite{LBLL}, the low-velocity transformations were stated
incorrectly in such textbooks as Purcell~\cite{purcell} and Landau and
Lifshchitz~\cite{LandL}, demonstrating the confusion over the two separate
limits.

To show Galilean covariance, these relationships are placed into their
respective limiting forms of the Maxwell equations along with the Galilean
transformation of the time and space derivative:
\begin{eqaligned}
      \dthp{}   = & \dth{} + \vbetb \cdot \gradh= \dth{} + \vVht \cdot \gradh
                                                                \\
       \gradhtp = & \gradh .
                                                                \label{galileanTrans}
\end{eqaligned}
The algebra is straightforward, and the only equations of interest are those
with time derivatives.  To demonstrate how the algebra works, the
Maxwell-Amp\`ere law in the EQS limits is shown.   Substituting in these
relationships gives
 \begin{eqaligned}
       \gradhtp \times \vBhtp   = &  \dthp{\vEhtp} + \aht \vJhtp
                                                                \\
       \gradh \times \vBht - \curlh{\lp \vbetb \times \vEht \rp}   = &  
            \dth{\vEht} + \lp \vbetb \cdot \gradh \rp \vEht + \aht \vJht - \aht \rht \vbetb
                                                                \\
       \gradh \times \vBht - \vbetb \diveh{\vEht} + \vEht\lp \diveh{\vbetb} \rp
            - \lp \vEht \cdot \gradh\rp \vbetb + \lp \vbetb \cdot \gradh \rp \vEht  = &  
       \dth{\vEht} + \lp \vbetb \cdot \gradh \rp \vEht + \aht \vJht - \aht \rht \vbetb
                                                                \\
       \gradh \times \vBht   = &  \dth{\vEht} + \aht \vJht,
 \end{eqaligned}
where in the last step all derivatives on $\vbeta$ were dropped because it is an
inertial transformation, and Gauss's law was used to cancel the $\rht$ term.
There is also a cancellation of $\lp \vbeta \cdot \gradh \rp \vE$ term on both sides.
Both cancellations are required for Galilean covariance.
Similar algebra can be used to show that the entire set of EQS and MQS limits of
the Maxwell equations are Galilean covariant.

The same method also shows that the EMQS equations are {\em not} Galilean
covariant.  The cancellations required in the Maxwell-Amp\`ere law do not occur.
The underlying reason is that the decomposition of the fields into longitudinal
and transverse components is neither Lorentz nor Galilean invariant.  The
non-Galilean covariance of the EMQS will be revisited in \refsec{sec:intsol},
when the retarded potentials are discussed.

\section{Potential of the quasi-static Maxwell equations}
\label{sec:potentials}

Our goal in this section is to derive the quasi-static forms of the Maxwell
equations in terms of potentials.  An obvious method would be to substitute in
the expression for $\vE$ and $\vB$ in terms of potentials into the previously
derived equations.  However, here we repeat the ordering to give more insight
into the important role of the gauge conditions in the quasi-static limit.  We
start by writing the equations for the fields from potentials in dimensional
form:
\begin{subequations}
   \label{FieldsFromPotentialsDimensional}
   \begin{align}
     \vE & = -\grad \phi - k_3 \dt{\vA},
												\\
     \vB & = \curl{\vA}.
   \end{align}
 \end{subequations}

Putting these expressions into the generalized dimensional form of the Maxwell
equations (\refeqs{MaxwellEquations-Unit}) gives the following form for the
Maxwell equations
\begin{subequations}
   \label{MaxwellsEquations-potential}
   \begin{align}
     - \gradsq\phi(\vx,t) - k_3 \dt{}\dive{\vA(\vx,t)} &= 4 \pi k_1 \rho(\vx,t)
												\\
     - \gradsq\vA(\vx,t) + \grad\dive{\vA(\vx,t)}  &= -\frac{1}{c^2 k_3} 
              \lb \dt{\grad \phi(\vx,t)} + k_3 \dtsq{\vA(\vx,t)} 
              - 4 \pi k_1 \vJ(\vx,t) \rb,
   \end{align}
\end{subequations}
with \refeqs{DivBUnit} and (\ref{FaradaysLawUnit}) being satisfied automatically
by the forms in \refeq{FieldsFromPotentialsDimensional}.
Following the same procedure for de-dimensionalization and ordering 
(and using $k_3=1/c$ as before) gives
 \begin{eqaligned}
   \label{MaxwellsEquations-potential-ordered}
     \gradsqh \phiht(\vxht,\tht) &= -\aht \rhoht(\vxht,\tht) - \beta \Aop
     \dth{}\diveh{\vAht(\vxht,\tht)},
												\\
     \gradsqh\vAht(\vxht,\tht) &= -\aht \Jor \poA \vJht 
                  + \gradh \lb \beta \poA \dth{\phiht(\vxht,\tht)}
                  +\diveh{\vAht(\vxht,\tht)} \rb 
                  + \beta^2 \dtsqh{\vAht(\vxht,\tht)},
 \end{eqaligned}
where, in accordance with \refeq{FieldsFromPotentialsDimensional}, 
we define $\phibr = \vEbr \lbr$ and $\vAbr = \vBbr \lbr$.
The displacement current is responsible for the second and fourth terms in the
last equation.  We therefore see that in the $\beta \rightarrow 0$ limit, the
vector potential never contributes to the displacement current.

Before taking the quasi-static ordering, we consider the fully
relativistic case.
The dimensionless form can be seen by taking $\beta \rightarrow 1$ and also
setting the ratios $\phibr/\vAbr$ and $c \rbr/\vJbr$ to unity.
Using the Lorenz gauge allows writing the equations in the pleasingly
symmetric and uncoupled form
\begin{subequations}
   \label{MaxwellEquationsPot-Dimensionless}
   \begin{align}
       \lp \gradsqh - \dtsqh{} \rp \phiht(\vxht,\tht) &= -\aht \rht(\vxht,\tht),
                            \label{MaxwellsEqsScapotEM} \\
     \lp \gradsqh - \dtsqh{} \rp \vAht(\vxht,\tht) &= -\aht \vJht(\vxht,\tht),
                            \label{MaxwellsEqsVecpotEM}
   \end{align}
\end{subequations}
where the Lorenz gauge is
\begin{equation}
      \dth{\phiht(\vxht,\tht)} +\diveh{\vAht(\vxht,\tht)} = 0.
      \label{LorenzGauge}
\end{equation}
The previous form, \refeqs{MaxwellsEquations-potential}, is more general.

We now consider the quasi-static orderings.
The electro-quasi-static (EQS) regime can be defined as the ordering regime of
$\vAbr/\phibr \sim \beta$, in addition to $c \rbr/\vJbr \sim 1/\beta$ as
discussed before, to give to lowest order
\begin{subequations}
   \label{MaxwellsEquations-potential-EQS}
   \begin{align}
     \gradsqh \phiht(\vxht,\tht) &= -\aht \rhoht(\vxht,\tht),
												\\
     \gradsqh\vAht(\vxht,\tht) &= -\aht \vJht(\vxht,\tht)
                  + \gradh \lb \dth{\phiht} +\diveh{\vAht} \rb.
   \end{align}
\end{subequations}
The Lorenz gauge is required  to place these equations into standard form
involving only a Laplacian, which is the same form as the electrostatic (ES) and
magnetostatic (MS) equations.  Along with the Maxwell equations, the ordered
equations for calculating the fields in the EQS limit are
\begin{subequations}
   \label{FieldsFromPotentialsEQS}
   \begin{align}
     \vEht & = -\gradh \phiht,
												\\
     \vBht & = \curlh{\vAht}.
   \end{align}
\end{subequations}
Looking at the potential equations, the EQS equations look like the electrostatic
and magnetostatic equations. The time dependence is implicit in the fields, and
in the the use of Lorenz gauge for the EQS approximation.

The magneto-quasi-static (MQS) regime can be defined as the ordering regime of 
$\phibr/\vAbr \sim \beta$ to give to lowest order
\begin{subequations}
   \label{MaxwellsEquations-potential-MQS}
   \begin{align}
         \gradsqh \phiht &= -\aht \rhoht - \dth{\diveh{\vA}},
												\\
     \gradsqh\vAht &= -\aht \vJht + \gradh \lp \diveh{\vAht} \rp.
   \end{align}
\end{subequations}
The Coulomb gauge is required to place these equations into 
the same form as the electrostatic (ES) and magnetostatic (MS) equations.
The ordered equations for calculating the fields from potentials are
\begin{subequations}
   \label{FieldsFromPotentialsMQS}
   \begin{align}
     \vEht & = -\gradh \phiht - \dth{\vAht},
												\\
     \vBht & = \curlh{\vAht}.
   \end{align}
\end{subequations}
In this limit, the equations relating the fields and potentials are the same as
in the full EM limit.

In summary, the Maxwell equations in potential form for both the EQS and MQS
limits can be written as
\begin{subequations}
   \label{MaxwellsEquations-potential-EQSMQS}
   \begin{align}
     \gradsqh \phiht(\vxht,\tht) &= -\aht \rhoht(\vxht,\tht)
												\\
     \gradsqh\vAht(\vxht,\tht) &= -\aht \vJht(\vxht,\tht)
   \end{align}
\end{subequations}
using the Lorenz and Coulomb gauges respectively.
The obvious question arises: If the potential forms for the EQS and MQS limits
are the same, then how do different equations sets for the Maxwell equations
arise?  In the EQS limit, the use of the Lorenz gauge condition instead of the
Coulomb gauge condition adds the effect of the displacement current.  In the
MQS limit, the calculations of the $\vE, \vB$ fields from the potentials given
by \refeq{FieldsFromPotentialsMQS} include an explicit time dependence in the
calculation of $\vE$. This gives the inductive physics from Faraday's law.  This
is shown in more detail in \refsec{sec:jefimenko}.  The subtleties of the
quasi-static as opposed to static equations are discussed further in 
\refsec{sec:bound}.

The previous derivation of the EMQS equations,
\refeqs{MaxwellsEquations-darwin}, used the fact that the transverse and
longitudinal components of the electric field have different quasi-static
limits.  The derivation of the EMQS equations in the potential formulation
involves the gauge conditions.  In the EQS and MQS limits (Galilean EM), the
Lorenz gauge is ordered the same way as the continuity equation,
\refeq{ContinuityEquation-ordered}, to obtain the nice Laplacian form of the
potential form of the Maxwell equations: the Lorenz gauge is used in the
time-like EQS limit (where we have the full continuity equation), and the Coulomb
gauge is used in the space-like MQS limit (where we have $\dive{\vJ}=0$).

For the EMQS limit, we take advantage of the gauge freedom to order
$\vAbr/\phibr \sim \beta$, like EQS, but specify the Coulomb gauge, like
MQS~\footnote{Ordering like MQS but keeping the Lorenz gauge gives an
inconsistent ordering.}.  This ordering and gauge in
\refeq{MaxwellsEquations-potential-ordered} gives 
\begin{subequations}
   \label{EMQSPot1}
   \begin{align} 
    \gradsqh \phiht(\vxht,\tht) &= -\aht \rhoht(\vxht,\tht)
                                        \label{MaxwellsEqsScapotEMQS} \\ 
     \gradsqh\vAht(\vxht,\tht) &= -\aht \vJht + \gradh \dth{\phiht(\vxht,\tht)}.  
                                    \label{MaxwellsEqsVecpotEMQS}
   \end{align} 
\end{subequations} 
This form is not convenient because $\phi$ appears in the equation for $\vA$.
To write this equation in a more useful form, we again look at the longitudinal
and transverse components.  The right side of \refeq{MaxwellsEqsVecpotEMQS} is
divergence free, as can easily be seen using the continuity equation.  As a
consequence, the right side can be written as proportional to the transverse
current density, $\vJ_T$. As a consequence, although not proven here, only the 
transverse vector potential, $\vA_T$, is specified by this equation.

The formal proof that the right-side of \refeq{MaxwellsEqsVecpotEMQS} is the
transverse current density is given in Appendix~\ref{sec:helmholtz}. There it is
shown that the last term is equivalent to $\aht \vJht_L$, which allows writing
the right-side as $\aht$ times the transverse current.  This gives the EMQS
equations in potential form as
\begin{subequations}
   \label{EMQSpot}
   \begin{align}
     \gradsqh \phiht(\vxht,\tht) &= -\aht \rht(\vxht,\tht),
                            \label{MaxwellsEqsScapotEMQS2} \\
     \gradsqh\vAht_T(\vxht,\tht) &= -\aht \vJht_T(\vxht,\tht).
                            \label{MaxwellsEqsVecpotEMQS2}
   \end{align}
\end{subequations}
A detailed derivation of this form is also given by Jackson~\cite{Jackson2002},
along with an illuminating discussion of the implications of using the Coulomb
gauge.

By definition and because of the Coulomb gauge, the longitudinal vector
potential satisfies
\begin{eqaligned}
       \label{CoulombGauge}
         \dive{\vA_L}&=0, \\
         \curl{\vA_L}&=0.
\end{eqaligned}
If the boundary conditions on $\vA_L$ vanish at infinity, then this
leads to $\vA_L=0$.  In any case, $\vA_L$ is small and hence
the longitudinal part of the electric field, $\vE_L$, is given by the scalar
potential.  Contrast this with the Lorenz gauge that when ordered gives
\begin{equation}
     \frac{\phibr}{\vAbr_L} \sim \frac{1}{\beta}.
\end{equation}
Because the EQS ordering is a statement of the ratio of the magnitude of $E/B$
and $\vB$ is from $\vA_T$, the EQS ordering also implies $A_T \sim A_L$.  By
choosing the EQS ordering but the Coulomb gauge, one is effectively ordering
$\phi \gg A_T \gg A_L$.  In Jackson's book, the Darwin approximation is
explained as an ordering of the retarded potentials.  To show equivalency
between our forms of the EMQS equations and the Darwin approximation, we discuss
the retarded potentials next.

\section{Integral solutions of the potential equations}
\label{sec:intsol}

The full Maxwell equations in dimensionless potential form
using the Lorenz gauge are given by
\footnote{See, for example, \cite{Griffiths}, Sect.~7.4.3; \cite{JacksonSecond}, Sect.~6.4; \cite{JacksonThird}, Sect.~6.3.}
\begin{subequations}
  \label{MaxwellEqsPot}
  \begin{align}
   \gradsqh \phiht - \dtsqh{\phiht} &= -\aht \rht,
                            \label{MaxwellsEqsScapot} \\
   \gradsqh\vAht - \dtsqh{\vAht} &= -\aht \vJht.
                            \label{MaxwellsEqsVecpot}
 \end{align}
\end{subequations}
The solution to these equations are given by the Lorenz integral forms
\footnote{See, for example, \cite{Griffiths}, Sect.~9.9.1; \cite{JacksonSecond}, Sec.~6.5; \cite{JacksonThird}, Sect.~6.5.}
\begin{subequations}
 \label{PotRetarded}
 \begin{align}
   \phiht(\vxht,\tht)&= \frac{\aht}{4\pi} \int \frac{\lb \rht(\xph,\tht) \rbret }{R} \dVp,
                                                                \label{scaPotRetarded} \\
   \vAht(\vxht,\tht) &= \frac{\aht}{4\pi} \int \frac{\lb \vJht(\xph,\tht) \rbret}{R} \dVp,
                                                                \label{vecPotRetarded}
 \end{align}
\end{subequations}
where $R=|\vxht-\xph|$, and $\lb \ \ \ \rbret$ indicates that the argument is
evaluated at the retarded time $\thtp=\tht-R$. Here $R$ is unitless, but we do
not use the hat notation for convenience.  Dropping the time dependence
gives the electrostatic and magnetostatic solutions respectively.

Because the quasi-static equations are the same as the static equations but with
implicit time dependence, the solutions are the same as the static solutions
with implicit time dependence:
\begin{subequations}
 \label{PotRetardedQS}
 \begin{align}
   \phiht(\vxht,\tht)&= \frac{\aht}{4\pi} \int \frac{\rht(\xp,t)}{R} \dVp,
                                                                \label{scaPotQS} \\
   \vAht(\vxht,\tht) &= \frac{\aht}{4\pi} \int \frac{\vJht(\xp,t)}{R} \dVp.
                                                                \label{vecPotQS}
\end{align}
\end{subequations}
The lack of retardation in the potentials explains why the quasi-static limits are
sometimes referred to as the $c \rightarrow \infty$ limit.  The solutions are
indeed instantaneous, but for the EQS limit, the use of the Lorenz gauge condition
and the full continuity equation indicates that some causal effects remain.  The ordering of the EQS
limit says that the Lorenz gauge condition and continuity equation may be used for
describing slow motions.%

The integral solutions to the EMQS equations, \refeqs{EMQSpot}, are evidently
\begin{subequations}
 \label{PotRetardedEMQS}
 \begin{align}
   \phiht(\vxht,\tht)&= \frac{\aht}{4\pi} \int \frac{\rht(\xp,t)}{R} \dVp,
                                                                \label{scaPotEMQS} \\
   \vAht_T(\vxht,\tht) &= \frac{\aht}{4\pi} \int \frac{\vJht_T(\xp,t)}{R} \dVp.
                                                                \label{vecPotEMQS}
 \end{align}
\end{subequations}
In this form, the only difference between the EMQS and other quasi-static limits
is the use of the transverse current to give a transverse vector potential.
Recall from 
\refeq{MaxwellsEqsVecpotEMQS} and the discussion thereafter, the transverse
current includes the time derivative of the scalar potential, and thus has an 
implicit time dependence.
The implications of this will be discussed in the next section.
After a bit of algebra~\cite{Jackson2002,Krause:2007fa}, the
transverse current of the last expression can be rewritten and the vector
potential solution of the EMQS equation is
\begin{align}
    \vAht_T= \frac{\aht}{4\pi} 
          \int &\frac{1}{R} \lb \vJht(\xp,t) -\vRht\,\vRht \cdot \vJht(\xp,t) \rb \dVp,
                                                                \label{AC_Darwin}
\end{align}
where the unit vector $\vRht$ points from the source to the location where
$\vAht_T$ is calculated.
This is the form presented Jackson's book~\cite{JacksonSecond,JacksonThird} that
derived the Darwin approximation as a quasi-static approximation of
\refeq{vecPotRetarded}.  In this view of the EMQS equations, the non-Galilean
covariance can be seen as a result of taking a specific direction; i.e.,
the form here is obviously not rotationally symmetric.

The integral solutions are useful for studying simple radiating systems.
In these studies, it is useful to classify the different regions
as~\footnote{See, for example, Ref.~\cite{JacksonSecond}, Sec. 9.3.}:

\begin{tabular}{lll}
      The near (static) zone:    &  $ d \ll \lbr \ll \lambda$  &  $ \lbr/\lambda \ll 1$  
      \\
      The intermediate (induction) zone:    \ \ \ \ \ \ 
                                      &  $ d \ll \lbr \sim \lambda$\ \ \ \ \ \ 
                                      & $ \lbr/\lambda \sim 1$  
      \\
      The far (radiation) zone:    &  $ d \ll \lambda \ll \lbr$ &  $ \lbr/\lambda \gg 1$,
\end{tabular}
\noindent
where $d$ is the source dimension and $\lbr$ is the region of interest.  The
last column expresses orderings in terms of the electrical length (that is
$\beta_{osc}$ defined in \refeq{order-osc} with $N_{osc}=1$).  Although the
first region is termed the static zone, it is more properly described as the
quasi-static zone as there are, necessarily, time-dependent sources.  As shown
in Jackson, the difference between the near-field part of the solution and the
radiation part of the solution is that the near-field solution exponentially
decays.  Again, exponentially decaying solutions should always indicate a
solution that could be described with a first-order in time partial differential
equation, and hence, a quasi-static limit.


\section{Generalized Coulomb and Biot-Savart Laws in the quasi-static limit}
\label{sec:jefimenko}

The generalized Coulomb and Biot-Savart Laws are derived using the solutions
in the previous section.  For full electromagnetics, these are known as
Jefimenko's equations~\cite{jefimenko1966}, and they are discussed in Griffith's
textbook~\cite{Griffiths} and the third edition of Jackson~\cite{JacksonThird}.
The derivation of the generalized Coulomb and Biot-Savart
equations is straightforward by substituting the field-potential
relationships in \refeqs{FieldsFromPotentialsEQS} and
(\ref{FieldsFromPotentialsMQS}) into \refeqs{scaPotQS} and (\ref{vecPotQS}), and
the equivalent EMQS relationships.
The following standard identities are useful in the derivation:
\begin{equation}
   \label{Identities}
        \gradh\lp \frac{1}{R} \rp = -\frac{\vRht}{R^2};\ \ \ \ 
        \diveh{\frac{\vRht}{R}} = \frac{1}{R^2};\ \ \ \ 
        \diveh{\frac{\vRht}{R^2}} = 4 \pi \delta;\ \ \ \ 
        \curlh{\frac{\vRht}{R^n}} = 0.
\end{equation}
Here $\vRht=\vR/R$, and as defined previously, both $\vR$ and $R$ are unitless.
The $\delta$ function is also unitless.  These identities and their derivations
are found in most standard classical EM textbooks. 

For the EQS equations, we obtain
\begin{subequations}
 \label{JefimenkoEQS}
 \begin{align}
     \vEht(\vxht,\tht) & = \frac{\aht}{4\pi} \int \lb 
              \rht(\xp,t) \frac{\vRht}{R^2} \rb \dVp
                                     \label{JefimenkoEQS-E} \\
     \vBht(\vxht,\tht) & = \frac{\aht}{4\pi} \int
              \lb \vJht(\xp,t) \times \frac{\vRht}{R^2} \rb \dVp.
                                     \label{JefimenkoEQS-B}
 \end{align}
\end{subequations}
These have the same form as the Coulomb and Biot-Savart laws; only the
implicit time dependency is added to the fields.
In the MQS limit, we obtain
\begin{subequations}
 \label{JefimenkoMQS}
 \begin{align}
     \vEht(\vxht,\tht) & = \frac{\aht}{4\pi} \int \lb 
              \rht(\xp,t) \frac{\vRht}{R^2} 
            - \frac{\vJdt(\xp,t)}{R} \rb \dVp,
                                      \label{JefimenkoMQS-E} \\
     \vBht(\vxht,\tht) & = \frac{\aht}{4\pi} \int
              \lb \vJ(\xp,t) \times \frac{\vRht}{R^2} \rb \dVp,
                                      \label{JefimenkoMQS-B}
 \end{align}
\end{subequations}
where the dot denotes a time derivative.  The presence of the time derivative of
$\vJ$ in the generalized Coulomb law is responsible for the induction term in
Faraday's law.

To derive the Maxwell equations in the quasi-static limits, the divergence and the
curl of the above equations is taken.  The algebra is straightforward using
\refeqs{Identities}; however, the equations of most interest are those with time
dependence  (the curl of $\vE$ in the MQS limit, and the curl of $\vB$ in the
EQS limit), so that algebra is shown in more detail.  The curl of
the MQS limit of the electric field equation, \refeq{JefimenkoMQS-E}, is
 \begin{align}
   \curlh{\vEht} & = \frac{\aht}{4\pi} \int \lb 
                  \rht(\xp,t) \curlh{\frac{\vRht}{R^2}}
                  -  \curlh{\frac{\vJdt(\xp,t)}{R}}\rb \dVp
                                      \nonumber \\
         & = - \dt{}\lp \frac{\aht}{4\pi} \int
                 \vJht(\xp,t) \times \frac{\vRht}{R^2} \dVp \rp
                                      \nonumber \\
         & = - \dt{\vBht},
                                      \nonumber
 \end{align}
giving Faraday's law in the MQS limit.

We next derive the Maxwell-Amp\`ere law in the EQS limit by taking the curl
of the magnetic field in \refeq{JefimenkoEQS-B}.
Before taking the curl of $\vBht$, it is useful to recall this vector identity:
\begin{equation}
      \label{vectorIdentity-curlcurl}
      \curl{\curl{\vA}} = - \gradsq\vA+\grad\dive{\vA}.
\end{equation}
When we take the curl of $\vBht$ from this generalized Biot-Savart equation, we
are effectively taking the curl-curl of the integral solutions for
$\vA$.  However, the integral solution was derived by inverting the Laplacian
for $\vA$, thus there is a difference.  That difference is the last term in that
equation, and the divergence of $\vA$ will bring in the gauge condition.  With
this background, 
the curl of \refeq{JefimenkoEQS-B} is most easily done by starting with
\refeq{vecPotQS}:
 \begin{align}
   \curlh{\vBht(\vxht,\tht)} & = \frac{\aht}{4\pi} \int \lb 
                  \curlh{} \curlh{}  \frac{\vJht(\xp,t)}{R} 
                  \rb \dVp
                                      \nonumber \\
              & = \frac{\aht}{4\pi} \int \lb
                  - \gradsqh \frac{\vJht}{R} + \gradh \diveh{\frac{\vJht(\xp,t)}{R}}
                  \rb \dVp
                                      \nonumber \\
              & = \frac{\aht}{4\pi} \int \lb \vJht 4\pi \delta(R) \rb \dVp
                  + \gradh \diveh{\vA(\vxht,\tht)}
                                      \nonumber \\
              & = \aht \vJht(\vxht,\tht) - \gradh \dt{\phiht(\vxht,\tht)}
                                      \nonumber \\
              & = \aht \vJht(\vxht,\tht) + \dt{\vEht(\vxht,\tht)},
                                      \nonumber 
 \end{align}
recovering the full the Maxwell-Amp\`ere law.  

The generalized Coulomb and Biot-Savart equations for the EMQS limit are
\begin{align}
      \vEht(\vxht,\tht) & = \frac{\aht}{4\pi} \int \lb 
              \rht(\xp,t) \frac{\vRht}{R^2} 
            - \frac{\vJdt_T(\xp,t)}{R} \rb \dVp.
                                      \label{JefimenkoEMQS-E} \\
      \vBht(\vxht,\tht) & = \frac{\aht}{4\pi} \int
              \lb \vJht_T(\xp,t) \times \frac{\vRht}{R^2} \rb \dVp.
                                      \label{JefimenkoEMQS-B}
\end{align}
The recovery of the EMQS limit of the Maxwell equations is straightforward.  The induction
term is recovered from the vector potential contribution to the electric field,
as in the MQS limit.  For the longitudinal component of the displacement
current, the fact that the current is only transverse implies that the
scalar potential contributes, as seen in \refeq{MaxwellsEqsVecpotEMQS}.  This
differs from the EQS limit, where the full displacement current appears from the
use of the Lorenz gauge.
\section{Maxwell equations for macroscopic media} 
\label{sec:bound}

In this section, we discuss the quasi-static limits as they apply to the Maxwell
equations for macroscopic media with bound electrons.  This area is thoroughly
studied by Rousseaux and
colleagues~\cite{Montigny:2006dq,Rousseaux:2008ij,Rousseaux:2008ev,Rousseaux:2013ft,Rapetti:2014ha}
so this discussion does not re-derive all of the results, but summarizes them
and places into the context of the current work.  

Macroscopic media are materials that have a different field within them because of
the response of the material to an external field.  For electrically-active media, a
polarization field arises, and for magnetically-active materials a magnetization
field arises.  These fields are then added to the original applied fields to
produce the displacement field, $\vD$, and the auxiliary magnetic field, $\vH$:
\begin{eqaligned}
     \vD &= \frac{\vE}{4\pi k_1} + \lambda \vP
												\\
     \vH &= \frac{c^2 k_3}{4\pi k_1} \vB - \lambda^\prime \vM,
\end{eqaligned}
where $\lambda$ and $\lambda^\prime$ are either unity or $4 \pi$ depending on
whether the unit system is rationalized or not. 
Thankfully, all unit systems have the same form for the constitutive 
relations (also the same for dimensionless forms):
\begin{eqaligned}
     \vD &= \eps \vE                        \\
     \vH &= \frac{1}{\mu} \vB.
\end{eqaligned}
In taking the Galilean limit of the macroscopic Maxwell equations, it is
important to understand their Lorentz transformation properties.  As discussed
extensively in Rousseaux's review paper, Ref.~\cite{Rousseaux:2013ft}, the key
is the Minkowski hypothesis that the constitutive constants are invariant in any
inertial frame:
\begin{eqaligned}
     \vDp &= \epsilon \vEp           \\
     \vHp &= \frac{1}{\mu} \vBp.
\end{eqaligned}
The Maxwell equations with macroscopic media are written as
 \begin{eqaligned}
   \label{MaxwellsEquations-macroscopicMedia}
     \diveh{\vDht} &= \rhoht                     \\
     \diveh{\vBht} &=0                           \\
     \curlh{\vEht} &= - \dth{\vBht}              \\
     \curlh{\vHht} &=   \dth{\vDht} + \vJht.
 \end{eqaligned}
These will have the same form whether dimensional or dimensionless.  In
generalizing the ordered Maxwell equations, the dimensions of $\vD$ are
$\bar{D}=\epsbr \vEbr$, and of $\vH$ are $\bar{H}=\vBbr/\mubr$.

The other constitutive relationship that will be used is $\vJ = \sigma \vE =
\vE/\eta$.  Although this form is not Galilean covariant, the generalizations
add an algebraic complexity that is unimportant.  The dimensional units are
straightforward.  To match onto circuit theory, resistivity will be used instead
of conductivity.

Using these relationships, the generalization of the ordered Maxwell equations
in \refeqs{MaxwellsEquations-ordered} is
\begin{eqaligned}
   \label{MaxwellsEquations-macroordered}
     \diveh{\vDht}& = \rht                                   \\
     \diveh{\vBht}& = 0                                           \\
     \curlh{\vEht}& =-\lp \beta \BoE \rp \dth{\vBht}              \\
     \curlh{\vHht}& = \lp \beta \EoB \cocm \rp \dth{\vDht} +
               \lp \EoB \cocm \aht_\eta \rp  \vEht,
\end{eqaligned}
where $c_m^2=1/(\mu\eps)$ is the speed of light in the media.
The ratio $c/c_m$ does not significantly change the results of ordering the
vacuum Maxwell equations.
However, the last term does cause a significant change.
The factor of $\aht_\eta$ can be expressed in multiple ways (shown here in
$MKSA$ units):
\begin{eqaligned}
      \aht_\eta & = \frac{c \mu_0 \lbr}{\eta}
												\\
                & = \frac{c}{\lbr} \frac{\lbr^2}{D_m} = \frac{\taum}{\tauem}
												\\
                & = \frac{\lbr}{c} \frac{1}{\epso\eta} = \frac{\tauem}{\taue},
\end{eqaligned}
where $D_m=\eta/\mu_0$ is the resistive diffusivity.  As shown below, the ratio
$\tauem/\taum$ is related to the $L/R$ time of LR circuits, while the ratio
$\taue/\tauem$ is related to the $RC$ time of RC circuits.
The constitutive relationship for $\vJ$, however, means that the appropriate 
asymptotic regime 
depends on the value of the parameter, $\eta$. With values ranging
from zero to infinity, this will put the range of parameters outside of the
orderings used in the vacuum equations.  The regions outside of the range of
validity are the static limits, and the full EM equations.  This area of
validity for the regimes will be discussed in \refsec{sec:regimes}, but first
the relationship of the Maxwell equations and circuit theory is briefly reviewed
to provide greater intuition.

\subsection{Circuit equations with simple elements}
\label{sec:circuit}

The relationship between the Maxwell equations and circuit theory is
well-understood, with Carson 1927 being an early reference~\cite{carson1927}.  In
Carson, the derivation used the retarded potentials,~\refeqs{PotRetarded}, to
show that the derivation of Kirchoff's ``laws''%
~\footnote{The quotes around 
      Kirchoff's ``laws'' are because they can be derived from Maxwell's
equations with a large number of assumptions, and hence, are not laws.}
requires assuming that the electrical length is small, and that
radiation effects are negligible (see~\refsec{sec:intsol}).  The derivation here
is more heuristic and follows directly from the EQS and MQS limits.  To derive
these circuit equations, it is assumed that the circuit elements are small
relative to the wavelength, time-dependence is important, and that the full
Maxwell equations are not needed.  In the discussion after \refeq{order-osc},
these assumptions were discussed more quantitatively for normal U.S. household
circuits to show that these assumptions are easily satisfied.

\paragraph{RC Circuits} First consider a direct current (DC) circuit with a single capacitor and
resistor connected in a loop.  Because of
the charge on the capacitor and the relatively low magnetic field energy of the
resistor, $c \rbr \gg \vJbr$, we are in the EQS regime.  The time dependence
is given by the Maxwell-Amp\`ere equation for the EQS equation.  We integrate that equation
across the cross sectional area, to calculate the effective total current (from
$I = \int \vJ \cdot d\vS$).  Doing so is equivalent to Kirchoff's ``Current
Law'', which is a reduced form of charge conservation:
\begin{eqaligned}
      \int \curl{\vH} \cdot d\vS  & =  \int \dt{\vD} \cdot d\vS  + \int \frac{1}{\eta} \vE \cdot d\vS,
										           \\
      \int \vH \cdot d\vl = 0 & \approx \frac{\eps A_C}{d} \Dt{V} + \frac{A_R}{\eta l} V
										           \\
                              & \approx C \Dt{V} + \frac{V}{R},
 \label{RCderiv}
\end{eqaligned}
where $V$ is the voltage, $R$ is the resistance of the resistor, and $C$ is the
capacitance of the capacitor.  We approximate $E \approx V/d$ for a capacitor,
where $d$ is the gap of the capacitor, and $E \approx V/l$ for a resistor, where $l$ is the
length of the resistor.  If there is a source, this equation becomes
\begin{equation}
      \tau_{RC} \frac{dV}{dt} + V = V_\text{source},
    \label{RCcircuit}
\end{equation}
where $\tau_{RC} = RC$.  This is a first-order differential equation with
exponential solutions, as expected from the first-order EQS equations.  
As $\tau_{RC} \rightarrow 0$,  the capacitor charges (or discharges) more
quickly.  Eventually this speed will violate the quasi-static assumption made in
deriving this equation.  At the other limit, the limit of $\tau_{RC} \rightarrow
\infty$ means that eventually the discharge is so slow that the electrostatic limit can be
used.

The heuristic nature of the derivation is evident: the inductance of the loop is
ignored; the resistance of the capacitor is ignored; and the complicated
geometry of the circuit elements in the integrals is also ignored.
For alternating current (AC) circuits, the relationship to
$\betaosc$, \refeq{order-osc}, is set by the frequency of the current.  
As discussed in the Introduction, the wavelength of \SI{60}{Hz} is 
\SI{5E6}{m}, and the size of most circuit elements is on the order of a
centimeters, resulting in a large electrical length.  In this case, the relevant
$\tbr$ used in $\Nosc$ would be the RC time scale modified for the impedance
of the capacitor, leading to a very small $\betaosc$.

Because in the EQS regime $\vE = - \grad \phi$, the electromotive
force is equivalent to the potential drop across a segment:
\begin{equation}
     \mathcal{E} \equiv \int_{\vx_0}^{\vx_1} \vE \cdot d\vl 
     \stackrel{EQS}{=} \phi(\vx_1) - \phi(\vx_0).
  \label{EMF}
\end{equation}
This is not true in the MQS regime.

\paragraph{LR Circuits}  Here we consider a simple loop with an inductor, a resistor,
and a voltage source.  Because inductors store magnetic fields, 
$\vBbr  \gg \vEbr$ and we are in the MQS regime.  The time dependence
term is in Faraday's equation.  Considering an integral over the loop, we have
\begin{eqaligned}
      \int \curl{\vE} \cdot d\vS  & = - \int \dt{\vB} \cdot d\vS 
										           \\
      \oint \eta \vJ \cdot d\vl & =  - \dt{} \oint \vA \cdot d\vl 
										           \\
      IR & =  - \frac{\aht}{r \pi} \dt{I} \int d\vl \cdot \int \frac{1}{R} d\vl^\prime
										           \\
      IR & =  - L \dt{I},
 \label{RLderiv}
\end{eqaligned}
\noindent
where an approximate form for the integral solution for $\vA$ in the MQS regime
was used,~\refeq{vecPotQS}, and $L$ uses the definition of inductance.
With a voltage source, this equation becomes
\begin{align}
      \tau_{RL} \frac{dI}{dt} + I = \frac{1}{R} V_\text{source},
      \label{LRcircuit}
\end{align}
where $\tau_{RL} = L/R$.
Here, we are ignoring both the capacitive effects of the inductor geometry,
and the resistance of the inductor itself.  In the MQS approximation, the
$\mathcal{E}$ is not the same as the voltage drop because the calculation of
$\vE$ includes the vector potential.  Nonetheless, the form in the second line
of \refeq{RLderiv} shows that the derivation is equivalent to calculating the
$\mathcal{E}$ across each element and effectively using Kirchoff's voltage law.
The static version of this equation corresponds to $\tau_{RL} = 0$, which
requires either the inductance  to go to zero or the resistance to go to
infinity.

\paragraph{RLC Circuits}
There are two ways to derive the equations for a simple RLC circuit.  As alluded
to above for a simple circuit, each element of the circuit should be considered
piecewise.  In this case, we use the fact that the solutions of the individual
EQS and MQS equations can be additive, as discussed by LBLL~\cite{LBLL}. The
appropriate $\mathcal{E}$ can be calculated for each element (including the
capacitor) and sum the segments.  The result is
\begin{align}
      \frac{d^2I}{dt^2} +  \frac{1}{\tau_{RL}} \frac{dI}{dt} + \frac{1}{LC} I =
      \frac{1}{\tau_{RL} R} \frac{d V_\text{source}}{dt}.
   \label{RLCcircuit}
\end{align}
Another derivation method uses the EMQS approximation.  The key question is
whether the $\mathcal{E}$ contribution from the capacitor, which comes from the
displacement current in \refeq{RCderiv}, is primarily longitudinal.  Because the
normal derivation capacitive effects uses Gauss's law, it is seen that it is.
For more complicated geometries where the separation of capacitive, inductive,
and resistive effects are not possible, the EMQS approximation provides a more
useful approximation than the additive properties of EQS and MQS.  This is
discussed in more detail next.

\subsection{Static and Quasi-static regimes of validity}
\label{sec:regimes}


%
%
%
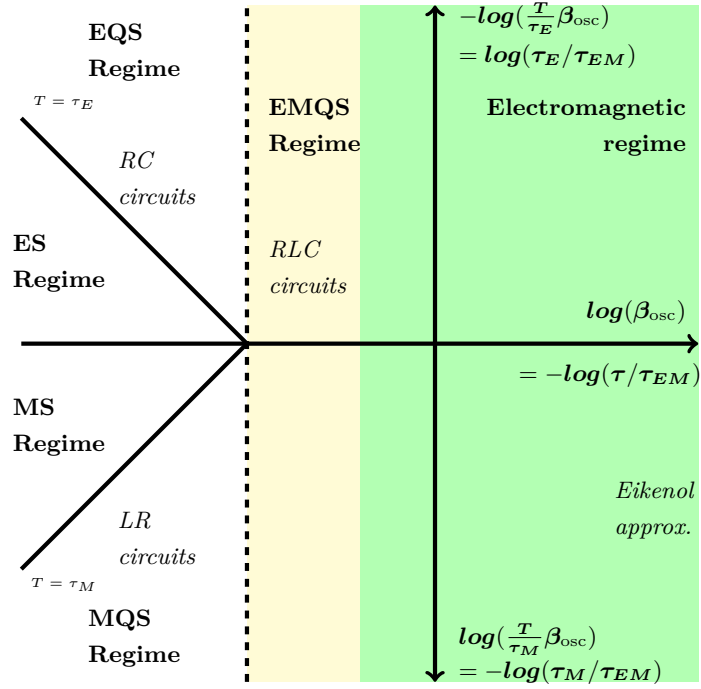
\begin{figure}
\begin{center}
\begin{tikzpicture}[ ]

  \draw[->,ultra thick] (-4.5,0) -- (4.5,0);
  \draw[<->,ultra thick] (1,-4.5) -- (1,4.5);
  \draw[ultra thick,dashed] (-1.5,-4.5) -- (-1.5, 4.5);
  \draw[ultra thick] (-1.5,0) -- (-4.5, 3.0);
  \draw[ultra thick] (-1.5,0) -- (-4.5,-3.0);
  \begin{scope}[on background layer]
        \fill[fill=yellow!20!white] (-1.50,-4.50) rectangle (-0.0,4.50);
  \end{scope}
  \begin{scope}[on background layer]
        \fill[fill=green!30!white] (+4.50,-4.50) rectangle (-0.00,4.50);
  \end{scope}

  \node (n00) [anchor=south west] at (2.9, 0.15)%
               {
                     $\mathbold{log(\betaosc)}$
               };
  \node (n00a)[anchor=north west] at (2.00,-0.15)%
               {
                     $\mathbold{=-log(\tau/\tauem)}$
               };
  \node (n01) [anchor=north west] at (1.2,-3.60)%
               {
                     $\mathbold{log(\frac{\tunit}{\taum}\betaosc)}$
               };
  \node (n01a) [anchor=north west] at (1.2, -4.09)%
               {
                     $\mathbold{=-log(\taum/\tauem)}$
               };
  \node (n02) [anchor=south west] at (1.2, +4.00)%
               {
                     $\mathbold{-log(\frac{\tunit}{\taue}\betaosc)}$
               };
  \node (n02a) [anchor=south west] at (1.2, +3.55)%
               {
                     $\mathbold{=log(\taue/\tauem)}$
               };

  \node (n11) [anchor=north west] at (+1.5, 3.5)%
               {
                  \begin{tabular}{r}
                    \bf Electromagnetic \\
                    \bf regime
                  \end{tabular}
               };
  \node (n12) [anchor=north west] at (-1.4, 3.5)%
               {
                  \begin{tabular}{l}
                    \bf EMQS \\
                    \bf Regime
                  \end{tabular}
               };
  \node (n32) [anchor=north west] at (-1.4, +1.6)%
               {
                  \begin{tabular}{l}
                        {\em \small RLC} \\
                        {\em \small circuits}
                  \end{tabular}
               };
  \node (n13) [anchor=north east] at (-2.2, -3.3)%
               {
                  \begin{tabular}{l}
                    \bf MQS \\
                    \bf Regime
                  \end{tabular}
               };
  \node (n23) [anchor=north east] at (-2.0, -2.0)%
               {
                  \begin{tabular}{l}
                        {\em \small LR} \\
                        {\em \small circuits}
                  \end{tabular}
               };
  \node (n03) [anchor=north east] at (-3.2, -0.5)%
               {
                  \begin{tabular}{l}
                    \bf MS \\
                    \bf Regime
                  \end{tabular}
               };
  \node (n03a) [anchor=north east] at (-2.2, -3.0)%
               {
                 \begin{minipage}{92pt}
                        \begingroup\makeatletter\def\f@size{5}\check@mathfonts
                        \def\maketag@@@#1{\hbox{\m@th\large\normalfont#1}}%
                    $\tunit = \taum$ 
                    \endgroup
                 \end{minipage}
               };

  \node (n14) [anchor=south east] at (-2.2, +3.3)%
               {
                  \begin{tabular}{l}
                    \bf EQS \\
                    \bf Regime
                  \end{tabular}
               };
  \node (n33) [anchor=south east] at (-2.0, +1.6)%
               {
                  \begin{tabular}{l}
                        {\em \small RC} \\
                        {\em \small circuits}
                  \end{tabular}
               };
  \node (n15) [anchor=south east] at (-3.2, +0.5)%
               {
                  \begin{tabular}{l}
                    \bf ES \\
                    \bf Regime
                  \end{tabular}
               };

  \node (n15a) [anchor=south east] at (-2.2, +3.0)%
               {
                 \begin{minipage}{92pt}
                        \begingroup\makeatletter\def\f@size{5}\check@mathfonts
                        \def\maketag@@@#1{\hbox{\m@th\large\normalfont#1}}%
                    $\tunit = \taue$ 
                    \endgroup
                 \end{minipage}
               };

  \node (n33) [anchor=north west] at (3.2, -1.6)%
               {
                  \begin{tabular}{l}
                        {\em \small Eikenol} \\
                        {\em \small approx.}
                  \end{tabular}
               };

\end{tikzpicture}
\end{center}
\caption{ 
      The Rousseaux-Rapetti diagram drawn in terms of $\betaosc$ shows the
      regions of validity for the various approximations heuristically.  The
      x-axis is in terms of the small parameter.  The upper vertical axis is
      proportional to $\tau_{RC}^{-1}$, and the lower vertical axis is
      proportional to $\tau_{LR}^{-1}$.
}
\label{fig:EMregimes}
\end{figure}

There are many situations where simple circuit assumptions are violated;
the most important of these occur where the size of the devices are
such that each individual device has capacitive, resistive, and inductive
contributions.  Figuring out the appropriate approximation to use can be
complicated.  As seen above, geometry is an important consideration;
however, Refs.~\cite{melcher-haus1989,Rapetti:2011uq,Rapetti:2014ha} provide
useful heuristics based on the ordering parameters above:
\begin{eqaligned}
   \label{ResistiveTimeScalesOrdering}
      0 \le \tbr \le \tauem            &\ \ \ \text{Electromagnetic\ regime}
										\\
      \tauem \ll \tbr \ll \taue, \taum &\ \ \ \text{Quasi-static\ regime}
												\\
      \taue, \taum \ll \tbr            &\ \ \ \text{Static\ regime}.
\end{eqaligned}
The relationship $\tauem \ll \tbr$ is equivalent to the $\beta \ll 1$ ordering
in this paper.  The relationship to the static regimes has not been previously
discussed here.
This can be cast into a form that is closer to this paper by relating the key
parameter $\tauem$ to $\beta_{osc}$, \refeq{order-osc}, used in this paper:
 \begin{equation}
   \label{MyScales}
   \tauem = \tbr \beta_{osc}.
 \end{equation}
Using these relationships, a diagram heuristically defining the regimes may be
drawn.  This follows those in Refs.~\cite{Rapetti:2011uq,Rapetti:2014ha} where
the key insight was to take the log of the axis such that order unity defines the
origin of the axis, and factors less than unity become the negative regions.
This allows the $y=0$ line to correspond to the $\eta = \infty$ for the
$ES/EQS$ regime and $\eta = 0$ for the  $MS/MQS$ regimes.  The discontinuity of
the regimes is a consequence of the separation of the time-like and space-like
regions of spacetime.
Graphically, the parameter space is divided into six different regimes, as seen in
Fig.~\ref{fig:EMregimes}.~\footnote{We term this diagram the Rousseaux-Rapetti
      diagram because the first appearance of the figure is in an unpublished
      Rousseaux document, and published versions are in two Rapetti-Rousseaux
      papers.  Rapetti and Rousseaux define the $x$-axis as
      $\log(\tbr/\tauem)$.  Our choice of $\log(\beta_{osc})$ as the
      $x$-axis means we are flipped relative to the published Rousseaux-Rapetti
      diagrams.   In addition to expressing the diagram in terms of
      $\beta_{osc}$, our diagram shifts things away from the y-axis to show that
      $\beta \ll 1$ is needed for the ordering.}.  
This provides an intuitive picture of when each regime is valid.
\section{Conclusions} 
\label{sec:conclusions}

\begin{table}
\begin{tabular}{lCCC}
\toprule
\textbf{Regime} & \textbf{Electro-QuasiStatic} 
                & \textbf{Magneto-QuasiStatic} 
                & \textbf{EM-QuasiStatic}
                                                                  \\ 
                & \textbf{(EQS)} 
                & \textbf{(MQS)} 
                & \textbf{(EMQS)}
                                                                  \\ \midrule
Ordering        & \bigO{\EoB, \roJ} \sim \frac{1}{\beta} 
                & \bigO{\EoB, \roJ} \sim \beta       
                & \bigO{\EToB} \sim \beta,\  
                  \bigO{\frac{\vELbr}{\vBbr}, \roJ} \sim \frac{1}{\beta}
                                                                  \\ \midrule
Maxwell Eqns. & \dive{\vE}=\aht \rho 
                    & \dive{\vE}=\aht \rho  
                    & \dive{\vE}=\aht \rho  
                                                                  \\ 
                    & \dive{\vB}=0 
                    & \dive{\vB}=0
                    & \dive{\vB}=0
                                                                  \\ 
                    & \curl{\vE}=0 
                    & \curl{\vE}=-\dt{\vB}
                    & \curl{\vE}=-\dt{\vB}
                                                                  \\ 
                   & \curl{\vB}= \dt{\vE} + \aht \vJ 
                   & \curl{\vB}= \aht \vJ
                   & \curl{\vB}= \dt{\vEL} + \aht \vJ 
                                                      \\ \midrule
Continuity Eqn. &  \dt{\rho} + \dive{\vJ}=0
                    &  \dive{\vJ}=0
                    &  \dt{\rho} + \dive{\vJ}=0
                                                      \\ \midrule
Galilean            &  \vEp=\vE
                    &  \vEp =   \vE + \vV \times \vB
                    & \textbf{Not}
                                                      \\          
Transform           & \vBp =  \vB - \vV \times \vE
                    & \vBp = \vB
                    & \textbf{Applicable}
                                                      \\          
                    &  \rhop=\rho
                    & \rhop   = \rho - \vV \cdot \vJ
                    &  
                                                      \\          
                    & \vJp = \vJ - \vV \rho
                    & \vJp = \vJ
                    &  
                                                      \\ \midrule
Fields from         &  \vE = -\grad \phi
                    &  \vE = -\grad \phi - \dt{\vA} 
                    &  \vE = -\grad \phi - \dt{\vA} 
                                                      \\ 
Potentials          &  \vB = \curl{\vA}
                    &  \vB = \curl{\vA}
                    &  \vB = \curl{\vA}
                                                      \\ \midrule
Potential ordering  & \bigO{\poA} \sim \frac{1}{\beta}                  
                    & \bigO{\poA} \sim  \beta              
                    &  \bigO{\frac{\phibr}{\vAbr_L} } \sim \beta,\  
                       \bigO{\frac{\phibr}{\vAbr_T} } \sim \frac{1}{\beta}
                                                                  \\  \midrule
Maxwell Eqns. & \gradsq \phi = -\aht \rho
                    & \gradsq \phi = -\aht \rho
                    & \gradsq \phi = -\aht \rho
                                                                  \\ 
(Potential Form)    & \gradsq\vA   = -\aht \vJ
                    & \gradsq\vA   = -\aht \vJ
                    & \gradsq\vA_T = -\aht \vJ_T
                                                      \\ 
                                                      \\ 
Gauge Condition     &  \dt{\phi} + \dive{\vA}=0
                    &  \dive{\vA}=0
                    &  \dive{\vA}=0
                                                                  \\ \midrule
Potential Solutions & \phi = \frac{\aht}{4 \pi} \int \frac{\rhopdep}{R} \dVp
                    & \phi = \frac{\aht}{4 \pi} \int \frac{\rhopdep}{R} \dVp
                    & \phi = \frac{\aht}{4 \pi} \int \frac{\rhopdep}{R} \dVp
                                                                  \\ 
(Integral form)     & \vA   = \frac{\aht}{4 \pi} \int \frac{\vJpdep}{R} \dVp
                    & \vA   = \frac{\aht}{4 \pi} \int \frac{\vJpdep}{R} \dVp
                    & \vA_T = \frac{\aht}{4 \pi} \int \frac{\vJTpdep}{R} \dVp
                                                                  \\ \midrule
Generalized         & \vE = \frac{\aht}{4 \pi} \int \rhopdep\frac{\vRht}{R^2} \dVp
                    & \vE = \frac{\aht}{4 \pi} \int \lb \rhopdep\frac{\vRht}{R^2} \right.
                    & \vE = \frac{\aht}{4 \pi} \int \lb \rhopdep\frac{\vRht}{R^2}\right. 
                                                                  \\ 
Coulomb and         &                                                                
                    &  \left. -\vJdpdep \frac{1}{R}\rb \dVp
                    &  \left. -\dot{\vJ_T}(\xp,t) \frac{1}{R}\rb \dVp
                                                                  \\ 
 Biot-Savart Laws   & \vB = \frac{\aht}{4 \pi} \int \vJ \times \frac{\vRht}{R^2} \dVp
                    & \vB = \frac{\aht}{4 \pi} \int \vJ \times \frac{\vRht}{R^2} \dVp
                    & \vB = \frac{\aht}{4 \pi} \int \vJ_T \times \frac{\vRht}{R^2} \dVp
                                                      \\ \midrule
Equiv. Circuit Eqn.  &  \tau_{RC} \frac{dV}{dt} + V = 0
                    &  \tau_{RL} \frac{dI}{dt} + I = 0
                    & \frac{d^2I}{dt^2} +  \frac{1}{\tau_{RL}} \frac{dI}{dt} + \frac{1}{LC} I =0
                                                      \\ \bottomrule
\end{tabular}
\caption{Summary of the results of this paper enables a comparison of the three
      quasi-static regimes of the Maxwell equations in all of their forms.  In
      this table, the equations are written in dimensionless form, but the hats
      are not used, and in referring to the units in the ordering, $k_3=1/c$ is
      used to place $\vE$ and $\vB$ in the same units for simplicity.}
\label{table:summary}
\end{table}

In this paper, three quasi-static limits of the Maxwell equations -- the
electro-quasi-static (EQS), the magneto-quasi-static (MQS), and the
electromagnetic-quasi-static (EMQS) -- are derived using an ordering analysis
with the small parameter given by the ratio of the characteristic length scale
to  $c$ times the time scale.  The formal perturbation technique is more precise
and illuminating than phrasing the quasi-static as taking the $c \rightarrow
\infty$ limit.  The three limits are expressed in four different forms of the
Maxwell equations: the local differential equations in terms of fields and
potentials, and the global integral solutions in terms of fields and potentials.
Summaries of the results are presented in \reftab{table:summary} and
Appendix~\ref{sec:sumfigs}.  The derivation for EMQS, or Darwin's approximation,
in this paper uses the Maxwell equations, and the integral solutions use to show
equivalence to the derivation in Jackson's book that cast it as a truncation of
the retarded potentials.  The EMQS limit is shown to not be Galilean covariant.
This approximation is obviously useful given it's many applications, but care
must be taken due to the lack of Galilean relativity.

This paper reviews the unit systems and issues that guide the choices of units.
Whether a system is unrationalized or rationalized is a choice of whether
factors of $4 \pi$ appear in the differential equations (left side of
Fig.~\ref{fig:EMstudy}) or the integral equations (right side of
Fig.~\ref{fig:EMstudy})~\footnote{The dimensionless equations presented here are
rationalized because the author agrees that it is more rational.}.  Unit systems
that have the same units for $\vE$ and $\vB$ ($k_3=1/c$) are generally useful in
regimes where Lorentz covariance is important, while having different units
($k_3=1$) provides more useful units for the Galilean, or engineering, limit.
There is only one parameter that truly controls the unit system as given by the
Buckingham $\Pi$ theorem.  The MKSA unit system makes it appear as if two
constants, with mass in their base unit definition, are needed. This, as well as
using current as a base unit rather than charge,  seems designed to create
confusion for physicists.  This system feels natural only in the engineering
limit where everyday usage involves macroscopic media, the quasi-static limit,
and force calculations involving Newton's law.


In the original LBLL paper, the ordering parameter $\beta$ was viewed strictly
as the ratio of an inertial frame velocity to $c$.  This paper broadens that
view with more general definitions of $\beta$.  With the introduction of
$\beta_{osc}$, this ordering parameter is explicitly related to the more
commonly used electrical length, $L/\lambda$ (as used in
Carson~\cite{carson1927} for example).  For the case of rotating media -- the
Wilson-Wilson experiment~\cite{wilson1913} or the 
Barnett effect~\cite{barnett1915} -- the slow rotation speed serves as the expansion
parameter.  Finally, in deriving the Lagrangian of a single particle with a
velocity small relative to the speed of light, this perturbation expansion
applies regardless of whether the particle is moving inertially.

A major advantage of the quasi-static limits is the elimination of light waves:
yielding significant analytical and computational simplicity.  In the EQS and
MQS limits, this is the result of changing the order of the equations from
second-order to first order.  Decaying EM fields suggest the applicability of
the quasi-static equations as seen in such examples as non-propagating waveguide
modes, near-field solutions from radiating source, and the $RC$ and $LR$ circuit
equations.  When studying a regime in which a quasi-static limit is valid, the
ability to eliminate the fast light-wave time scale yields valuable numerical
benefits.  If solving for the EM fields with frequency-domain or implicit
time-domain methods, then the resultant matrices will be better conditioned due
to the analytic elimination of a faster time scale: the largest eigenvalue
becomes smaller.  If solving explicitly with time domain methods, then time
steps can be made larger because of the larger Courant-Friedrichs-Levy (CFL)
limit.  The use of a formal perturbation theory enables estimates of the errors
associated with using this approximation. These estimates can then be included
within the numerical solution as an {\em a posteriori} check on validity.
Although this paper provides a framework for such developments, further work is
needed.

Studying quasi-static limits is also beneficial for pedagogical reasons.  Here,
the generalized Coulomb and Biot-Savart laws for the EQS limit are derived for
the first time.  This derivation illuminates the close relationship between the
Lorenz gauge condition and the displacement current, as well as the induction
term and the electric field contributions from the time derivative of the
magnetic vector potential.  As another simple example, consider the role of
displacement current in a parallel plate capacitor with a time-dependent
current.  In the original LBLL paper, they state that the EQS equations
\emph{cannot} model capacitors.  Their logic is that the induction is needed to
cause a time-dependent voltage (i.e., current) on the opposite capacitor plate.
However, this can be viewed as an EQS problem where the key is developing
boundary conditions of the voltage, or electrostatic potential, consistent with
the entire circuit~\cite{verboncoeur1993}.  In other words, the $RC$ circuit can
be calculated from the EQS equations and the induction term is not needed, as
discussed in \refsec{sec:circuit}.


Important experiments~\cite{Rousseaux:2013ft} such as the Wilson-Wilson
experiment~\cite{Wilson:1930io} or those describing the Barnett
effect~\cite{barnett1915} are in the MQS regime.  Applications of the EQS regime
are fewer, but one example is electrohydrodynamics~\cite{castellanos2014}.
Another application is in the the area of plasma physics where the EQS limit is
used for modeling the plasma sheath region~\cite{turner2013,verboncoeur1993}.
Even though it called the electrostatic model in that field, the oscillating
voltage boundary conditions mean that it is electro-quasi-static.  The results
here point towards a method for self-consistently including the magnetic field
effects in the particle motion in plasma physics simulations.  Although not
discussed in this paper, but discussed by LBLL, the Lorentz force equation in
the EQS limit contains only the electric field to leading order (absent a
zeroth-order, static, externally applied magnetic field which is consistent with
this theory, as discussed in \refsec{sec:MEorder}). LBLL discuss ways of taking
advantage of this theory to combine solutions.  This will be the subject of
future work.

The unitless version of the inhomogeneous vacuum Maxwell equations in
Lorentz-covariant form are
\begin{equation}
      \partial_{\mu} \hat{F}^{\mu\nu} = \aht \hat{J}^\mu.
\end{equation}
By providing the ordering of the vacuum Maxwell equations, we show that the
three quasi-static limits of the fields depend on the properties of the source
equation, as is perhaps unsurprising given this form of the equation.  But every
day experience with the Maxwell's equations in the quasi-static limit, such as
circuit equations or lab experiments, typically involve macroscopic media.  When
studying macroscopic media, the distinct nature of the electric and magnetic
fields in the quasi-static limit is even more distinct --- the electric field in
a capacitor appears unrelated to the magnetic field in an inductor.  This paper
follows the work of Rapetti and Rousseaux, and Melcher and Haus, in using
circuit equations to aid in providing intuition on when the quasi-static
equations are useful.  Using these equations allows the definition of 6 regions:
EM, EMQS, EQS, MQS, ES, and MS.  For more detailed understanding of how the
quasi-static equations leads to quantitative comparisons with experiments, the
review paper of Rousseaux~\cite{Rousseaux:2013ft} serves as an excellent
starting point.  With the exception of the textbooks of Melcher and
colleagues~\cite{woodson-melcher1968,melcher-haus1989}, which are engineering
textbooks, no textbook that we are aware of discusses the EQS regime, likely
because of its relatively limited practical use due to the Lorentz force
limitations discussed above.  However, a complete understanding of
electrodynamics benefits from a complete view of all three quasi-static limits.


\begin{acknowledgments} 

The author especially thanks Dan Abell for a very thorough reading of the
manuscript, for the physics discussions they engendered, and for the Helmholtz
decomposition derivation shown in Appendix~\ref{sec:helmholtz}, which greatly
improves the paper.  The author would like to thank Drs.~Kristian Beckwith,
Thomas Jenkins, Jacob King, David Smithe, and Peter Stoltz for discussions
related to this work and Thomas Jenkins for also reviewing this paper.  This
material is based on work supported by US Department of Energy, Office of
Science, Office of Fusion Energy Sciences under award numbers DE-SC0018313 and
DE-SC0019067.

\end{acknowledgments}

\bibliography{merged}

\begin{thebibliography}{45}%
\makeatletter
\providecommand \@ifxundefined [1]{%
 \@ifx{#1\undefined}
}%
\providecommand \@ifnum [1]{%
 \ifnum #1\expandafter \@firstoftwo
 \else \expandafter \@secondoftwo
 \fi
}%
\providecommand \@ifx [1]{%
 \ifx #1\expandafter \@firstoftwo
 \else \expandafter \@secondoftwo
 \fi
}%
\providecommand \natexlab [1]{#1}%
\providecommand \enquote  [1]{``#1''}%
\providecommand \bibnamefont  [1]{#1}%
\providecommand \bibfnamefont [1]{#1}%
\providecommand \citenamefont [1]{#1}%
\providecommand \href@noop [0]{\@secondoftwo}%
\providecommand \href [0]{\begingroup \@sanitize@url \@href}%
\providecommand \@href[1]{\@@startlink{#1}\@@href}%
\providecommand \@@href[1]{\endgroup#1\@@endlink}%
\providecommand \@sanitize@url [0]{\catcode `\\12\catcode `\$12\catcode
  `\&12\catcode `\#12\catcode `\^12\catcode `\_12\catcode `\%12\relax}%
\providecommand \@@startlink[1]{}%
\providecommand \@@endlink[0]{}%
\providecommand \url  [0]{\begingroup\@sanitize@url \@url }%
\providecommand \@url [1]{\endgroup\@href {#1}{\urlprefix }}%
\providecommand \urlprefix  [0]{URL }%
\providecommand \Eprint [0]{\href }%
\providecommand \doibase [0]{http://dx.doi.org/}%
\providecommand \selectlanguage [0]{\@gobble}%
\providecommand \bibinfo  [0]{\@secondoftwo}%
\providecommand \bibfield  [0]{\@secondoftwo}%
\providecommand \translation [1]{[#1]}%
\providecommand \BibitemOpen [0]{}%
\providecommand \bibitemStop [0]{}%
\providecommand \bibitemNoStop [0]{.\EOS\space}%
\providecommand \EOS [0]{\spacefactor3000\relax}%
\providecommand \BibitemShut  [1]{\csname bibitem#1\endcsname}%
\let\auto@bib@innerbib\@empty
\bibitem [{\citenamefont {Barnett}(1915)}]{barnett1915}%
  \BibitemOpen
  \bibfield  {author} {\bibinfo {author} {\bibnamefont {Barnett}, \bibfnamefont
  {S.~J.}}} (\bibinfo {year} {1915}),\ \href {\doibase 10.1103/PhysRev.6.239}
  {\bibfield  {journal} {\bibinfo  {journal} {Phys. Rev.}\ }\textbf {\bibinfo
  {volume} {6}},\ \bibinfo {pages} {239}}\BibitemShut {NoStop}%
\bibitem [{\citenamefont {Carson}(1927)}]{carson1927}%
  \BibitemOpen
  \bibfield  {author} {\bibinfo {author} {\bibnamefont {Carson}, \bibfnamefont
  {J.~R.}}} (\bibinfo {year} {1927}),\ \href@noop {} {\bibfield  {journal}
  {\bibinfo  {journal} {Bell System Technical Journal}\ }\textbf {\bibinfo
  {volume} {6}}~(\bibinfo {number} {1}),\ \bibinfo {pages} {1}}\BibitemShut
  {NoStop}%
\bibitem [{\citenamefont {Castellanos}(2014)}]{castellanos2014}%
  \BibitemOpen
  \bibfield  {author} {\bibinfo {author} {\bibnamefont {Castellanos},
  \bibfnamefont {A.}}} (\bibinfo {year} {2014}),\ \href@noop {} {\emph
  {\bibinfo {title} {Electrohydrodynamics}}},\ Vol.\ \bibinfo {volume} {380}\
  (\bibinfo  {publisher} {Springer})\BibitemShut {NoStop}%
\bibitem [{\citenamefont {Darwin}(1920)}]{darwin}%
  \BibitemOpen
  \bibfield  {author} {\bibinfo {author} {\bibnamefont {Darwin}, \bibfnamefont
  {C.}}} (\bibinfo {year} {1920}),\ \href@noop {} {\bibfield  {journal}
  {\bibinfo  {journal} {Phil. Mag.}\ }\textbf {\bibinfo {volume} {39}},\
  \bibinfo {pages} {537}}\BibitemShut {NoStop}%
\bibitem [{\citenamefont {De~Montigny}\ \emph {et~al.}(2003)\citenamefont
  {De~Montigny}, \citenamefont {Khanna},\ and\ \citenamefont
  {Santana}}]{demontigny2003}%
  \BibitemOpen
  \bibfield  {author} {\bibinfo {author} {\bibnamefont {De~Montigny},
  \bibfnamefont {M.}}, \bibinfo {author} {\bibfnamefont {F.}~\bibnamefont
  {Khanna}}, \ and\ \bibinfo {author} {\bibfnamefont {A.}~\bibnamefont
  {Santana}}} (\bibinfo {year} {2003}),\ \href@noop {} {\bibfield  {journal}
  {\bibinfo  {journal} {International Journal of Theoretical Physics}\ }\textbf
  {\bibinfo {volume} {42}}~(\bibinfo {number} {4}),\ \bibinfo {pages}
  {649}}\BibitemShut {NoStop}%
\bibitem [{\citenamefont {Eichenwald}(1903)}]{Eichenwald1903}%
  \BibitemOpen
  \bibfield  {author} {\bibinfo {author} {\bibnamefont {Eichenwald},
  \bibfnamefont {A.}}} (\bibinfo {year} {1903}),\ \href@noop {} {\bibfield
  {journal} {\bibinfo  {journal} {Ann. Phys.}\ }\textbf {\bibinfo {volume}
  {11}},\ \bibinfo {pages} {421}}\BibitemShut {NoStop}%
\bibitem [{\citenamefont {Eichenwald}(1904)}]{Eichenwald1904}%
  \BibitemOpen
  \bibfield  {author} {\bibinfo {author} {\bibnamefont {Eichenwald},
  \bibfnamefont {A.}}} (\bibinfo {year} {1904}),\ \href@noop {} {\bibfield
  {journal} {\bibinfo  {journal} {Ann. Phys.}\ }\textbf {\bibinfo {volume}
  {13}},\ \bibinfo {pages} {919}}\BibitemShut {NoStop}%
\bibitem [{\citenamefont {Einstein}\ and\ \citenamefont
  {Laub}(1908{\natexlab{a}})}]{EinsteinLaub1908c}%
  \BibitemOpen
  \bibfield  {author} {\bibinfo {author} {\bibnamefont {Einstein},
  \bibfnamefont {A.}}, \ and\ \bibinfo {author} {\bibfnamefont
  {J.}~\bibnamefont {Laub}}} (\bibinfo {year} {1908}{\natexlab{a}}),\
  \href@noop {} {\bibfield  {journal} {\bibinfo  {journal} {Ann. Phys.}\
  }\textbf {\bibinfo {volume} {26}},\ \bibinfo {pages} {232}}\BibitemShut
  {NoStop}%
\bibitem [{\citenamefont {Einstein}\ and\ \citenamefont
  {Laub}(1908{\natexlab{b}})}]{EinsteinLaub1908d}%
  \BibitemOpen
  \bibfield  {author} {\bibinfo {author} {\bibnamefont {Einstein},
  \bibfnamefont {A.}}, \ and\ \bibinfo {author} {\bibfnamefont
  {J.}~\bibnamefont {Laub}}} (\bibinfo {year} {1908}{\natexlab{b}}),\
  \href@noop {} {\bibfield  {journal} {\bibinfo  {journal} {Ann. Phys.}\
  }\textbf {\bibinfo {volume} {26}},\ \bibinfo {pages} {445}}\BibitemShut
  {NoStop}%
\bibitem [{\citenamefont {Einstein}\ and\ \citenamefont
  {Laub}(1908{\natexlab{c}})}]{EinsteinLaub1908a}%
  \BibitemOpen
  \bibfield  {author} {\bibinfo {author} {\bibnamefont {Einstein},
  \bibfnamefont {A.}}, \ and\ \bibinfo {author} {\bibfnamefont
  {J.}~\bibnamefont {Laub}}} (\bibinfo {year} {1908}{\natexlab{c}}),\
  \href@noop {} {\bibfield  {journal} {\bibinfo  {journal} {Ann. Phys.}\
  }\textbf {\bibinfo {volume} {26}},\ \bibinfo {pages} {541}}\BibitemShut
  {NoStop}%
\bibitem [{\citenamefont {Einstein}\ and\ \citenamefont
  {Laub}(1908{\natexlab{d}})}]{EinsteinLaub1908b}%
  \BibitemOpen
  \bibfield  {author} {\bibinfo {author} {\bibnamefont {Einstein},
  \bibfnamefont {A.}}, \ and\ \bibinfo {author} {\bibfnamefont
  {J.}~\bibnamefont {Laub}}} (\bibinfo {year} {1908}{\natexlab{d}}),\
  \href@noop {} {\bibfield  {journal} {\bibinfo  {journal} {Ann. Phys.}\
  }\textbf {\bibinfo {volume} {26}},\ \bibinfo {pages} {532}}\BibitemShut
  {NoStop}%
\bibitem [{\citenamefont {Griffiths}(2005)}]{Griffiths}%
  \BibitemOpen
  \bibfield  {author} {\bibinfo {author} {\bibnamefont {Griffiths},
  \bibfnamefont {D.~J.}}} (\bibinfo {year} {2005}),\ \href@noop {} {\emph
  {\bibinfo {title} {Introduction to electrodynamics}}}\ (\bibinfo  {publisher}
  {AAPT})\BibitemShut {NoStop}%
\bibitem [{\citenamefont {Haus}\ and\ \citenamefont
  {Melcher}(1989)}]{melcher-haus1989}%
  \BibitemOpen
  \bibfield  {author} {\bibinfo {author} {\bibnamefont {Haus}, \bibfnamefont
  {H.~A.}}, \ and\ \bibinfo {author} {\bibfnamefont {J.~R.}\ \bibnamefont
  {Melcher}}} (\bibinfo {year} {1989}),\ \href@noop {} {\emph {\bibinfo {title}
  {Electromagnetic fields and energy}}}\ (\bibinfo  {publisher} {Prentice
  Hall})\BibitemShut {NoStop}%
\bibitem [{\citenamefont {Heras}(2010{\natexlab{a}})}]{Heras:2010hi}%
  \BibitemOpen
  \bibfield  {author} {\bibinfo {author} {\bibnamefont {Heras}, \bibfnamefont
  {J.~A.}}} (\bibinfo {year} {2010}{\natexlab{a}}),\ \href@noop {} {\bibfield
  {journal} {\bibinfo  {journal} {European Journal of Physics}\ }\textbf
  {\bibinfo {volume} {31}}~(\bibinfo {number} {5}),\ \bibinfo {pages}
  {1177}}\BibitemShut {NoStop}%
\bibitem [{\citenamefont {Heras}(2010{\natexlab{b}})}]{Heras:2010jw}%
  \BibitemOpen
  \bibfield  {author} {\bibinfo {author} {\bibnamefont {Heras}, \bibfnamefont
  {J.~A.}}} (\bibinfo {year} {2010}{\natexlab{b}}),\ \href@noop {} {\bibfield
  {journal} {\bibinfo  {journal} {American Journal of Physics}\ }\textbf
  {\bibinfo {volume} {78}}~(\bibinfo {number} {10}),\ \bibinfo {pages}
  {1048}}\BibitemShut {NoStop}%
\bibitem [{\citenamefont {Heras}\ and\ \citenamefont
  {B{\'a}ez}(2008)}]{Heras:2008dj}%
  \BibitemOpen
  \bibfield  {author} {\bibinfo {author} {\bibnamefont {Heras}, \bibfnamefont
  {J.~A.}}, \ and\ \bibinfo {author} {\bibfnamefont {G.}~\bibnamefont
  {B{\'a}ez}}} (\bibinfo {year} {2008}),\ \href@noop {} {\bibfield  {journal}
  {\bibinfo  {journal} {European Journal of Physics}\ }\textbf {\bibinfo
  {volume} {30}}~(\bibinfo {number} {1}),\ \bibinfo {pages} {23}}\BibitemShut
  {NoStop}%
\bibitem [{\citenamefont {Jackson}(1975)}]{JacksonSecond}%
  \BibitemOpen
  \bibfield  {author} {\bibinfo {author} {\bibnamefont {Jackson}, \bibfnamefont
  {J.~D.}}} (\bibinfo {year} {1975}),\ \href@noop {} {\emph {\bibinfo {title}
  {Classical Electrodynamics}}},\ \bibinfo {edition} {2nd}\ ed.\ (\bibinfo
  {publisher} {Wiley},\ \bibinfo {address} {New York})\BibitemShut {NoStop}%
\bibitem [{\citenamefont {Jackson}(1999)}]{JacksonThird}%
  \BibitemOpen
  \bibfield  {author} {\bibinfo {author} {\bibnamefont {Jackson}, \bibfnamefont
  {J.~D.}}} (\bibinfo {year} {1999}),\ \href@noop {} {\emph {\bibinfo {title}
  {Classical Electrodynamics}}},\ \bibinfo {edition} {3rd}\ ed.\ (\bibinfo
  {publisher} {Wiley},\ \bibinfo {address} {New York})\BibitemShut {NoStop}%
\bibitem [{\citenamefont {Jackson}(2002)}]{Jackson2002}%
  \BibitemOpen
  \bibfield  {author} {\bibinfo {author} {\bibnamefont {Jackson}, \bibfnamefont
  {J.~D.}}} (\bibinfo {year} {2002}),\ \href@noop {} {\bibfield  {journal}
  {\bibinfo  {journal} {American Journal of Physics}\ }\textbf {\bibinfo
  {volume} {70}}~(\bibinfo {number} {9}),\ \bibinfo {pages} {917}}\BibitemShut
  {NoStop}%
\bibitem [{\citenamefont {Jefimenko}(1966)}]{jefimenko1966}%
  \BibitemOpen
  \bibfield  {author} {\bibinfo {author} {\bibnamefont {Jefimenko},
  \bibfnamefont {O.~D.}}} (\bibinfo {year} {1966}),\ \href@noop {} {\emph
  {\bibinfo {title} {Electricity and magnetism}}}\ (\bibinfo  {publisher}
  {Appleton-Century-Crofts})\BibitemShut {NoStop}%
\bibitem [{\citenamefont {Krause}\ \emph {et~al.}(2007)\citenamefont {Krause},
  \citenamefont {Apte},\ and\ \citenamefont {Morrison}}]{Krause:2007fa}%
  \BibitemOpen
  \bibfield  {author} {\bibinfo {author} {\bibnamefont {Krause}, \bibfnamefont
  {T.~B.}}, \bibinfo {author} {\bibfnamefont {A.}~\bibnamefont {Apte}}, \ and\
  \bibinfo {author} {\bibfnamefont {P.~J.}\ \bibnamefont {Morrison}}} (\bibinfo
  {year} {2007}),\ \href@noop {} {\bibfield  {journal} {\bibinfo  {journal}
  {Physics of plasmas}\ }\textbf {\bibinfo {volume} {14}}~(\bibinfo {number}
  {10}),\ \bibinfo {pages} {102112}}\BibitemShut {NoStop}%
\bibitem [{\citenamefont {Landau}\ and\ \citenamefont
  {Lifschitz}(1959)}]{LandL}%
  \BibitemOpen
  \bibfield  {author} {\bibinfo {author} {\bibnamefont {Landau}, \bibfnamefont
  {L.}}, \ and\ \bibinfo {author} {\bibfnamefont {E.}~\bibnamefont
  {Lifschitz}}} (\bibinfo {year} {1959}),\ \href@noop {} {\emph {\bibinfo
  {title} {The Classical Theory of Fields}}}\ (\bibinfo  {publisher} {Pergamon
  Press},\ \bibinfo {address} {New York})\BibitemShut {NoStop}%
\bibitem [{\citenamefont {Le~Bellac}\ and\ \citenamefont
  {L{\'e}vy-Leblond}(1973)}]{LBLL}%
  \BibitemOpen
  \bibfield  {author} {\bibinfo {author} {\bibnamefont {Le~Bellac},
  \bibfnamefont {M.}}, \ and\ \bibinfo {author} {\bibfnamefont {J.-M.}\
  \bibnamefont {L{\'e}vy-Leblond}}} (\bibinfo {year} {1973}),\ \href@noop {}
  {\bibfield  {journal} {\bibinfo  {journal} {Il Nuovo Cimento B (1971-1996)}\
  }\textbf {\bibinfo {volume} {14}}~(\bibinfo {number} {2}),\ \bibinfo {pages}
  {217}}\BibitemShut {NoStop}%
\bibitem [{\citenamefont {L{\'e}vy-Leblond}(1965)}]{levy1965}%
  \BibitemOpen
  \bibfield  {author} {\bibinfo {author} {\bibnamefont {L{\'e}vy-Leblond},
  \bibfnamefont {J.-M.}}} (\bibinfo {year} {1965}),\ \href@noop {} {\bibfield
  {journal} {\bibinfo  {journal} {Ann. Inst. H. Poincar{\'e}}\ }\textbf
  {\bibinfo {volume} {3}},\ \bibinfo {pages} {1}}\BibitemShut {NoStop}%
\bibitem [{\citenamefont {Manfredi}(2013)}]{manfredi2013}%
  \BibitemOpen
  \bibfield  {author} {\bibinfo {author} {\bibnamefont {Manfredi},
  \bibfnamefont {G.}}} (\bibinfo {year} {2013}),\ \href@noop {} {\bibfield
  {journal} {\bibinfo  {journal} {European Journal of Physics}\ }\textbf
  {\bibinfo {volume} {34}}~(\bibinfo {number} {4}),\ \bibinfo {pages}
  {859}}\BibitemShut {NoStop}%
\bibitem [{\citenamefont {de~Montigny}\ and\ \citenamefont
  {Rousseaux}(2007)}]{deMontigny:2007kg}%
  \BibitemOpen
  \bibfield  {author} {\bibinfo {author} {\bibnamefont {de~Montigny},
  \bibfnamefont {M.}}, \ and\ \bibinfo {author} {\bibfnamefont
  {G.}~\bibnamefont {Rousseaux}}} (\bibinfo {year} {2007}),\ \href@noop {}
  {\bibfield  {journal} {\bibinfo  {journal} {American Journal of Physics}\
  }\textbf {\bibinfo {volume} {75}}~(\bibinfo {number} {11}),\ \bibinfo {pages}
  {984}}\BibitemShut {NoStop}%
\bibitem [{\citenamefont {Montigny}\ and\ \citenamefont
  {Rousseaux}(2006)}]{Montigny:2006dq}%
  \BibitemOpen
  \bibfield  {author} {\bibinfo {author} {\bibnamefont {Montigny},
  \bibfnamefont {M.~d.}}, \ and\ \bibinfo {author} {\bibfnamefont
  {G.}~\bibnamefont {Rousseaux}}} (\bibinfo {year} {2006}),\ \href@noop {}
  {\bibfield  {journal} {\bibinfo  {journal} {European Journal of Physics}\
  }\textbf {\bibinfo {volume} {27}}~(\bibinfo {number} {4}),\ \bibinfo {pages}
  {755}}\BibitemShut {NoStop}%
\bibitem [{\citenamefont {Purcell}(1965)}]{purcell}%
  \BibitemOpen
  \bibfield  {author} {\bibinfo {author} {\bibnamefont {Purcell}, \bibfnamefont
  {E.}}} (\bibinfo {year} {1965}),\ \href@noop {} {\emph {\bibinfo {title}
  {Electricity and Magnetism}}},\ \bibinfo {edition} {3rd}\ ed.\ (\bibinfo
  {publisher} {Berkeley Physics Course},\ \bibinfo {address} {Reading,
  Massachusetts})\BibitemShut {NoStop}%
\bibitem [{\citenamefont {Rapetti}\ and\ \citenamefont
  {Rousseaux}(2011)}]{Rapetti:2011uq}%
  \BibitemOpen
  \bibfield  {author} {\bibinfo {author} {\bibnamefont {Rapetti}, \bibfnamefont
  {F.}}, \ and\ \bibinfo {author} {\bibfnamefont {G.}~\bibnamefont
  {Rousseaux}}} (\bibinfo {year} {2011})\ (\bibinfo {organization} {{IET 8th
  International Conference on Computation in Electromagnetics}})\BibitemShut
  {NoStop}%
\bibitem [{\citenamefont {Rapetti}\ and\ \citenamefont
  {Rousseaux}(2014)}]{Rapetti:2014ha}%
  \BibitemOpen
  \bibfield  {author} {\bibinfo {author} {\bibnamefont {Rapetti}, \bibfnamefont
  {F.}}, \ and\ \bibinfo {author} {\bibfnamefont {G.}~\bibnamefont
  {Rousseaux}}} (\bibinfo {year} {2014}),\ \href@noop {} {\bibfield  {journal}
  {\bibinfo  {journal} {Applied Numerical Mathematics}\ }\textbf {\bibinfo
  {volume} {79}},\ \bibinfo {pages} {92}}\BibitemShut {NoStop}%
\bibitem [{\citenamefont {Roentgen}(1888)}]{Roentgen1888}%
  \BibitemOpen
  \bibfield  {author} {\bibinfo {author} {\bibnamefont {Roentgen},
  \bibfnamefont {W.~C.}}} (\bibinfo {year} {1888}),\ \href@noop {} {\bibfield
  {journal} {\bibinfo  {journal} {Ann. Phys.}\ }\textbf {\bibinfo {volume}
  {35}},\ \bibinfo {pages} {264}}\BibitemShut {NoStop}%
\bibitem [{\citenamefont {Roentgen}(1890)}]{Roentgen1890}%
  \BibitemOpen
  \bibfield  {author} {\bibinfo {author} {\bibnamefont {Roentgen},
  \bibfnamefont {W.~C.}}} (\bibinfo {year} {1890}),\ \href@noop {} {\bibfield
  {journal} {\bibinfo  {journal} {Ann. Phys.}\ }\textbf {\bibinfo {volume}
  {40}},\ \bibinfo {pages} {93}}\BibitemShut {NoStop}%
\bibitem [{\citenamefont {Rousseaux}(2003)}]{rousseaux2003}%
  \BibitemOpen
  \bibfield  {author} {\bibinfo {author} {\bibnamefont {Rousseaux},
  \bibfnamefont {G.}}} (\bibinfo {year} {2003}),\ in\ \href@noop {} {\emph
  {\bibinfo {booktitle} {Annales de la fondation de Broglie}}},\ Vol.~\bibinfo
  {volume} {28},\ pp.\ \bibinfo {pages} {261--270}\BibitemShut {NoStop}%
\bibitem [{\citenamefont {Rousseaux}(2008)}]{Rousseaux:2008ij}%
  \BibitemOpen
  \bibfield  {author} {\bibinfo {author} {\bibnamefont {Rousseaux},
  \bibfnamefont {G.}}} (\bibinfo {year} {2008}),\ \href@noop {} {\bibfield
  {journal} {\bibinfo  {journal} {Europhysics Letters (EPL)}\ }\textbf
  {\bibinfo {volume} {84}}~(\bibinfo {number} {2}),\ \bibinfo {pages}
  {20002}}\BibitemShut {NoStop}%
\bibitem [{\citenamefont {Rousseaux}(2013)}]{Rousseaux:2013ft}%
  \BibitemOpen
  \bibfield  {author} {\bibinfo {author} {\bibnamefont {Rousseaux},
  \bibfnamefont {G.}}} (\bibinfo {year} {2013}),\ \href@noop {} {\bibfield
  {journal} {\bibinfo  {journal} {The European Physical Journal Plus}\ }\textbf
  {\bibinfo {volume} {128}}~(\bibinfo {number} {8}),\ \bibinfo {pages}
  {105207}}\BibitemShut {NoStop}%
\bibitem [{\citenamefont {Rousseaux}\ and\ \citenamefont
  {Domps}(2004)}]{rousseaux2004}%
  \BibitemOpen
  \bibfield  {author} {\bibinfo {author} {\bibnamefont {Rousseaux},
  \bibfnamefont {G.}}, \ and\ \bibinfo {author} {\bibfnamefont
  {A.}~\bibnamefont {Domps}}} (\bibinfo {year} {2004}),\ \href@noop {}
  {\bibfield  {journal} {\bibinfo  {journal} {Bulletin de l’Union des
  Physiciens}\ }\textbf {\bibinfo {volume} {863}}}\BibitemShut {NoStop}%
\bibitem [{\citenamefont {Rousseaux}\ \emph {et~al.}(2008)\citenamefont
  {Rousseaux}, \citenamefont {Kofman},\ and\ \citenamefont
  {Minazzoli}}]{Rousseaux:2008ev}%
  \BibitemOpen
  \bibfield  {author} {\bibinfo {author} {\bibnamefont {Rousseaux},
  \bibfnamefont {G.}}, \bibinfo {author} {\bibfnamefont {R.}~\bibnamefont
  {Kofman}}, \ and\ \bibinfo {author} {\bibfnamefont {O.}~\bibnamefont
  {Minazzoli}}} (\bibinfo {year} {2008}),\ \href@noop {} {\bibfield  {journal}
  {\bibinfo  {journal} {The European Physical Journal D}\ }\textbf {\bibinfo
  {volume} {49}}~(\bibinfo {number} {2}),\ \bibinfo {pages} {249}}\BibitemShut
  {NoStop}%
\bibitem [{\citenamefont {Turner}\ \emph {et~al.}(2013)\citenamefont {Turner},
  \citenamefont {Derzsi}, \citenamefont {Donk\'{o}}, \citenamefont {Eremin},
  \citenamefont {Kelly}, \citenamefont {Lafleur},\ and\ \citenamefont
  {Mussenbrock}}]{turner2013}%
  \BibitemOpen
  \bibfield  {author} {\bibinfo {author} {\bibnamefont {Turner}, \bibfnamefont
  {M.~M.}}, \bibinfo {author} {\bibfnamefont {A.}~\bibnamefont {Derzsi}},
  \bibinfo {author} {\bibfnamefont {Z.}~\bibnamefont {Donk\'{o}}}, \bibinfo
  {author} {\bibfnamefont {D.}~\bibnamefont {Eremin}}, \bibinfo {author}
  {\bibfnamefont {S.~J.}\ \bibnamefont {Kelly}}, \bibinfo {author}
  {\bibfnamefont {T.}~\bibnamefont {Lafleur}}, \ and\ \bibinfo {author}
  {\bibfnamefont {T.}~\bibnamefont {Mussenbrock}}} (\bibinfo {year} {2013}),\
  \href {\doibase 10.1063/1.4775084} {\bibfield  {journal} {\bibinfo  {journal}
  {\href{http://dx.doi.org/10.1063/1.4775084}{Phys. Plasmas}}\ }\textbf
  {\bibinfo {volume} {20}},\ \bibinfo {pages} {013507}}\BibitemShut {NoStop}%
\bibitem [{\citenamefont {Verboncoeur}\ \emph {et~al.}(1993)\citenamefont
  {Verboncoeur}, \citenamefont {Alves}, \citenamefont {Vahedi},\ and\
  \citenamefont {Birdsall}}]{verboncoeur1993}%
  \BibitemOpen
  \bibfield  {author} {\bibinfo {author} {\bibnamefont {Verboncoeur},
  \bibfnamefont {J.~P.}}, \bibinfo {author} {\bibfnamefont {M.~V.}\
  \bibnamefont {Alves}}, \bibinfo {author} {\bibfnamefont {V.}~\bibnamefont
  {Vahedi}}, \ and\ \bibinfo {author} {\bibfnamefont {C.~K.}\ \bibnamefont
  {Birdsall}}} (\bibinfo {year} {1993}),\ \href@noop {} {\bibfield  {journal}
  {\bibinfo  {journal} {Journal of Computational Physics}\ }\textbf {\bibinfo
  {volume} {104}}~(\bibinfo {number} {2}),\ \bibinfo {pages} {321}}\BibitemShut
  {NoStop}%
\bibitem [{\citenamefont {Weber}(1997)}]{weber1997}%
  \BibitemOpen
  \bibfield  {author} {\bibinfo {author} {\bibnamefont {Weber}, \bibfnamefont
  {T.~A.}}} (\bibinfo {year} {1997}),\ \href@noop {} {\bibfield  {journal}
  {\bibinfo  {journal} {American Journal of Physics}\ }\textbf {\bibinfo
  {volume} {65}}~(\bibinfo {number} {10}),\ \bibinfo {pages} {946}}\BibitemShut
  {NoStop}%
\bibitem [{\citenamefont {Wilson}(1905)}]{wilson1905}%
  \BibitemOpen
  \bibfield  {author} {\bibinfo {author} {\bibnamefont {Wilson}, \bibfnamefont
  {H.~A.}}} (\bibinfo {year} {1905}),\ \href@noop {} {\bibfield  {journal}
  {\bibinfo  {journal} {Phil. Trans. R. Soc. Lond. A}\ }\textbf {\bibinfo
  {volume} {204}}~(\bibinfo {number} {372-386}),\ \bibinfo {pages}
  {121}}\BibitemShut {NoStop}%
\bibitem [{\citenamefont {Wilson}\ \emph {et~al.}(1913)\citenamefont {Wilson},
  \citenamefont {Wilson} \emph {et~al.}}]{wilson1913}%
  \BibitemOpen
  \bibfield  {author} {\bibinfo {author} {\bibnamefont {Wilson}, \bibfnamefont
  {M.}}, \bibinfo {author} {\bibfnamefont {H.~A.}\ \bibnamefont {Wilson}},
  \emph {et~al.}} (\bibinfo {year} {1913}),\ \href@noop {} {\bibfield
  {journal} {\bibinfo  {journal} {Proc. R. Soc. Lond. A}\ }\textbf {\bibinfo
  {volume} {89}}~(\bibinfo {number} {608}),\ \bibinfo {pages} {99}}\BibitemShut
  {NoStop}%
\bibitem [{\citenamefont {Wilson}\ and\ \citenamefont
  {Espenschied}(1930)}]{Wilson:1930io}%
  \BibitemOpen
  \bibfield  {author} {\bibinfo {author} {\bibnamefont {Wilson}, \bibfnamefont
  {W.}}, \ and\ \bibinfo {author} {\bibfnamefont {L.}~\bibnamefont
  {Espenschied}}} (\bibinfo {year} {1930}),\ \href@noop {} {\bibfield
  {journal} {\bibinfo  {journal} {Bell Labs Technical Journal}\ }\textbf
  {\bibinfo {volume} {9}}~(\bibinfo {number} {3}),\ \bibinfo {pages}
  {407}}\BibitemShut {NoStop}%
\bibitem [{\citenamefont {Woodson}\ and\ \citenamefont
  {Melcher}(1968)}]{woodson-melcher1968}%
  \BibitemOpen
  \bibfield  {author} {\bibinfo {author} {\bibnamefont {Woodson}, \bibfnamefont
  {H.~H.}}, \ and\ \bibinfo {author} {\bibfnamefont {J.~R.}\ \bibnamefont
  {Melcher}}} (\bibinfo {year} {1968}),\ \href@noop {} {\emph {\bibinfo {title}
  {Electromechanical dynamics}}}\ (\bibinfo  {publisher} {Wiley})\BibitemShut
  {NoStop}%
\bibitem [{\citenamefont {Yee}(1966)}]{yee-fdtd}%
  \BibitemOpen
  \bibfield  {author} {\bibinfo {author} {\bibnamefont {Yee}, \bibfnamefont
  {K.}}} (\bibinfo {year} {1966}),\ \href@noop {} {\bibfield  {journal}
  {\bibinfo  {journal} {IEEE Transactions on Antennas and Propagation}\
  }\textbf {\bibinfo {volume} {14}}~(\bibinfo {number} {3}),\ \bibinfo {pages}
  {302}}\BibitemShut {NoStop}%
\end{thebibliography}%

\clearpage
\newpage
\appendix
\section{The Helmholtz Theorem}
\label{sec:helmholtz}

In this appendix, we review the Helmholtz decomposition of a vector and how it
is used in the main text of the paper.  The basic material here essentially the
same as Appendix B of Griffiths' book~\cite{Griffiths}, but in a proof that we
find simpler.%
~\footnote{See also~\cite{JacksonSecond}, Sect.~6.5;~\cite{JacksonThird}, Sect.~6.3} 
The Helmoltz Theorem says that \emph{any vector field $\vF(\vx)$ can be decomposed into
longitudinal (irrotational) and transverse (solenoidal) components as}
\begin{eqaligned}
   \label{HelmholtzDecomp}
      \vF(\vx) =& \vF_L(\vx) + \vF_T(\vx)
                                      \\
      =& -\grad{U(\vx)} + \curl{\vV(\vx)},
\end{eqaligned}
\emph{and where the fields $U$ and $\vV$ are given by}
\begin{eqaligned}
      U(\vx) =& \frac{1}{4\pi} \int \frac{\divep{\vF(\xp)}}{R} \dVp
                 - \frac{1}{4\pi} \oint \frac{\vF(\xp)}{R} \cdot \dSp
                                      \\
      \vV(\vx) =& \frac{1}{4\pi} \int \frac{\curlp{\vF(\xp)}}{R} \dVp
                 + \frac{1}{4\pi} \oint \frac{\vF(\xp)}{R} \times \dSp,
\end{eqaligned}
\emph{where} $R=|\vx-\xp|$.

We begin the proof by expressing $\vF$ as a global integral using the Green's
function kernel of $1/R$:
\begin{eqaligned}
      \vF(\vx) =&  \int \vF(\xp) \delta(\vx - \xp) \dVp
                                                                         \\
                =&  -\frac{1}{4\pi}\int \vF(\xp) \gradsq\frac{1}{R} \dVp
                                                                         \\
                =&  -\gradsq \lp \frac{1}{4\pi}\int \frac{\vF(\xp)}{R} \dVp \rp,
\end{eqaligned}
where we have used relationships in \refeq{Identities} to convert the delta
function to the Laplacian operator.  Using \refeq{vectorIdentity-curlcurl}, we
can then write:
\begin{equation}
     \vF(\vx) = -\grad \lp \dive{\frac{1}{4\pi}\int \frac{\vF(\xp)}{R} \dVp \rp}
                +\curl \lp \curl{\frac{1}{4\pi}\int \frac{\vF(\xp)}{R} \dVp \rp}.
   \label{F1}
\end{equation}
This gives a form for $U$ and $\vV$, but not yet the desired form.  We need to
move the divergence and curl operators to inside of the integrals and have them
operate over the primed variables rather than the unprimed variables.  
For the divergence operator, we convert as follows:
\begin{eqaligned}
      \dive{\frac{\vF(\xp)}{R}} =& \vF(\xp) \cdot \grad{\frac{1}{R}} 
                                = -\vF(\xp) \cdot \gradp{\frac{1}{R}} 
                                                                         \\
                                =& - \divep{ \frac{\vF(\xp)}{R}}
                                   + \frac{1}{R} \divep{\vF(\xp)},
\end{eqaligned}
where we have used $\dive{\vF(\xp)}=0$ in the first step.  
This result with Gauss's theorem gives
\begin{equation}
       \dive{\lp \int \frac{\vF(\xp)}{R} \dVp \rp}
                                =  \int \frac{\divep{\vF(\xp)}}{R}  \dVp
                                  - \oint \frac{\vF(\xp)}{R} \cdot \dSp.
\end{equation}
Similarly for the term with the curl:
\begin{eqaligned}
      \curl{\frac{\vF(\xp)}{R}} =& -\vF(\xp) \times \grad{ \frac{1}{R}} 
                                =   \vF(\xp) \times \gradp{\frac{1}{R}}
                                                                         \\
                                =& - \curlp{ \frac{\vF(\xp)}{R}}
                                   + \frac{1}{R} \curlp{\vF(\xp)},
\end{eqaligned}
and then converting the volume integral to a surface integral:
\begin{equation}
      \curl{\lp \int \frac{\vF(\xp)}{R} \dVp \rp}
                                =  \int \frac{\curlp{\vF(\xp)}}{R}  \dVp
                                  + \oint \frac{\vF(\xp)}{R} \times \dSp.
\end{equation}
Using these results, we can write \refeq{F1} as
\begin{eqaligned}
      \vF(\vx) =& -\grad \lb\frac{1}{4\pi} \int \frac{\divep{\vF(\xp)}}{R}  \dVp
                       - \frac{1}{4\pi}\oint \frac{\vF(\xp)}{R} \cdot \dSp \rb
                                                                         \\
                & +\curl{\lb \frac{1}{4\pi}\int \frac{\curlp{\vF(\xp)}}{R}  \dVp
                    + \frac{1}{4\pi}\oint \frac{\vF(\xp)}{R} \times \dSp \rb}.
   \label{F2}
\end{eqaligned}
This effectively proves the theorem.  Assuming that the surface terms vanish,
this allows us to write the longitudinal and transverse components in the more
commonly written form
\begin{subequations}
   \label{LongitudinalTransverse-Helmholtz}
   \begin{align}
      \vF_L(\vx) =& -\grad U 
            =  -\grad \lb\frac{1}{4\pi}\int \frac{\divep{\vF(\xp)}1}{R} \dVp \rb
   \label{Flong}
                                                                         \\
      \vF_T(\vx) =& \curl{V} 
            = \curl{\lb\frac{1}{4\pi} \int \frac{\curlp{\vF(\xp)}}}{R} \dVp \rb.
   \label{Ftrans}
   \end{align}
\end{subequations}
For the electrostatic, EQS, and MQS limits, these expression to calculate $\phi
= U$ based on $\vF = \vE$ yield
\begin{equation}
    \phi =  \frac{1}{4\pi} \int \frac{1}{R} \divep{\vE(\xp)} \dVp =
    \frac{\aht}{4\pi} \int \frac{1}{R} \rho(\xp) \dVp,
\end{equation}
which is the same expressions as \refeq{scaPotRetarded}.  For the magnetostatic
and MQS limit, calculating $\vA = \vV$ based on $\vF = \vB$ yields 
\begin{equation}
    \vA =  \frac{1}{4\pi} \int \frac{1}{R} \curlp{\vB(\xp)} \dVp =
    \frac{\aht}{4\pi} \int \frac{1}{R} \vJ(\xp) \dVp,
    \label{A_from_Helmholtz}
\end{equation}
which is the same expression as \refeq{vecPotRetarded}.

For the EMQS equation, the last term in \refeq{MaxwellsEqsVecpotEMQS}, can be
written as
\begin{eqaligned}
      \gradh \dth{\phiht(\vxht,\tht)} =& 
             -\frac{\aht}{4\pi} \gradh \int \dth{}\frac{\rhoht}{R} \dVp
                                                                         \\
      =& -\frac{\aht}{4\pi} \gradh \int \dth{}\frac{\divep{\vJ}}{R} \dVp
                                                                         \\
      =& \aht \vJht_L,
      \label{dphidt_is_JL}
\end{eqaligned}
which is the form needed to prove that the EMQS equation uses the transverse
current.

The astute reader may notice that we did include the EQS limit in the set of
equations that \refeq{A_from_Helmholtz} applies.  That does not work because the
curl of $\vB$ will bring in the time derivative of $\vE$.  By taking the divergence
of $\vA$ in that case, one can show that it is zero using the continuity
equation, similar to the derivation in \refeq{dphidt_is_JL}.  Hence, this
method of calculating $\vA$ only gives the transverse current.  As shown in the
text, the inversion of the Laplacian as given by \refeq{vecPotRetarded}  gives
both the transverse and longitudinal components of $\vA$ required to recover the
full set of EQS equations as discussed in \refsec{sec:jefimenko}.
\newpage
\clearpage
\section{Figures summarizing results}
\label{sec:sumfigs}

The textbook of Griffiths~\cite{Griffiths} uses ``triangle diagrams'' to show the relationship
between the sources, the fields, and the potentials for electrostatics and
magnetostatics.  As an alternative, these ``electromagnetic quad diagrams'' as a more useful
pedagogical tool.  These are shown for all six regions discussed in the paper: the
two static limits, the three quasi-static limits, and the full Maxwell equations.
\begin{figure*}
\begin{tikzpicture}[thick,scale=0.80, every node/.style={scale=0.80}]

  \node (n99) [above] at (0, 4.500)%
               { \large \bf
                   Electromagnetics (EM)
               };
  \draw[ultra thick] (-4.5,0) -- (5.5,0);
  \draw[ultra thick] (0,-3.8) -- (0,4.2);

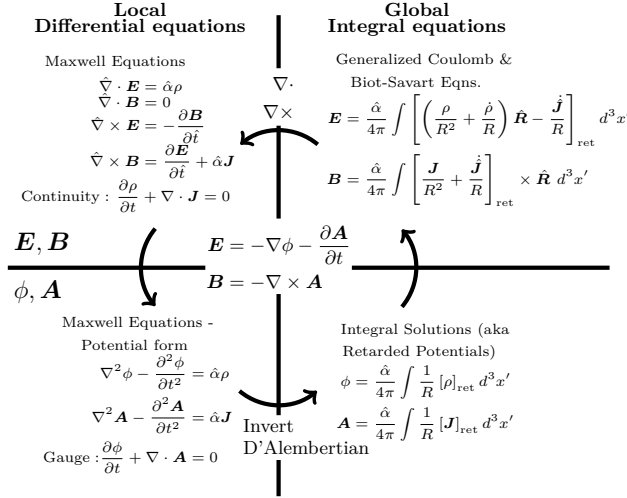
  \node (n11) [anchor=south east] at (-0.6, 0.8)%
               {
                 \begin{minipage}{92pt}
                       {\scriptsize Maxwell Equations}
                       \vspace{-0.1in}
                        \begingroup\makeatletter\def\f@size{7}\check@mathfonts
                        \def\maketag@@@#1{\hbox{\m@th\large\normalfont#1}}%
                     \begin{flalign}
                       \diveh{\vE} &= \aht \rho
                                                                  \nonumber \\[-2\jot]  
                       \diveh{\vB} &=0
                                                                  \nonumber \\[-2\jot]
                       \curlh{\vE} &= - \dth{\vB}
                                                                  \nonumber \\[-1\jot]
                       \curlh{\vB} &=   \dth{\vE} + \aht \vJ
                                                                  \nonumber \\[-1\jot]
                       {\rm Continuity:\ } \dt{\rho}&+ \dive{\vJ} = 0
                                                                      \nonumber 
                   \end{flalign}\endgroup
                 \end{minipage}
               };
  \node (n12) [anchor=north east] at (-0.6, -0.7)%
               {
                 \begin{minipage}{92pt}
                       {\scriptsize Maxwell Equations - Potential form}
                       \vspace{-0.1in}
                        \begingroup\makeatletter\def\f@size{7}\check@mathfonts
                        \def\maketag@@@#1{\hbox{\m@th\large\normalfont#1}}%
                     \begin{flalign}
                       \gradsq\phi - \dtsq{\phi} &= \aht \rho
                                                                  \nonumber \\
                       \gradsq\vA - \dtsq{\vA} &= \aht \vJ
                                                                  \nonumber \\
                       {\rm Gauge:} \dt{\phi}+ \dive{\vA} &= 0
                                                                  \nonumber
                   \end{flalign}\endgroup
                 \end{minipage}
               };
  \node (n13) [anchor=north west] at (0.7, -0.8)%
               {
                 \begin{minipage}{92pt}
                       {\scriptsize  Integral Solutions (aka Retarded Potentials) }
                       \vspace{-0.1in}
                        \begingroup\makeatletter\def\f@size{7}\check@mathfonts
                        \def\maketag@@@#1{\hbox{\m@th\large\normalfont#1}}%
                     \begin{flalign}
   \phi&= \frac{\aht}{4\pi} \int \frac{1}{R}\lb \rho \rbret \dVp
                                                                  \nonumber \\
   \vA &= \frac{\aht}{4\pi} \int \frac{1}{R}\lb \vJ \rbret \dVp 
                                                                  \nonumber
                   \end{flalign}\endgroup
                 \end{minipage}
               };
  \node (n14) [anchor=south west] at (0.68, 0.9)%
               {
                 \begin{minipage}{92pt}
                       {\scriptsize  Generalized Coulomb \& Biot-Savart Eqns. 
                        }
                       \vspace{-0.1in}
                        \begingroup\makeatletter\def\f@size{7}\check@mathfonts
                        \def\maketag@@@#1{\hbox{\m@th\large\normalfont#1}}%
                     \begin{flalign}
                       \vE & = \frac{\aht}{4\pi} \int \lb 
                                \lp \frac{\rho}{R^2}
                                  + \frac{\rdt}{R} \rp  \vRht
                                  - \frac{\vJdt}{R} 
                                   \rbret \dVp
                                   \nonumber \\
                       \vB & = \frac{\aht}{4\pi} \int
                                 \lb \frac{\vJ}{R^2} 
                                 + \frac{\vJdt}{R} \rbret 
                                  \times \vRht\  \dVp
                                                                  \nonumber
                   \end{flalign}\endgroup
                 \end{minipage}
               };

  \draw [->,ultra thick] (-2.05, 0.65) arc [radius=1.0, start angle=140, end angle= 220];
  \draw [->,ultra thick] ( 2.05,-0.65) arc [radius=1.0, start angle=-40, end angle=  40];
  \draw [->,ultra thick] ( 0.65, 2.05) arc [radius=1.0, start angle=50,  end angle= 130];
  \draw [->,ultra thick] (-0.65,-2.05) arc [radius=1.0, start angle=230, end angle= 310];

  \draw [fill=white,white] (-1.20,-0.42) rectangle (1.2,0.8);
  \node (n00) [] at (0.0, 0.29)%
               {
                 \begin{minipage}{0pt}
                   \begin{align}
                       \vE & = -\grad \phi - \dt{\vA}
                                                                  \nonumber \\
                       \vB & = -\curl{\vA}
                                                                  \nonumber 
                   \end{align}
                 \end{minipage}
               };
  \draw [fill=white,white] (-0.28, 3.30) rectangle (0.28,2.38);
  \node (n00) [] at (0.0, 3.00)%
               {
                 \begin{minipage}{0pt}
                   \begin{align}
                     \dive{} 
                                                                  \nonumber \\
                     \curl{}
                                                                  \nonumber 
                   \end{align}
                 \end{minipage}
               };
  \draw [fill=white,white] (-0.80, -2.47) rectangle (0.50,-3.18);
  \node (n00) [] at (-0.59, -2.80)%
               {
                 \begin{minipage}{0pt}
                   Invert \\
                   D'Alembertian
                 \end{minipage}
               };

  \node (n00) [] at (-4.4, 0.40)%
               {
                 \begin{minipage}{0pt}
                   \large
                     $\vE, \vB$
                 \end{minipage}
               };
  \node (n00) [] at (-4.4, -0.40)%
               {
                 \begin{minipage}{0pt}
                   \large
                     $\phi, \vA$
                 \end{minipage}
               };
  \node (n980) [] at (-2.3, 4.30)%
               {
                   \bf Local
               };
  \node (n980) [] at (2.3, 4.30)%
               {
                   \bf Global
               };
  \node (n98) [] at (-2.3, 4.00)%
               {
                   \bf
                   Differential equations
               };
  \node (n99) [] at ( 2.3, 4.00)%
               {
                   \bf
                   Integral equations
               };

\end{tikzpicture}
\begin{tikzpicture}[thick,scale=0.80, every node/.style={scale=0.80}]

  \node (n99) [above] at (0, 4.500)%
               { \large \bf
                   EM Quasistatics (EMQS)
               };

  \draw[ultra thick] (-4.5,0) -- (4.5,0);
  \draw[ultra thick] (0,-3.8) -- (0,4.2);

  \node (n11) [anchor=south east] at (-0.6, 1.0)%
               {
                 \begin{minipage}{92pt}
                       {\scriptsize Maxwell Equations}
                       \vspace{-0.1in}
                        \begingroup\makeatletter\def\f@size{7}\check@mathfonts
                        \def\maketag@@@#1{\hbox{\m@th\large\normalfont#1}}%
                     \begin{flalign}
                       \diveh{\vE} &= \aht \rho
                                                                  \nonumber \\[-2\jot]  
                       \diveh{\vB} &=0
                                                                  \nonumber \\[-2\jot]
                       \curlh{\vE} &= - \dth{\vB}
                                                                  \nonumber \\[-1\jot]
                       \curlh{\vB} &=   \dth{\vE_L} + \aht \vJ
                                                                  \nonumber \\[-1\jot]
                       {\rm Continuity:\ } \dt{\rho}&+ \dive{\vJ} = 0
                                                                      \nonumber 
                   \end{flalign}\endgroup
                 \end{minipage}
               };
  \node (n12) [anchor=north east] at (-0.6, -0.8)%
               {
                 \begin{minipage}{92pt}
                       {\scriptsize Maxwell Equations - Potential form}
                       \vspace{-0.1in}
                        \begingroup\makeatletter\def\f@size{7}\check@mathfonts
                        \def\maketag@@@#1{\hbox{\m@th\large\normalfont#1}}%
                     \begin{flalign}
                       \gradsq\phi &= \aht \rho
                                                                  \nonumber \\
                       \gradsq\vA_T &= \aht \vJ_T
                                                                  \nonumber \\
                       {\rm Gauge:} \dive{\vA} &= 0
                                                                  \nonumber
                   \end{flalign}\endgroup
                 \end{minipage}
               };
  \node (n13) [anchor=north west] at (0.68, -0.8)%
               {
                 \begin{minipage}{92pt}
                       {\scriptsize  Integral Solutions (aka Retarded Potentials) }
                       \vspace{-0.1in}
                        \begingroup\makeatletter\def\f@size{7}\check@mathfonts
                        \def\maketag@@@#1{\hbox{\m@th\large\normalfont#1}}%
                     \begin{flalign}
   \phi&= \frac{\aht}{4\pi} \int \frac{1}{R}\lb \rho \rb \dVp
                                                                  \nonumber \\
   \vA_T &= \frac{\aht}{4\pi} \int \frac{1}{R}\lb \vJ_T \rb \dVp 
                                                                  \nonumber
                   \end{flalign}\endgroup
                 \end{minipage}
               };
  \node (n14) [anchor=south west] at (0.7, 0.9)%
               {
                 \begin{minipage}{92pt}
                       {\scriptsize  Generalized Coulomb \& Biot-Savart Eqns. 
                        }
                       \vspace{-0.1in}
                        \begingroup\makeatletter\def\f@size{7}\check@mathfonts
                        \def\maketag@@@#1{\hbox{\m@th\large\normalfont#1}}%
                     \begin{flalign}
                        \vE & = \frac{\aht}{4\pi} \int \lb 
                                \rho \frac{\vRht}{R^2} 
                              + \frac{\vJdt_T}{R} \rb \dVp.
                                                        \nonumber \\
                        \vB & = \frac{\aht}{4\pi} \int
                                \lb \vJ_T \times \frac{\vRht}{R^2} \rb \dVp
                                                                  \nonumber
                   \end{flalign}\endgroup
                 \end{minipage}
               };

  \draw [->,ultra thick] (-2.05, 0.65) arc [radius=1.0, start angle=140, end angle= 220];
  \draw [->,ultra thick] ( 2.05,-0.65) arc [radius=1.0, start angle=-40, end angle=  40];
  \draw [->,ultra thick] ( 0.65, 2.05) arc [radius=1.0, start angle=50,  end angle= 130];
  \draw [->,ultra thick] (-0.65,-2.05) arc [radius=1.0, start angle=230, end angle= 310];

  \draw [fill=white,white] (-1.20,-0.42) rectangle (1.2,0.8);
  \node (n00) [] at (0.0, 0.29)%
               {
                 \begin{minipage}{0pt}
                   \begin{align}
                       \vE & = -\grad \phi - \dt{\vA_T}
                                                                  \nonumber \\
                       \vB & = -\curl{\vA}
                                                                  \nonumber 
                   \end{align}
                 \end{minipage}
               };
  \draw [fill=white,white] (-0.28, 3.30) rectangle (0.28,2.38);
  \node (n00) [] at (0.0, 3.00)%
               {
                 \begin{minipage}{0pt}
                   \begin{align}
                     \dive{} 
                                                                  \nonumber \\
                     \curl{}
                                                                  \nonumber 
                   \end{align}
                 \end{minipage}
               };
  \draw [fill=white,white] (-1.06, -2.47) rectangle (0.50,-3.18);
  \node (n00) [] at (-0.59, -2.80)%
               {
                 \begin{minipage}{0pt}
                   Invert \\
                   Laplacian
                 \end{minipage}
               };

  \node (n00) [] at (-4.4, 0.40)%
               {
                 \begin{minipage}{0pt}
                   \large
                     $\vE, \vB$
                 \end{minipage}
               };
  \node (n00) [] at (-4.4, -0.40)%
               {
                 \begin{minipage}{0pt}
                   \large
                     $\phi, \vA$
                 \end{minipage}
               };
  \node (n980) [] at (-2.3, 4.30)%
               {
                   \bf Local
               };
  \node (n990) [] at (2.3, 4.30)%
               {
                   \bf Global
               };
  \node (n00) [] at (-2.3, 4.00)%
               {
                   \bf
                   Differential equations
               };
  \node (n00) [] at ( 2.3, 4.00)%
               {
                   \bf
                   Integral equations
               };

\end{tikzpicture}
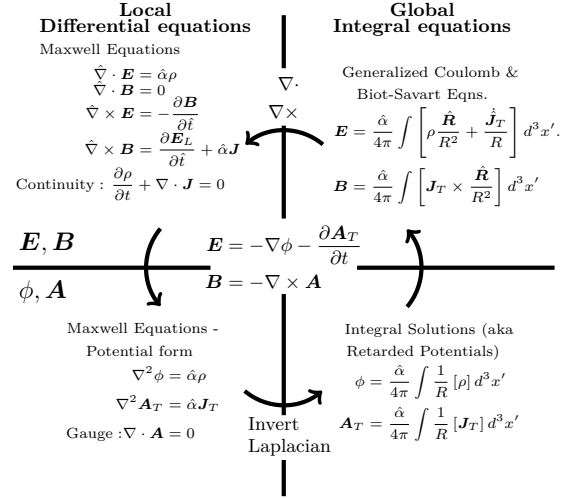
\begin{tikzpicture}[thick,scale=0.80, every node/.style={scale=0.80}]

  \node (n99) [above] at (0, 4.500)%
               { \large \bf
                   Electro-Quasistatics (EQS)
               };
  \draw[ultra thick] (-4.5,0) -- (4.5,0);
  \draw[ultra thick] (0,-3.8) -- (0,4.2);

  \node (n11) [anchor=south east] at (-0.6, 1.0)%
               {
                 \begin{minipage}{92pt}
                       {\scriptsize Maxwell Equations}
                       \vspace{-0.1in}
                        \begingroup\makeatletter\def\f@size{7}\check@mathfonts
                        \def\maketag@@@#1{\hbox{\m@th\large\normalfont#1}}%
                     \begin{flalign}
                       \diveh{\vE} &= \aht \rho
                                                                  \nonumber \\[-2\jot]  
                       \diveh{\vB} &=0
                                                                  \nonumber \\[-2\jot]
                       \curlh{\vE} &= 0
                                                                  \nonumber \\[-1\jot]
                       \curlh{\vB} &=   \dth{\vE} + \aht \vJ
                                                                  \nonumber \\[-1\jot]
                       {\rm Continuity:\ } \dt{\rho}&+ \dive{\vJ} = 0
                                                                      \nonumber 
                   \end{flalign}\endgroup
                 \end{minipage}
               };
  \node (n12) [anchor=north east] at (-0.6, -0.8)%
               {
                 \begin{minipage}{92pt}
                       {\scriptsize Maxwell Equations - Potential form}
                       \vspace{-0.1in}
                        \begingroup\makeatletter\def\f@size{7}\check@mathfonts
                        \def\maketag@@@#1{\hbox{\m@th\large\normalfont#1}}%
                     \begin{flalign}
                       \gradsq\phi &= \aht \rho
                                                                  \nonumber \\
                       \gradsq\vA &= \aht \vJ
                                                                  \nonumber \\
                       {\rm Gauge:} \dt{\phi}+ \dive{\vA} &= 0
                                                                  \nonumber
                   \end{flalign}\endgroup
                 \end{minipage}
               };
  \node (n13) [anchor=north west] at (0.7, -0.8)%
               {
                 \begin{minipage}{92pt}
                       {\scriptsize  Integral Solutions (aka Retarded Potentials) }
                       \vspace{-0.1in}
                        \begingroup\makeatletter\def\f@size{7}\check@mathfonts
                        \def\maketag@@@#1{\hbox{\m@th\large\normalfont#1}}%
                     \begin{flalign}
   \phi&= \frac{\aht}{4\pi} \int \frac{1}{R}\lb \rho \rb \dVp
                                                                  \nonumber \\
   \vA &= \frac{\aht}{4\pi} \int \frac{1}{R}\lb \vJ \rb \dVp 
                                                                  \nonumber
                   \end{flalign}\endgroup
                 \end{minipage}
               };
  \node (n14) [anchor=south west] at (0.7, 0.9)%
               {
                 \begin{minipage}{92pt}
                       {\scriptsize  Generalized Coulomb \& Biot-Savart Eqns. 
                        }
                       \vspace{-0.1in}
                        \begingroup\makeatletter\def\f@size{7}\check@mathfonts
                        \def\maketag@@@#1{\hbox{\m@th\large\normalfont#1}}%
                     \begin{flalign}
                       \vE & = \frac{\aht}{4\pi} \int \lb 
                                    \rho \frac{\vRht}{R^2} \rb \dVp
                                   \nonumber \\
                       \vB & = \frac{\aht}{4\pi} \int
                         \lb \vJ \times \frac{\vRht}{R^2} \rb \dVp
                                                                  \nonumber
                   \end{flalign}\endgroup
                 \end{minipage}
               };

  \draw [->,ultra thick] (-2.05, 0.65) arc [radius=1.0, start angle=140, end angle= 220];
  \draw [->,ultra thick] ( 2.05,-0.65) arc [radius=1.0, start angle=-40, end angle=  40];
  \draw [->,ultra thick] ( 0.65, 2.05) arc [radius=1.0, start angle=50,  end angle= 130];
  \draw [->,ultra thick] (-0.65,-2.05) arc [radius=1.0, start angle=230, end angle= 310];

  \draw [fill=white,white] (-1.20,-0.42) rectangle (1.2,0.8);
  \node (n00) [] at (0.0, 0.29)%
               {
                 \begin{minipage}{0pt}
                   \begin{align}
                       \vE & = -\grad \phi
                                                                  \nonumber \\
                       \vB & = -\curl{\vA}
                                                                  \nonumber 
                   \end{align}
                 \end{minipage}
               };
  \draw [fill=white,white] (-0.28, 3.30) rectangle (0.28,2.38);
  \node (n00) [] at (0.0, 3.00)%
               {
                 \begin{minipage}{0pt}
                   \begin{align}
                     \dive{} 
                                                                  \nonumber \\
                     \curl{}
                                                                  \nonumber 
                   \end{align}
                 \end{minipage}
               };
  \draw [fill=white,white] (-1.00, -2.47) rectangle (0.50,-3.18);
  \node (n00) [] at (-0.59, -2.80)%
               {
                 \begin{minipage}{0pt}
                   Invert \\
                   Laplacian
                 \end{minipage}
               };

  \node (n00) [] at (-4.4, 0.40)%
               {
                 \begin{minipage}{0pt}
                   \large $\vE, \vB$
                 \end{minipage}
               };
  \node (n00) [] at (-4.4, -0.40)%
               {
                 \begin{minipage}{0pt}
                   \large $\phi, \vA$
                 \end{minipage}
               };
  \node (n980) [] at (-2.3, 4.30)%
               {
                   \bf Local
               };
  \node (n990) [] at (2.3, 4.30)%
               {
                   \bf Global
               };
  \node (n00) [] at (-2.3, 4.00)%
               {
                   \bf Differential equations
               };
  \node (n00) [] at ( 2.3, 4.00)%
               {
                   \bf Integral equations
               };

\end{tikzpicture}
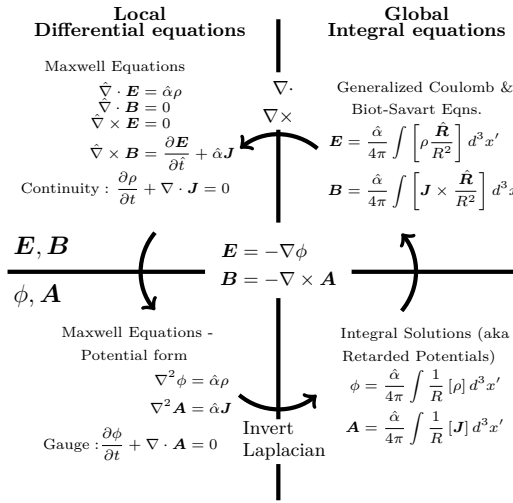
\begin{tikzpicture}[thick,scale=0.80, every node/.style={scale=0.80}]

  \node (n99) [above] at (0, 4.500)%
               { \large \bf
                   Magneto-Quasistatics (MQS)
               };

  \draw[ultra thick] (-4.5,0) -- (4.5,0);
  \draw[ultra thick] (0,-3.8) -- (0,4.2);

  \node (n11) [anchor=south east] at (-0.6, 1.0)%
               {
                 \begin{minipage}{92pt}
                       {\scriptsize Maxwell Equations}
                       \vspace{-0.1in}
                        \begingroup\makeatletter\def\f@size{7}\check@mathfonts
                        \def\maketag@@@#1{\hbox{\m@th\large\normalfont#1}}%
                     \begin{flalign}
                       \diveh{\vE} &= \aht \rho
                                                                  \nonumber \\[-2\jot]  
                       \diveh{\vB} &=0
                                                                  \nonumber \\[-2\jot]
                       \curlh{\vE} &= - \dth{\vB}
                                                                  \nonumber \\[-1\jot]
                       \curlh{\vB} &=   \aht \vJ
                                                                  \nonumber \\[-1\jot]
                       {\rm Continuity:\ } \dive{\vJ} &= 0
                                                                      \nonumber 
                   \end{flalign}\endgroup
                 \end{minipage}
               };
  \node (n12) [anchor=north east] at (-0.6, -0.8)%
               {
                 \begin{minipage}{92pt}
                       {\scriptsize Maxwell Equations - Potential form}
                       \vspace{-0.1in}
                        \begingroup\makeatletter\def\f@size{7}\check@mathfonts
                        \def\maketag@@@#1{\hbox{\m@th\large\normalfont#1}}%
                     \begin{flalign}
                       \gradsq\phi &= \aht \rho
                                                                  \nonumber \\
                       \gradsq\vA &= \aht \vJ
                                                                  \nonumber \\
                       {\rm Gauge:} \dive{\vA} &= 0
                                                                  \nonumber
                   \end{flalign}\endgroup
                 \end{minipage}
               };
  \node (n13) [anchor=north west] at (0.7, -0.8)%
               {
                 \begin{minipage}{92pt}
                       {\scriptsize  Integral Solutions (aka Retarded Potentials) }
                       \vspace{-0.1in}
                        \begingroup\makeatletter\def\f@size{7}\check@mathfonts
                        \def\maketag@@@#1{\hbox{\m@th\large\normalfont#1}}%
                     \begin{flalign}
   \phi&= \frac{\aht}{4\pi} \int \frac{1}{R}\lb \rho \rb \dVp
                                                                  \nonumber \\
   \vA &= \frac{\aht}{4\pi} \int \frac{1}{R}\lb \vJ \rb \dVp 
                                                                  \nonumber
                   \end{flalign}\endgroup
                 \end{minipage}
               };
  \node (n14) [anchor=south west] at (0.7, 0.9)%
               {
                 \begin{minipage}{92pt}
                       {\scriptsize  Generalized Coulomb \& Biot-Savart Eqns. 
                        }
                       \vspace{-0.1in}
                        \begingroup\makeatletter\def\f@size{7}\check@mathfonts
                        \def\maketag@@@#1{\hbox{\m@th\large\normalfont#1}}%
                     \begin{flalign}
                       \vE & = \frac{\aht}{4\pi} \int \lb 
                                \rho \frac{\vRht}{R^2} 
                              + \frac{\vJdt}{R} \rb \dVp.
                                                        \nonumber \\
                       \vB & = \frac{\aht}{4\pi} \int
                                \lb \vJ \times \frac{\vRht}{R^2} \rb \dVp
                                                                  \nonumber
                   \end{flalign}\endgroup
                 \end{minipage}
               };

  \draw [->,ultra thick] (-2.05, 0.65) arc [radius=1.0, start angle=140, end angle= 220];
  \draw [->,ultra thick] ( 2.05,-0.65) arc [radius=1.0, start angle=-40, end angle=  40];
  \draw [->,ultra thick] ( 0.65, 2.05) arc [radius=1.0, start angle=50,  end angle= 130];
  \draw [->,ultra thick] (-0.65,-2.05) arc [radius=1.0, start angle=230, end angle= 310];

  \draw [fill=white,white] (-1.20,-0.42) rectangle (1.2,0.8);
  \node (n00) [] at (0.0, 0.29)%
               {
                 \begin{minipage}{0pt}
                   \begin{align}
                       \vE & = -\grad \phi - \dt{\vA}
                                                                  \nonumber \\
                       \vB & = -\curl{\vA}
                                                                  \nonumber 
                   \end{align}
                 \end{minipage}
               };
  \draw [fill=white,white] (-0.28, 3.30) rectangle (0.28,2.38);
  \node (n00) [] at (0.0, 3.00)%
               {
                 \begin{minipage}{0pt}
                   \begin{align}
                     \dive{} 
                                                                  \nonumber \\
                     \curl{}
                                                                  \nonumber 
                   \end{align}
                 \end{minipage}
               };
  \draw [fill=white,white] (-1.06, -2.47) rectangle (0.50,-3.18);
  \node (n00) [] at (-0.59, -2.80)%
               {
                 \begin{minipage}{0pt}
                   Invert \\
                   Laplacian
                 \end{minipage}
               };

  \node (n00) [] at (-4.4, 0.40)%
               {
                 \begin{minipage}{0pt}
                   \large $\vE, \vB$
                 \end{minipage}
               };
  \node (n00) [] at (-4.4, -0.40)%
               {
                 \begin{minipage}{0pt}
                   \large $\phi, \vA$
                 \end{minipage}
               };
  \node (n980) [] at (-2.3, 4.30)%
               {
                   \bf Local
               };
  \node (n990) [] at (2.3, 4.30)%
               {
                   \bf Global
               };
  \node (n00) [] at (-2.3, 4.00)%
               {
                   \bf Differential equations
               };
  \node (n00) [] at ( 2.3, 4.00)%
               {
                   \bf Integral equations
               };

\end{tikzpicture}
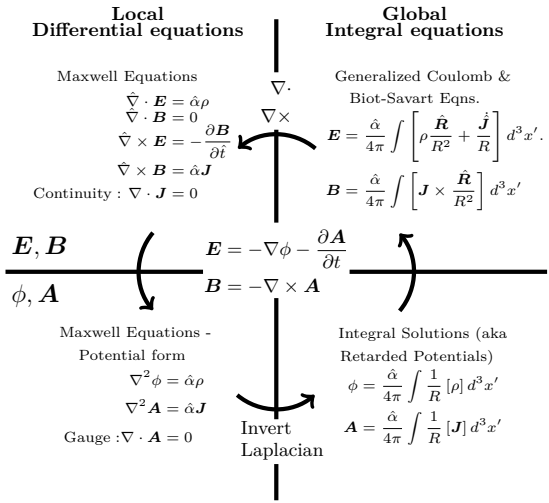
\begin{tikzpicture}[thick,scale=0.80, every node/.style={scale=0.80}]
  \node (n99) [above] at (0, 4.500)%
               { \large \bf Electro-Statics (ES) };
  \draw[ultra thick] (-4.5,0) -- (4.5,0);
  \draw[ultra thick] (0,-3.8) -- (0,4.2);

  \node (n11) [anchor=south east] at (-0.6, 1.0)%
               {
                 \begin{minipage}{92pt}
                       {\scriptsize Maxwell Equations}
                       \vspace{-0.1in}
                        \begingroup\makeatletter\def\f@size{7}\check@mathfonts
                        \def\maketag@@@#1{\hbox{\m@th\large\normalfont#1}}%
                     \begin{flalign}
                       \diveh{\vE} &= \aht \rho
                                                  \nonumber \\[-2\jot]  
                       \curlh{\vE} &= 0
                                                  \nonumber 
                   \end{flalign}\endgroup
                 \end{minipage}
               };
  \node (n12) [anchor=north east] at (-0.6, -0.8)%
               {
                 \begin{minipage}{92pt}
                       {\scriptsize Maxwell Equations - Potential form}
                       \vspace{-0.1in}
                        \begingroup\makeatletter\def\f@size{7}\check@mathfonts
                        \def\maketag@@@#1{\hbox{\m@th\large\normalfont#1}}%
                     \begin{flalign}
                       \gradsq\phi &= \aht \rho
                                                                  \nonumber \\
                   \end{flalign}\endgroup
                 \end{minipage}
               };
  \node (n13) [anchor=north west] at (0.7, -0.8)%
               {
                 \begin{minipage}{92pt}
                       {\scriptsize  Integral Solutions (aka Retarded Potentials) }
                       \vspace{-0.1in}
                        \begingroup\makeatletter\def\f@size{7}\check@mathfonts
                        \def\maketag@@@#1{\hbox{\m@th\large\normalfont#1}}%
                     \begin{flalign}
   \phi&= \frac{\aht}{4\pi} \int \frac{1}{R}\lb \rho \rb \dVp
                                                                  \nonumber
                   \end{flalign}\endgroup
                 \end{minipage}
               };
  \node (n14) [anchor=south west] at (0.7, 0.9)%
               {
                 \begin{minipage}{92pt}
                       {\scriptsize  Coulomb Eqn. 
                        }
                       \vspace{-0.1in}
                        \begingroup\makeatletter\def\f@size{7}\check@mathfonts
                        \def\maketag@@@#1{\hbox{\m@th\large\normalfont#1}}%
                     \begin{flalign}
                       \vE & = \frac{\aht}{4\pi} \int \lb 
                                    \rho \frac{\vRht}{R^2} \rb \dVp
                                                                  \nonumber
                   \end{flalign}\endgroup
                 \end{minipage}
               };

  \draw [->,ultra thick] (-2.05, 0.65) arc [radius=1.0, start angle=140, end angle= 220];
  \draw [->,ultra thick] ( 2.05,-0.65) arc [radius=1.0, start angle=-40, end angle=  40];
  \draw [->,ultra thick] ( 0.65, 2.05) arc [radius=1.0, start angle=50,  end angle= 130];
  \draw [->,ultra thick] (-0.65,-2.05) arc [radius=1.0, start angle=230, end angle= 310];

  \draw [fill=white,white] (-1.20,-0.42) rectangle (1.2,0.8);
  \node (n00) [] at (0.0, 0.29)%
               {
                 \begin{minipage}{0pt}
                   \begin{align}
                       \vE & = -\grad \phi                \nonumber 
                   \end{align}
                 \end{minipage}
               };
  \draw [fill=white,white] (-0.28, 3.30) rectangle (0.28,2.38);
  \node (n00) [] at (0.0, 3.00)%
               {
                 \begin{minipage}{0pt}
                   \begin{align}
                     \dive{} 
                                                                  \nonumber \\
                     \curl{}
                                                                  \nonumber 
                   \end{align}
                 \end{minipage}
               };
  \draw [fill=white,white] (-1.06, -2.47) rectangle (0.50,-3.18);
  \node (n00) [] at (-0.59, -2.80)%
               {
                 \begin{minipage}{0pt}
                   Invert \\
                   Laplacian
                 \end{minipage}
               };

  \node (n00) [] at (-4.4, 0.40)%
               {
                 \begin{minipage}{0pt}
                   \large $\vE, \vB$
                 \end{minipage}
               };
  \node (n00) [] at (-4.4, -0.40)%
               {
                 \begin{minipage}{0pt}
                   \large $\phi, \vA$
                 \end{minipage}
               };
  \node (n980) [] at (-2.3, 4.30)%
               {
                   \bf Local
               };
  \node (n990) [] at (2.3, 4.30)%
               {
                   \bf Global
               };
  \node (n00) [] at (-2.3, 4.00)%
               {
                   \bf Differential equations
               };
  \node (n00) [] at ( 2.3, 4.00)%
               {
                   \bf Integral equations
               };

\end{tikzpicture}
\begin{tikzpicture}[thick,scale=0.80, every node/.style={scale=0.80}]

  \node (n99) [above] at (0, 4.500)%
               { \large \bf
                   Magneto-Statics (MS)
               };

  \draw[ultra thick] (-4.5,0) -- (4.5,0);
  \draw[ultra thick] (0,-3.8) -- (0,4.2);

  \node (n11) [anchor=south east] at (-0.6, 1.0)%
               {
                 \begin{minipage}{92pt}
                       {\scriptsize Maxwell Equations}
                       \vspace{-0.1in}
                        \begingroup\makeatletter\def\f@size{7}\check@mathfonts
                        \def\maketag@@@#1{\hbox{\m@th\large\normalfont#1}}%
                     \begin{flalign}
                       \diveh{\vB} &=0
                                                                  \nonumber \\[-2\jot]
                       \curlh{\vB} &=   \aht \vJ
                                                                      \nonumber 
                   \end{flalign}\endgroup
                 \end{minipage}
               };
  \node (n12) [anchor=north east] at (-0.6, -0.8)%
               {
                 \begin{minipage}{92pt}
                       {\scriptsize Maxwell Equations - Potential form}
                       \vspace{-0.1in}
                        \begingroup\makeatletter\def\f@size{7}\check@mathfonts
                        \def\maketag@@@#1{\hbox{\m@th\large\normalfont#1}}%
                     \begin{flalign}
                       \gradsq\vA &= \aht \vJ
                                                                  \nonumber \\
                       {\rm Gauge:} \dive{\vA} &= 0
                                                                  \nonumber
                   \end{flalign}\endgroup
                 \end{minipage}
               };
  \node (n13) [anchor=north west] at (0.7, -0.8)%
               {
                 \begin{minipage}{92pt}
                       {\scriptsize  Integral Solutions (aka Retarded Potentials) }
                       \vspace{-0.1in}
                        \begingroup\makeatletter\def\f@size{7}\check@mathfonts
                        \def\maketag@@@#1{\hbox{\m@th\large\normalfont#1}}%
                     \begin{flalign}
   \vA &= \frac{\aht}{4\pi} \int \frac{1}{R}\lb \vJ \rb \dVp 
                                                                  \nonumber
                   \end{flalign}\endgroup
                 \end{minipage}
               };
  \node (n14) [anchor=south west] at (0.7, 0.9)%
               {
                 \begin{minipage}{92pt}
                       {\scriptsize  Biot-Savart Eqn.
                        }
                       \vspace{-0.1in}
                        \begingroup\makeatletter\def\f@size{7}\check@mathfonts
                        \def\maketag@@@#1{\hbox{\m@th\large\normalfont#1}}%
                     \begin{flalign}
                       \vB & = \frac{\aht}{4\pi} \int
                         \lb \vJ \times \frac{\vRht}{R^2} \rb \dVp
                                                                  \nonumber
                   \end{flalign}\endgroup
                 \end{minipage}
               };

  \draw [->,ultra thick] (-2.05, 0.65) arc [radius=1.0, start angle=140, end angle= 220];
  \draw [->,ultra thick] ( 2.05,-0.65) arc [radius=1.0, start angle=-40, end angle=  40];
  \draw [->,ultra thick] ( 0.65, 2.05) arc [radius=1.0, start angle=50,  end angle= 130];
  \draw [->,ultra thick] (-0.65,-2.05) arc [radius=1.0, start angle=230, end angle= 310];

  \draw [fill=white,white] (-1.20,-0.42) rectangle (1.2,0.8);
  \node (n00) [] at (0.0, 0.29)%
               {
                 \begin{minipage}{0pt}
                   \begin{align}
                       \vB & = -\curl{\vA}
                                                                  \nonumber 
                   \end{align}
                 \end{minipage}
               };
  \draw [fill=white,white] (-0.28, 3.30) rectangle (0.28,2.38);
  \node (n00) [] at (0.0, 3.00)%
               {
                 \begin{minipage}{0pt}
                   \begin{align}
                     \dive{} 
                                                                  \nonumber \\
                     \curl{}
                                                                  \nonumber 
                   \end{align}
                 \end{minipage}
               };
  \draw [fill=white,white] (-1.06, -2.47) rectangle (0.50,-3.18);
  \node (n00) [] at (-0.59, -2.80)%
               {
                 \begin{minipage}{0pt}
                   Invert \\
                   Laplacian
                 \end{minipage}
               };

  \node (n00) [] at (-4.4, 0.40)%
               {
                 \begin{minipage}{0pt}
                   \large $\vE, \vB$
                 \end{minipage}
               };
  \node (n00) [] at (-4.4, -0.40)%
               {
                 \begin{minipage}{0pt}
                   \large $\phi, \vA$
                 \end{minipage}
               };
  \node (n980) [] at (-2.3, 4.30)%
               {
                   \bf Local
               };
  \node (n990) [] at (2.3, 4.30)%
               {
                   \bf Global
               };
  \node (n00) [] at (-2.3, 4.00)%
               {
                   \bf Differential equations
               };
  \node (n00) [] at ( 2.3, 4.00)%
               {
                   \bf Integral equations
               };

\end{tikzpicture}
\label{EMsummary}
\end{figure*}

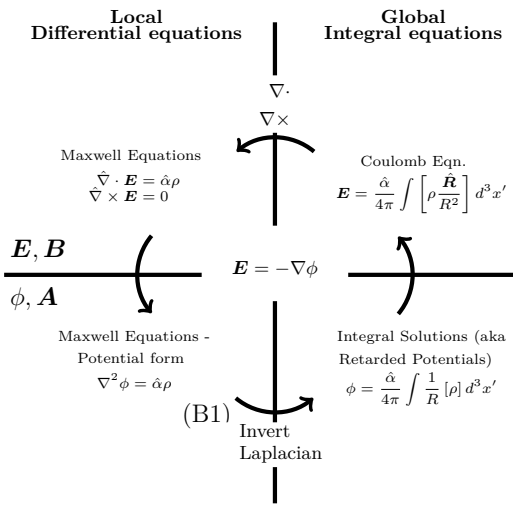
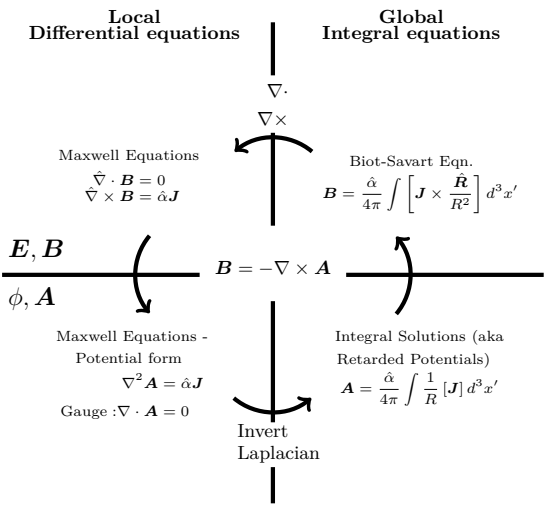
\end{document}